\documentclass[twocolumn,showpacs,amsfonts,aps,prc,nofootinbib,floatfix]{revtex4}

\usepackage{bm}
\usepackage{graphicx}
\usepackage{amsmath}

\def\la{\langle}
\def\ra{\rangle}

\def\ecc{{\varepsilon}}
\def\epart{{\ecc_{\mathrm{part}}}}

\usepackage{color}

\newcommand{\eq}{{\,=\,}}

\begin{document}


\title{Event-by-event shape and flow fluctuations of relativistic
heavy-ion collision fireballs}

\author{Zhi Qiu}
\author{Ulrich Heinz}
\affiliation{Department of Physics, The Ohio State University,
  Columbus, OH 43210-1117, USA}

\begin{abstract}
Heavy-ion collisions create deformed quark-gluon plasma (QGP) fireballs
which explode anisotropically. The viscosity of the fireball matter
determines its ability to convert the initial spatial deformation into
momentum anisotropies that can be measured in the final hadron spectra. A
quantitatively precise empirical extraction of the QGP viscosity thus
requires a good understanding of the initial fireball deformation. This
deformation fluctuates from event to event, and so does the finally
observed momentum anisotropy. We present a harmonic decomposition of the 
initial fluctuations in shape and orientation of the fireball and perform 
event-by-event (2+1)-dimensional ideal fluid dynamical simulations to 
extract the resulting fluctuations in the magnitude and direction of the 
corresponding harmonic components of the final anisotropic flow 
at midrapidity. The final harmonic flow coefficients are found to 
depend non-linearly on the initial harmonic eccentricity coefficients. We 
show that, on average, initial density fluctuations suppress the buildup 
of elliptic flow relative to what one obtains from a smooth initial profile 
of the same eccentricity, and discuss implications for the phenomenological 
extraction of the QGP shear viscosity from experimental elliptic flow data.  
\end{abstract}


\date{\today}

\maketitle

\section{Introduction}
\label{sec1}

In ultrarelativistic heavy-ion collision experiments, a fraction of the
incoming kinetic energy is converted into new matter deposited in the 
collision zone. The distribution of this matter in the plane transverse to 
the colliding beams is inhomogeneous and fluctuates from collision to 
collision. At the collision energies available at the Relativistic Heavy 
Ion Collider (RHIC) and the Large Hadron Collider (LHC), the produced
matter is sufficiently dense and strongly interacting that it quickly 
reaches a state of approximate local thermal equilibrium. Its subsequent 
evolution can thus be described by fluid dynamics until it eventually
becomes too dilute and breaks apart. Hydrodynamic forces (i.e. pressure 
gradients) convert the inhomogeneities and deformations of its initial 
spatial density distribution into anisotropies of the final hydrodynamic
flow. The latter can be extracted from the momentum distributions of
the finally emitted particles. The efficiency with which the geometric 
deformation and fluctuating inhomogeneities in the initial density
distribution are converted into final flow anisotropies is controlled
by the viscosity of the expanding fluid. For a given source deformation,
ideal fluid dynamics generates the largest flow anisotropy; it corresponds
to the limit of zero mean free path and instantaneous thermalization,
which allows for the largest possible collective response, via final 
state interactions, to irregularities in the geometric structure of the 
fireball \cite{Heinz:2001xi}. Viscosity accounts for finite interaction 
cross sections and non-zero mean free paths which reduce the amount of 
flow anisotropy that can be generated from a given geometric deformation.
In hydrodynamic language, viscous pressure components inhibit the 
development of flow anisotropies and tend to smoothen irregularities
in the flow distribution. By measuring the flow anisotropies and 
relating them to the initial geometric deformations (as calculated from
theoretical models for the collision geometry) one can, in principle,
determine the fluid's viscosity experimentally \cite{Romatschke:2007mq,%
Song:2008si,Lacey:2010fe,Song:2010mg,Song:2011hk}.

Until recently, most of the attention has been focussed on elliptic flow
$v_2\eq\left\la\cos(2\phi_p)\right\ra$ and its relation to the 
spatial eccentricity $\varepsilon_x\eq\frac{\la y^2{-}x^2\ra}
{\la y^2{+}x^2\ra}\eq\la r^2\cos(2\phi_s)\ra/\la r^2\ra$
(more precise definitions will be given in Sec.~\ref{sec2}) 
\cite{Romatschke:2007mq,Song:2008si,Lacey:2010fe,Song:2010mg,Song:2011hk,%
Teaney:2003kp,Song:2007fn,Song:2008hj,Hirano:2005xf,Hirano:2009ah,%
Hirano:2010jg}. Event-by-event fluctuations 
in the initial state were only treated on average, by taking into account
their effects on the average eccentricity of an ensemble of events 
\cite{Song:2010mg,Song:2011hk,Hirano:2009ah,Hirano:2010jg,Miller:2003kd} and
then propagating a smooth initial density profile corresponding to that 
ensemble average hydrodynamically. In this way one can only compute the 
average elliptic flow, but not its fluctuations from event to event
\cite{Alver:2007zz,Sorensen:2008zk,Ollitrault:2009ie}. If one assumes
that the elliptic flow is linearly proportional to the initial eccentricity,
one can distribute the elliptic flow fluctuations around this average
$v_2$ value in the same way as the initial eccentricity fluctuates around
its average \cite{Bhalerao:2006tp} and use this to predict flow
fluctuations from eccentricity fluctuations \cite{Drescher:2007ax}.
This ignores, however, the recently discovered \cite{Alver:2010gr} fact
that initial-state shape fluctuations of the collision region lead to
event-by-event fluctuations not only of the elliptic deformation 
$\varepsilon_x$, but simultaneously of all higher-order harmonic 
eccentricity coefficients \cite{Qin:2010pf,Lacey:2010hw}, and that the 
simultaneous presence of several harmonic eccentricity coefficients can 
lead to hydrodynamic cross-talk between anisotropic flows of different 
harmonic order \cite{Qin:2010pf}. Without the possibility to hydrodynamically 
evolve fluctuating initial conditions event-by-event, the assumption of a 
linear dependence of $v_2$ on $\varepsilon_x$ can thus not be rigorously 
tested. 

Experimentally, anisotropic flow coefficients $v_n$ are measured by
analyzing multiparticle correlations in azimuthal angle around the
beam axis \cite{Voloshin:2008dg}. Event-by-event flow fluctuations
and additional non-flow correlations influence the elliptic flow $v_2$
derived from different such measures in different ways 
\cite{Voloshin:2007pc,Alver:2008zza,Voloshin:2008dg,Yi:2011hs} and
affect the magnitude of the extracted flow. This has serious implications
for the determination of the quark-gluon plasma shear viscosity from 
elliptic flow data \cite{Song:2010mg,Song:2011hk}: quantitatively trustworthy 
results require a detailed understanding of the spectrum of fluctuations of 
the anisotropic flow coefficients in the experimentally observed final state 
and its relation to the fluctuations of the corresponding moments of the 
initial eccentricity distributions that are believed to drive the flows 
measured by various different methods. In addition, widely held beliefs
as to which moments of the initial eccentricity distribution are directly
related to which moments of the final flow distribution 
\cite{Bhalerao:2006tp,Alver:2010dn,Lacey:2010hw} must be tested by generating 
the final states hydrodyamically event by event, i.e. separately for each 
fluctuating initial condition \cite{Andrade:2006yh,Petersen:2010md,Qin:2010pf,%
Holopainen:2010gz,Schenke:2010rr,Chatterjee:2011dw}. These are the goals 
addressed in the present article. (For related work using non-hydrodynamic 
models for the transfer of initial eccentricities into final flows see 
\cite{Sorensen:2010zq}.)

To generate fluctuating initial conditions we use Monte Carlo versions of
the Glauber \cite{Miller:2007ri} and {\tt fKLN} \cite{KLN,Drescher:2006ca} 
models, in the implementation by Hirano and Nara \cite{Hirano:2009ah,%
Hirano:2010jg}. After defining and compiling in Sec.~\ref{sec2} several 
different definitions of initial-state eccentricities and final-state 
harmonic flow coefficients, we explore in Sec.~\ref{sec3} the centrality 
dependence of the average ellipticity, triangularity and a few higher-order 
harmonic coefficients in their various incarnations found in the literature.
In Sec.~\ref{sec4} we use ideal event-by-event hydrodynamics\footnote{For
  convenience we here use the longitudinally boost-invariant 
  (2+1)-dimensional viscous hydrodynamic code {\tt VISH2+1} 
  \cite{Song:2007fn} for zero viscosity, restricting our analysis
  to anisotropic transverse flow at midrapidity. This allows for easy 
  later inclusion of viscous effects. A (3+1)-d viscous hydrodynamic code
  has recently become available \cite{Schenke:2010rr}, and a comparison
  between (2+1)-d and (3+1)-d viscous evolution near midrapidity is
  in progress \cite{SSH}.}
to analyze the correlations between final state flow anisotropies and their
associated flow angles with the initial state eccentricity coefficients and 
their associated angles. We identify strong cross-talk between coefficients
of different harmonic order, especially in peripheral collisions with strong
elliptic flow. In Section~\ref{sec5} we compare the conversion efficiency
of initial-state ellipticity and triangularity into final-state elliptic 
and triangular flow in single-shot and event-by-event hydrodynamics. We 
find significant differences and discuss their implications. We summarize 
our results in Section~\ref{sec6} and discuss different radial weights for 
the eccentricity definitions in the Appendix.

\section{Definitions}
\label{sec2}

In this section we discuss different definitions for the harmonic flow 
and eccentricity coefficients and briefly describe the models used in 
computing the initial entropy and energy density profiles whose
 eccentricities are evaluated in Sec.~\ref{sec3} and which are evolved 
hydrodynamically in Sections~\ref{sec4} and \ref{sec5}.

\subsection{Ellipticity}
\label{sec2a}

Usually known simply as ``eccentricity'', we define the ``ellipticity''
$\ecc\equiv\ecc_2$ of a given matter distribution in the transverse 
$(x,y)$ plane in terms of its $r^2$-weighted second azimuthal moment 
\cite{Alver:2010gr,Alver:2010dn},
\begin{equation} 
\label{eq1}
  \ecc_2\, e^{i 2\psi_2^\mathrm{PP}} 
  = -\frac{\int dx\,dy\, r^2 e^{i 2\phi}\,e(x,y)}
          {\int dx\,dy\, r^2\,e(x,y)},
\end{equation}
where $x\eq{r}\cos\phi$, $y\eq{r}\sin\phi$. This formula assumes that the 
origin is the center of the distribution $e(x,y)$. In a Monte Carlo approach 
for generating the initial distribution $e(x,y)$ (see Sec.~\ref{sec2d}) 
this must be ensured by recentering each event before using Eq.~(\ref{eq1}). 
By default we characterize in Eq.~(\ref{eq1}) the matter distribution by 
its energy density $e(x,y)$ \cite{Kolb:2003dz}. Since some authors (e.g. 
\cite{Hirano:2009ah,Hirano:2010jg}) prefer defining the source ellipticity 
in terms of its entropy density distribution $s(x,y)$, we compare in 
Sec.~\ref{sec3b} energy- and entropy-weighted ellipticities.

In Eq.~(\ref{eq1}), $x$ and $y$ are ``reaction plane'' (RP) coordinates:
The reaction plane is the $(x,z)$ plane, with $z$ pointing along the beam 
and $x$ pointing along the direction of the impact parameter $\bm{b}$ 
between the colliding nuclei. $y$ is perpendicular to the reaction plane.
Because of the minus sign on the r.h.s. of Eq.~(\ref{eq1}), the angle 
$\psi_2$ on the l.h.s. of Eq.~(\ref{eq1}) points in the direction of the 
minor axis of the corresponding ellipse. For an elliptically deformed
Gaussian density distribution, this is the direction of the largest 
density gradient and thus of the largest hydrodynamic acceleration and also
of the finally observed elliptic flow. The direction of this minor axis
defines, together with the beam direction $z$, the ``participant plane''
(PP). It is tilted relative to the reaction plane by $\psi_2^\mathrm{PP}$. 
The label ``participant'' is motivated by the fact that the initial energy 
and entropy density distributions of the collision fireball reflect (more 
or less directly, depending on the model for secondary particle creation) 
the transverse distribution of the nucleons participating in the particle 
production process. The ellipticity $\ecc_2$ in Eq.~(\ref{eq1}) is 
correspondingly called {\em ``participant eccentricity''} and also denoted as
$\epart$.\footnote{Traditionally $\epart$ is defined in terms of the 
  transverse density of wounded nucleons, but since what matters for the 
  subsequent hydrodynamic evolution is not the distribution of wounded 
  nucleons themselves but of the matter generated by the wounded nucleons, 
  we use the name $\epart$ for the ellipticity characterizing the 
  thermalized matter.}  
It can be written as
\begin{eqnarray}
\label{eq2}
  \epart &\equiv& \ecc_2 = |\ecc_2\, e^{i 2 \psi_2^\mathrm{PP}}| 
\nonumber
\\
  &=& \frac{\sqrt{\{r^2\cos(2\phi)\} + \{r^2\sin(2\phi)\}}}
                 {\{r^2\}} 
\\\nonumber
  &=& \frac{\sqrt{\{y^2{-}x^2\}^2 + 4{\{xy\}^2}}}
           {\{y^2{+}x^2\}}.
\end{eqnarray}
Here $\{\dots\}\eq\int dx\,dy\,(\dots)\,e(x,y)$ defines the ``event average''
over the matter distribution $e(x,y)$ in a single collision event
\cite{Alver:2008zza}. Equivalently, the participant eccentricity can be 
written as 
\begin{equation}
\label{eq3}
 \ecc_\mathrm{part} = \frac{\{y^2{-}x^2\}'}{\{y^2{+}x^2\}'}
\end{equation}
where $\{\dots\}'\eq\int dx\,dy\,(\dots)\,e'(x,y)$ indicates the average
over a rotated event with energy density $e'(x,y)=e\bigl(x
\cos\psi_2^\mathrm{PP}{-}y\sin\psi_2^\mathrm{PP},x\sin\psi_2^\mathrm{PP}{+}y
\cos\psi_2^\mathrm{PP}\bigr)$ whose minor and major axes now align with $x$ 
and $y$. 

The event-average $\{\dots\}$ is to be distinguished from the the ``ensemble
average'' $\la\dots\ra\eq\frac{1}{N}\sum_{n\eq1}^N \{\dots\}_n$ where $N$ 
is the total number of events and $\{\dots\}_n$ is the event-average over 
the energy density $e_n(x,y)$ in event number $n$. The {\em average 
participant eccentricity} is thus defined as
\begin{equation}
\label{eq4}
  \la\epart\ra = \frac{1}{N}\sum_{n\eq1}^N (\epart)_n.
\end{equation}
This differs from the {\em mean eccentricity} $\bar{\ecc}_\mathrm{part}$
of the average (recentered and rotated by $\psi_2^\mathrm{PP}$) 
energy density $\bar{e}'(x,y)\eq\frac{1}{N}\sum_{n\eq1}^N 
e'_n(x,y)$ which can be written in the following equivalent ways: 
\begin{equation}
\label{eq5}
 \bar{\ecc}_\mathrm{part} = 
 \frac{\sqrt{\la\{y^2{-}x^2\}\ra^2 + 4\la\{xy\}\ra^2}}
            {\la\{y^2{+}x^2\}\ra} = 
 \frac{\la\{y^2{-}x^2\}'\ra}{\la\{y^2{+}x^2\}'\ra}.
\end{equation}
In contrast to (\ref{eq4}), one here ensemble-averages over numerator 
and denominator separately before forming the ratio. 

Nature performs heavy-ion collisions event by event, and hydrodynamic forces
generate in each event an elliptic component $v_2$ of the anisotropic flow
which is causally related to the specific initial ellipticity $\epart$ in
that event. Theorists often do not compute the hydrodynamic evolution of
the collision fireball event by event, but approximate Nature's procedure
by generating from a superposition of many fluctuating initial conditions 
a single smooth initial distribution $\bar{e}(x,y)$ which they then evolve 
hydrodynamically in a ``single shot'', extracting the {\em mean elliptic 
flow} $\bar{v}_2$ corresponding to the {\em mean eccentricity} 
$\bar{\ecc}_\mathrm{part}$ of that averaged source distribution. Obviously, 
$\bar{v}_2$ is a deterministic consequence of $\bar{\ecc}_\mathrm{part}$
and does not fluctuate at all; it can not be measured experimentally. What 
can (at least in principle, although not easily in practice) be measured 
experimentally \cite{Ollitrault:2009ie} is the {\em average elliptic flow} 
$\la v_2\ra$ of a large ensemble of collision events. This observable
is conceptually more closely related to $\la\epart\ra$ than to
$\bar{\ecc}_\mathrm{part}$; for an exactly linear hydrodynamic response
$v_2\sim\epart$, one has 
$\la v_2\ra/\la\epart\ra\eq\bar{v}_2/\bar{\ecc}_\mathrm{part}$ 
\cite{Bhalerao:2006tp}. We will explore the differences between 
$\bar{\ecc}_\mathrm{part}$ and $\la\epart\ra$ and discuss consequences
for the theoretically computed $\bar{v}_2$ as opposed to the measured
\cite{Ollitrault:2009ie} $\la v_2\ra$ in Secs.~\ref{sec3a} and \ref{sec5}.

In addition to these ``participant eccentricities'' one can also define
``reaction plane eccentricities''. For a single event, the {\em reaction 
plane eccentricity} $\ecc_\mathrm{RP}$ is defined by
\begin{equation}
\label{eq6}
 \ecc_\mathrm{RP} = \frac{\{y^2{-}x^2\}}{\{y^2{+}x^2\}}
\end{equation}
in terms of an event-average over the (properly centered) energy density
$e(x,y)$. The so-called {\em standard eccentricity} is defined as the
analogous ratio of expectation values taken with a smooth average
energy density $\bar{e}(x,y)\eq\frac{1}{N}\sum_{n\eq1}^N 
e_n(x,y)$ obtained by superimposing many events {\em without} rotating 
them from the participant to the reaction plane:
\begin{equation}
\label{eq7}
 \ecc_s \equiv \bar{\ecc}_\mathrm{RP} 
  = \frac{\la\{y^2{-}x^2\}\ra}{\la\{y^2{+}x^2\}\ra}.
\end{equation}
In other words, the standard eccentricity is the {\em mean reaction plane
eccentricity}. In contrast, the {\em average reaction plane eccentricity}
is defined by
\begin{equation}
\label{eq8}
 \la\ecc_\mathrm{RP}\ra 
  = \left\la\frac{\{y^2{-}x^2\}}{\{y^2{+}x^2\}}\right\ra.
\end{equation}

Contrary to what the reader may have been led to believe by our remarks 
above, experiments do {\em not} directly measure the average elliptic flow
$\la v_2\ra$ (which for linear $v_2\sim\epart$ would be directly related
to the average participant eccentricity $\la\epart\ra$ (\ref{eq3})).
Instead they measure quantities such as $v_2\{\mathrm{EP}\}$, $v_2\{2\}$,
and $v_2\{4\}$ that, even if so-called non-flow contributions could be
completely ignored, are affected by event-by-event $v_2$-fluctuations
and thus differ from $\la v_2\ra$. $\la v_2\ra$ can be reconstructed
from the experimental measurements with some additional assumptions 
\cite{Ollitrault:2009ie} which on the surface look harmless but should be
further tested. Motivated by the hypothesis of linear hydrodynamic response,
$v_2\sim\epart$, these $v_2$ measures motivate the definition of 
corresponding ellipticity measures \cite{Bhalerao:2006tp}, the so-called
2$^\mathrm{nd}$ and 4$^\mathrm{th}$ order cumulants:
\begin{equation} 
\label{eq9}
  \ecc\{2\} = \sqrt{\la \ecc_{\mathrm{part}}^2 \ra}
\end{equation}
and
\begin{equation} 
\label{eq10}
  \ecc\{4\} = \left(\la\ecc^2_\mathrm{part}\ra^2 
  - (\la\ecc^4_\mathrm{part}\ra{-}\la\ecc^2_\mathrm{part}\ra^2)\right)^{1/4}.
\end{equation}
Note that the last expression involves the difference of two positive
definite quantities which itself does not need to be positive definite.
If fluctuations get large, the expression under the fourth root can become
negative, leaving $\ecc\{4\}$ undefined. We will see that this can happen 
in the most central and the most peripheral centrality bins.

It was shown in \cite{Voloshin:2007pc} that in the MC-Glauber model
the real and imaginary parts of the complex ellipticity defined by 
Eq.~(\ref{eq1}), with the wounded nucleon density as weight function on 
the r.h.s., both have approximately Gaussian fluctuations, with equal
widths $\sigma_\ecc$. If this is the case, the magnitude $\ecc_2$ of
this ellipticity exhibits fluctuations of Bessel-Gaussian
type\footnote{This 
    takes into account that $\ecc_2$ can never fluctuate to negative 
    values.} 
\cite{Voloshin:1994mz}, leading to the identity \cite{Voloshin:2007pc} 
\begin{equation}
\label{eq11}
  \ecc\{4\} = \la\ecc_\mathrm{RP}\ra.
\end{equation}
For sufficiently large average ellipticities $\la\ecc_2\ra$ (i.e. 
sufficiently large impact parameters) one may hope to be able to ignore
the restriction that $\ecc_2$ can never fluctuate to negative values, 
and correspondingly assume the $\ecc_2$ exhibits Gaussian (instead of
Bessel-Gaussian) fluctuations. In this case one has \cite{Voloshin:2007pc}
\begin{eqnarray}
\label{eq11a}
\nonumber
  \ecc\{2\}^2 &=& \la\epart\ra^2 +\sigma^2_\ecc,
\\
  \ecc\{4\}^2 &=& \sqrt{(\la\epart\ra^2 - \sigma^2_\ecc)^2-2\sigma_\ecc^4},
\end{eqnarray}
from which it follows that $\la\ecc_2\eq\ecc_\mathrm{part}\ra^4$ is the 
arithmetic mean of $\ecc\{2\}^4$ and $\ecc\{4\}^4$:
\begin{equation}
\label{eq11b}
  \frac{\ecc\{2\}^4{+}\ecc\{4\}^4}{2\la\ecc_\mathrm{part}\ra^4}=1. 
\end{equation}
We will use Eqs.~(\ref{eq11}) and (\ref{eq11b}) (which hold irrespective 
of the fluctuation width $\sigma_\ecc$) in Sec.~\ref{sec3a}, and their 
analogues for the elliptic flow $v_2$ in Sec.~\ref{sec5c}, to test the 
assumptions of Bessel-Gaussian and Gaussian fluctuations of the 
event-by-event ellipticity and elliptic flow fluctuations in the 
Monte Carlo Glauber (MC-Glauber) and Monte Carlo fKLN (MC-KLN) models.

If the hydrodynamic response were indeed linear, $v_2\sim\epart$, and 
non-flow effects could be ignored, the following identities would hold: 
\begin{equation} 
\label{eq12}
  \frac{\la v_2\ra}{\la\epart\ra} = \frac{\bar{v}_2}{\bar{\ecc}_\mathrm{part}}
  = \frac{v_2\{2\}}{\ecc\{2\}} = \frac{v_2\{4\}}{\ecc\{4\}}.
\end{equation}
To test these theoretically one needs event-by-event hydrodynamics
which is the only possibility to properly account for event-by-event 
flow fluctuations. In the past, event-by-event hydrodynamical evolution 
of fluctuating initial conditions has been technologically out of reach.
Comparisons between theory and experiment have been based on ``single-shot 
hydrodynamic evolution'' which propagates a smooth initial profile obtained 
by either using the so-called optical versions of the Glauber and fKLN models
or averaging over many fluctuating initial profiles from their Monte Carlo
versions (MC-Glauber and MC-KLN, respectively). Assuming linear
hydrodynamic response, one can still compare the theoretically computed
$\la v_2\ra$ with the experimentally measured $v_2\{2\}$ or $v_2\{4\}$
if one normalizes the former by $\la\ecc_\mathrm{part}\ra$ and the latter 
by $\ecc\{2\}$ or $\ecc\{4\}$, respectively, calculated {\em from the same 
initial state model} \cite{Song:2010mg,Song:2011hk}. In this context the 
identity $\ecc\{4\}\eq\la\ecc_\mathrm{RP}\ra$ (which holds if the 
ellipticity fluctuations are Gaussian) becomes particularly useful 
because it suggests that the measured $v_2\{4\}$ can be directly
compared with a single-shot hydrodynamic $v_2$ obtained from a smooth 
reaction-plane averaged initial density of ellipticity 
$\la\ecc_\mathrm{RP}\ra$, without any corrections for flow fluctuations. 
Even better, $v_2\{4\}$ can be shown to be completely free of two-particle 
non-flow contributions \cite{Bhalerao:2006tp,Voloshin:2007pc}. These
arguments have been used in \cite{Hirano:2010jg} and provide a strong 
motivation for us to test the underlying assumptions (Gaussian ellipticity 
fluctuations and linear hydrodynamic elliptic flow response) in the 
present work.

We close this subsection by recalling the expression for the participant 
plane angle of a given event (see e.g. \cite{Alver:2008zza})
\begin{equation}
\label{eq13}
  \psi_2^\mathrm{PP}= \frac{1}{2}
  \tan^{-1}\left(\frac{2\{xy\}}{\{y^2{-}x^2\}}\right)
\end{equation}
and for its transverse area
\begin{equation}
\label{eq14}
   S = \pi \sqrt{\{x^2\}'\{y^2\}'}.
\end{equation}
Both expressions assume that the events are properly centered at the origin.

\subsection{Higher order eccentricity coefficients}
\label{sec2b}

The definition (\ref{eq1}) can be generalized to higher harmonic 
eccentricity coefficients \cite{Alver:2010gr,Alver:2010dn}:
\begin{equation} 
\label{eq15}
 \ecc_n\, e^{i n\psi_n^\mathrm{PP}} 
  = -\frac{\int dx\,dy\, r^2 e^{i n\phi}\,e(x,y)}
          {\int dx\,dy\, r^2\,e(x,y)}.
\end{equation}
Alternatively one can use $r^n$ instead of $r^2$ as radial weight 
on the right hand side \cite{Qin:2010pf}:
\begin{equation} 
\label{eq16}
 \ecc'_n\, e^{i n\psi_n^\mathrm{'PP}} 
  = -\frac{\int dx\,dy\, r^n e^{i n\phi}\,e(x,y)}
          {\int dx\,dy\, r^n\,e(x,y)}.
\end{equation}
Still another variant uses the entropy density $s(x,y)$ instead of
the energy density $e(x,y)$ as weight function:
\begin{eqnarray}
\label{eq17}
 \ecc_n(s)\, e^{i n\psi_n^\mathrm{PP}(s)} 
  &=& -\frac{\int dx\,dy\, r^2 e^{i n\phi}\,s(x,y)}
            {\int dx\,dy\, r^2\,s(x,y)},
\\
\label{eq18}
 \ecc'_n(s)\, e^{i n\psi_n^\mathrm{'PP}(s)} 
  &=& -\frac{\int dx\,dy\, r^n e^{i n\phi}\,s(x,y)}
            {\int dx\,dy\, r^n\,s(x,y)}.
\end{eqnarray}
We note that the $r^2$-weighted eccentricity coefficients $\ecc_n$ fall 
off faster with increasing harmonic order $n$ than the $r^n$-weighted 
eccentricities $\ecc'_n$ (see Appendix). Also, as in (\ref{eq1}), the minus 
sign in Eqs.~(\ref{eq15})-(\ref{eq18}) guarantees that, for a Gaussian density 
distribution that has only $n^\mathrm{th}$-order eccentricity $\ecc_n$, 
the angle $\psi_n^\mathrm{PP}$ points in the direction of the steepest 
density gradient, and thus in the direction of the corresponding 
hydrodynamically generated $n^\mathrm{th}$-order harmonic flow $v_n$ 
(see next subsection). It can be written as $-1\eq{e}^{-in(\pi/n)}$ and 
amounts to a rotation of $\psi_n^\mathrm{PP}$ by $\pi/n$. For example, 
if the profile is square-shaped, $\psi_4^\mathrm{PP}$ points to the 
sides instead of its corners.

A complete characterization of the fluctuating initial density profile, 
that captures all aspects of the location of ``hot-spots'' and their
gradients, uses an expansion of the initial (energy or entropy) density 
profile in terms of cumulants \cite{Teaney:2010vd}. We will postpone their
discussion to a future analysis.

As stated before, we will use the energy density as the default weight 
function; in cases of possible ambiguity, we will use the notations 
$\ecc_n(e)$, $\ecc_n(s)$ etc. to distinguish between energy and entropy 
density weighted eccentricity coefficients and angles. Eccentricities 
$\ecc$ without harmonic index $n$ denote ellipicities (i.e. in the 
absence of $n$, $n\eq2$ is implied).

The coefficients $\ecc_n$ and angles $\psi_n^\mathrm{PP}$ define the 
eccentricies and angles of the matter distribution in the participant plane. 
We note that the participant plane angles $\psi_n^\mathrm{PP}$ associated 
with eccentricity coefficients of different harmonic order $n$ do not, in 
general, agree (see Sec.~\ref{sec4a}). We will not study higher harmonic 
generalizations of the reaction-plane ellipticity (\ref{eq6}).

\subsection{Harmonic flow coefficients}
\label{sec2c}

We charaterize the finally observed momentum distribution $dN/(dy\,p_T
dp_T\,d\phi_p)$ by ``harmonic flow coefficients'' constructed in analogy
to Eq.~(\ref{eq15}), but without the extra minus sign:
\begin{eqnarray} 
\label{eq19}
  &&\!\!\!\!\!\!
  v_n(y,p_T)\, e^{i n \psi_n^\mathrm{EP}(y,p_T)} = 
  \frac{\int d\phi_p\, e^{i n\phi_p}\,\frac{dN}{dy\,p_Tdp_T\,d\phi_p}}
       {\frac{dN}{dy\,p_Tdp_T}},\quad
\\
\label{eq20}
  &&\!\!\!\!\!\!
  v_n(y)\, e^{i n \psi_n^\mathrm{EP}(y)} = 
  \frac{\int p_Tdp_T\,d\phi_p\, e^{i n\phi_p}\,\frac{dN}{dy\,p_Tdp_T\,d\phi_p}}
       {\frac{dN}{dy}}.\quad
\end{eqnarray}
In boost-invariant hydrodynamics they are rapidity-inde\-pen\-dent, so 
we drop the argument $y$ and keep in mind that we should only compare
with midrapidity data at $y\eq0$ where the assumption of boost-invariant 
longitudinal expansion is most justified. The spectra $\frac{dN}{dy\,p_T
dp_T\,d\phi_p}$ are computed from the hydrodynamic output with the
Cooper-Frye prescription \cite{Cooper:1974mv} along an isothermal 
kinetic decoupling surface of temperature $T_\mathrm{dec}\eq140$\,MeV. 
Equation~(\ref{eq19}) defines the $p_T$-differential harmonic flow 
$v_n(p_T)$ and flow angle $\psi_n^\mathrm{EP}(p_T)$, whereas 
Eq.~(\ref{eq20}) gives their $p_T$-integrated values $v_n$ and 
$\psi_n^\mathrm{EP}$. The orientation of the final momentum distribution 
defines the ``event plane'', indicated by superscript EP. Again, different 
harmonic flows are usually associated with differently oriented event 
planes. The first three harmonic flow coefficients are the directed flow 
($v_1$), elliptic flow ($v_2$), and triangular flow ($v_3$).

\subsection{Initial-state models}
\label{sec2d}

We use Monte Carlo versions \cite{Hirano:2009ah,Hirano:2010jg} of the 
Glauber \cite{Miller:2007ri} and {\tt fKLN} \cite{Drescher:2006ca} models 
to generate fluctuating initial conditions for the entropy density in 
$200\,A$\,GeV Au+Au collisions. For the MC-Glauber model we assume a 
two-component (soft+hard) model with a small hard fraction 
($\delta\eq0.14$ \cite{Hirano:2009ah}); we also use a Woods-Saxon profile 
for the distribution of nucleon centers whose radius and surface 
thickness parameters have been corrected for the finite nucleon size 
\cite{Hirano:2009ah}. The resulting entropy density profile is normalized 
to the final charged hadron multiplicity density $dN_\mathrm{ch}/dy$ in 
central collisions; after this normalization, the centrality dependence 
of the initial entropy production is fixed by the model (MC-Glauber or 
MC-KLN). To convert the initial entropy density to energy density, we 
use the equation of state (EOS) s95p-PCE which matches Lattice QCD data 
at high temperatures to a chemically frozen hadron resonance gas at low 
temperatures \cite{Huovinen:2009yb,Shen:2010uy}, using 
$T_\mathrm{chem}\eq165$\,MeV as chemical freeze-out temperature.

%
\begin{table}[ht!] 
\caption{Centrality table for Au+Au at $200\,A$\,GeV \cite{Hirano:2009ah}.
         \label{T1}}
\label{tab:1} 
\begin{center}
\begin{tabular}{ccccccccc}
  \hline\hline
  centrality &\vline & $b_\mathrm{min}$(fm) &\vline & $b_\mathrm{max}$(fm) 
  &\vline & $\bar{b}$ (fm) &\vline & $\bar{N}_{\mathrm{part}}$ \\
  \hline
  0-5\%   &\vline & 0.0	&\vline	 & 3.3	&\vline	& 2.2	&\vline	  & 352.2 \\
  5-10\%  &\vline & 3.3	&\vline  & 4.7	&\vline	& 4.04	&\vline	  & 294.7 \\
  10-15\% &\vline & 4.7	&\vline  & 5.8	&\vline	& 5.27	&\vline	  & 245.6 \\
  15-20\% &\vline & 5.8	&\vline  & 6.7	&\vline	& 6.26 	&\vline	  & 204.2 \\
  20-30\% &\vline & 6.7	&\vline  & 8.2	&\vline	& 7.48 	&\vline	  & 154.5 \\
  30-40\% &\vline & 8.2	&\vline	 & 9.4	&\vline	& 8.81 	&\vline	  & 103.8 \\
  40-50\% &\vline & 9.4	&\vline	 & 10.6	&\vline	& 10.01	&\vline	  & 64.9 \\
  50-60\% &\vline & 10.6 &\vline & 11.6	&\vline	& 11.11	&\vline	  & 36.6 \\
  60-70\% &\vline & 11.6 &\vline & 12.5	&\vline	& 12.06	&\vline	  & 18.8\\
  70-80\% &\vline & 12.5 &\vline & 13.4	&\vline	& 12.96	&\vline	  & 7.5\\
  80-90\% &\vline & 13.4 &\vline & 14.3	&\vline	& 13.85	&\vline	  & 4.4\\
  \hline
\end{tabular}
\end{center}
\end{table}
%

In the following we compute harmonic eccentricity and flow coefficients
as functions of impact parameter $b$ and collision centrality (\%).
The centrality classes are defined in terms of percentages of the total
inelastic cross section, calculated from the distribution of the number of 
wounded nucleons $dN_\mathrm{event}/dN_\mathrm{part}$ in the optical Glauber 
model (i.e. without accounting for fluctuations in $N_\mathrm{part}$
at given impact parameter). Each centrality class is thus characterized by 
a range of impact parameters $b_\mathrm{min}<b<b_\mathrm{max}$ and an 
average value $\bar{b}$, together with a mean number of wounded nucleons 
$\bar{N}_\mathrm{part}$. They are listed in Table~\ref{T1} 
\cite{Hirano:2009ah}.  

\subsection{Averaging procedures for the initial profiles}
\label{sec2e}

%
\begin{figure*}
\begin{center}
 \includegraphics[width=0.42\linewidth]{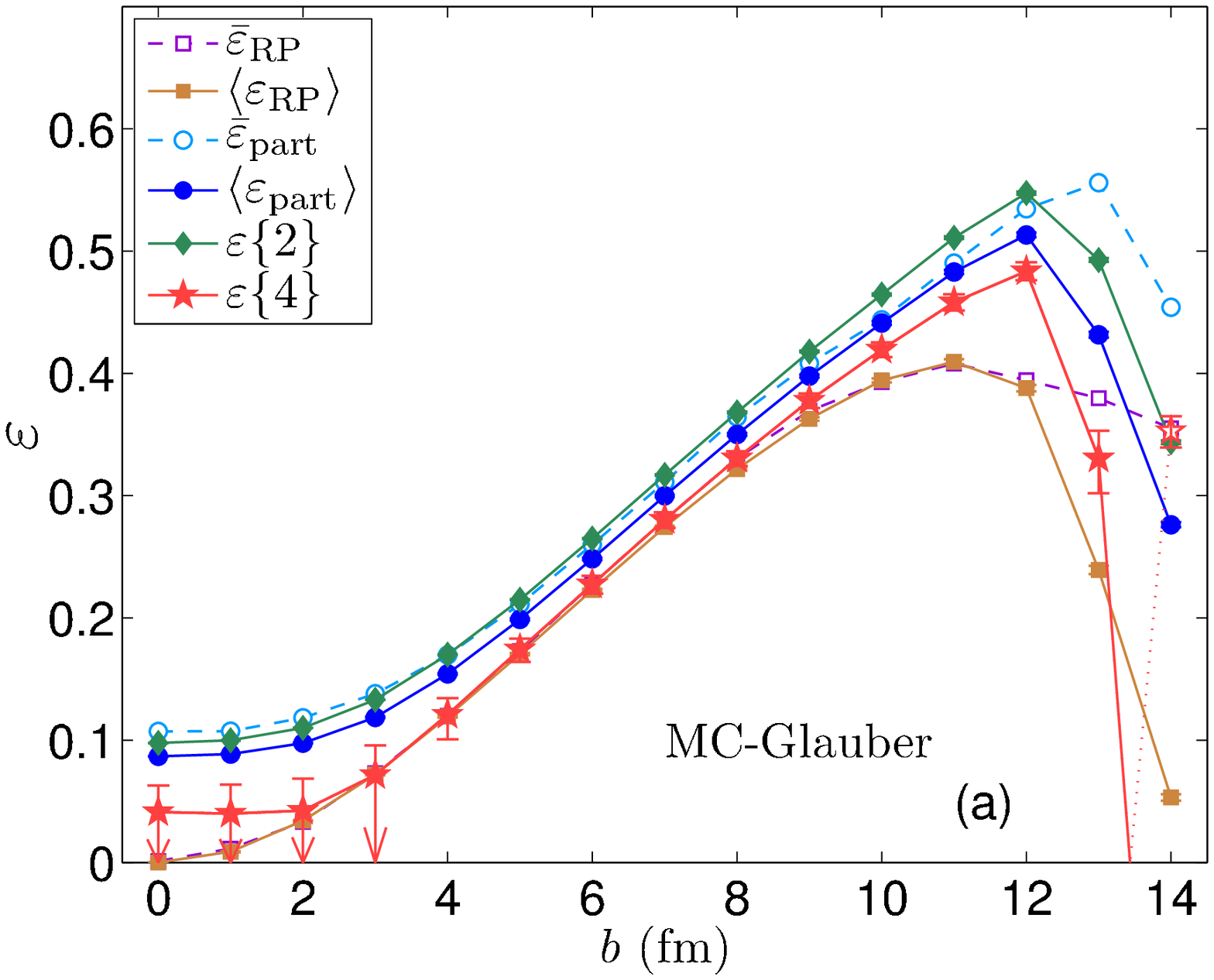}
 \includegraphics[width=0.42\linewidth]{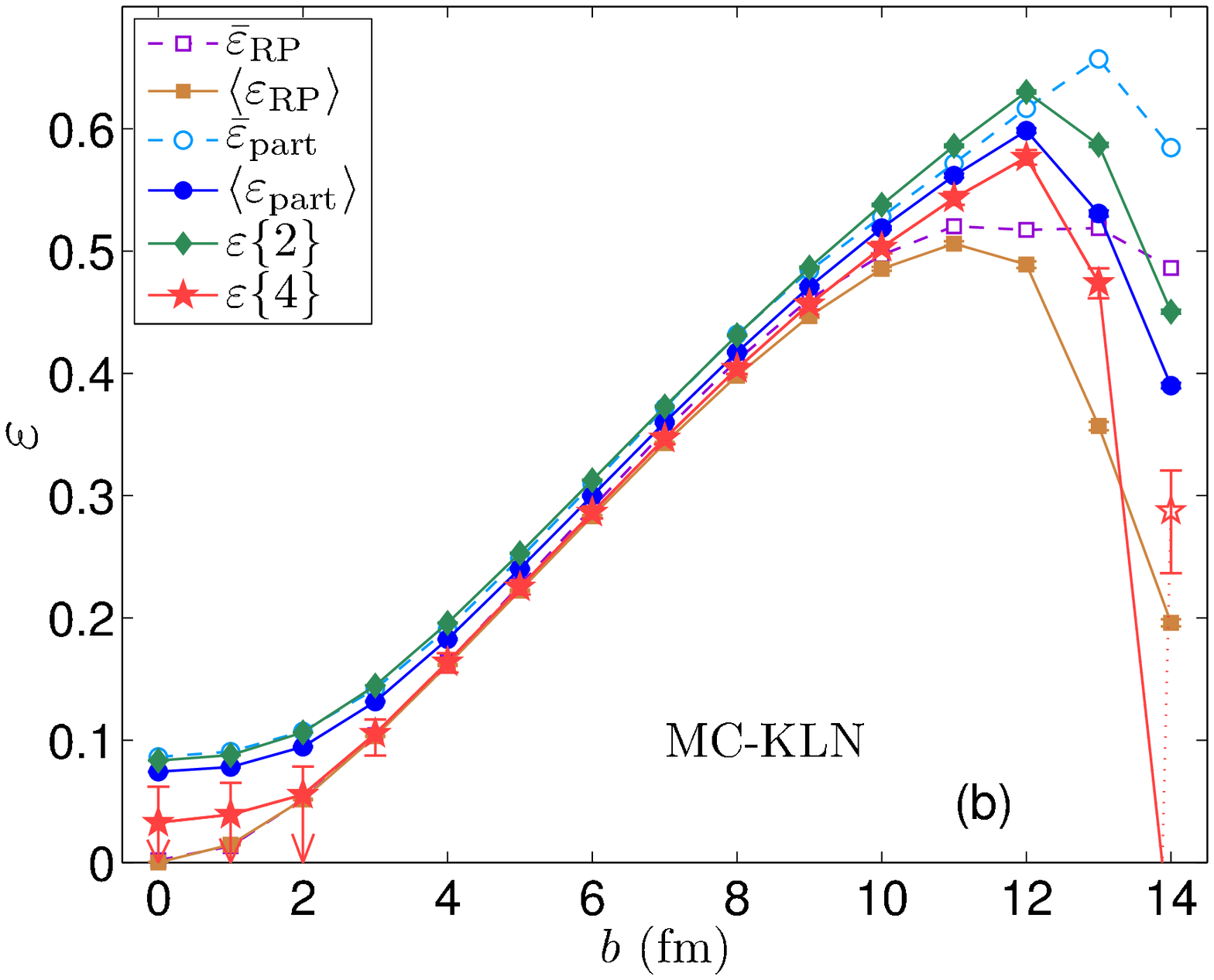}\\
 \includegraphics[width=0.42\linewidth]{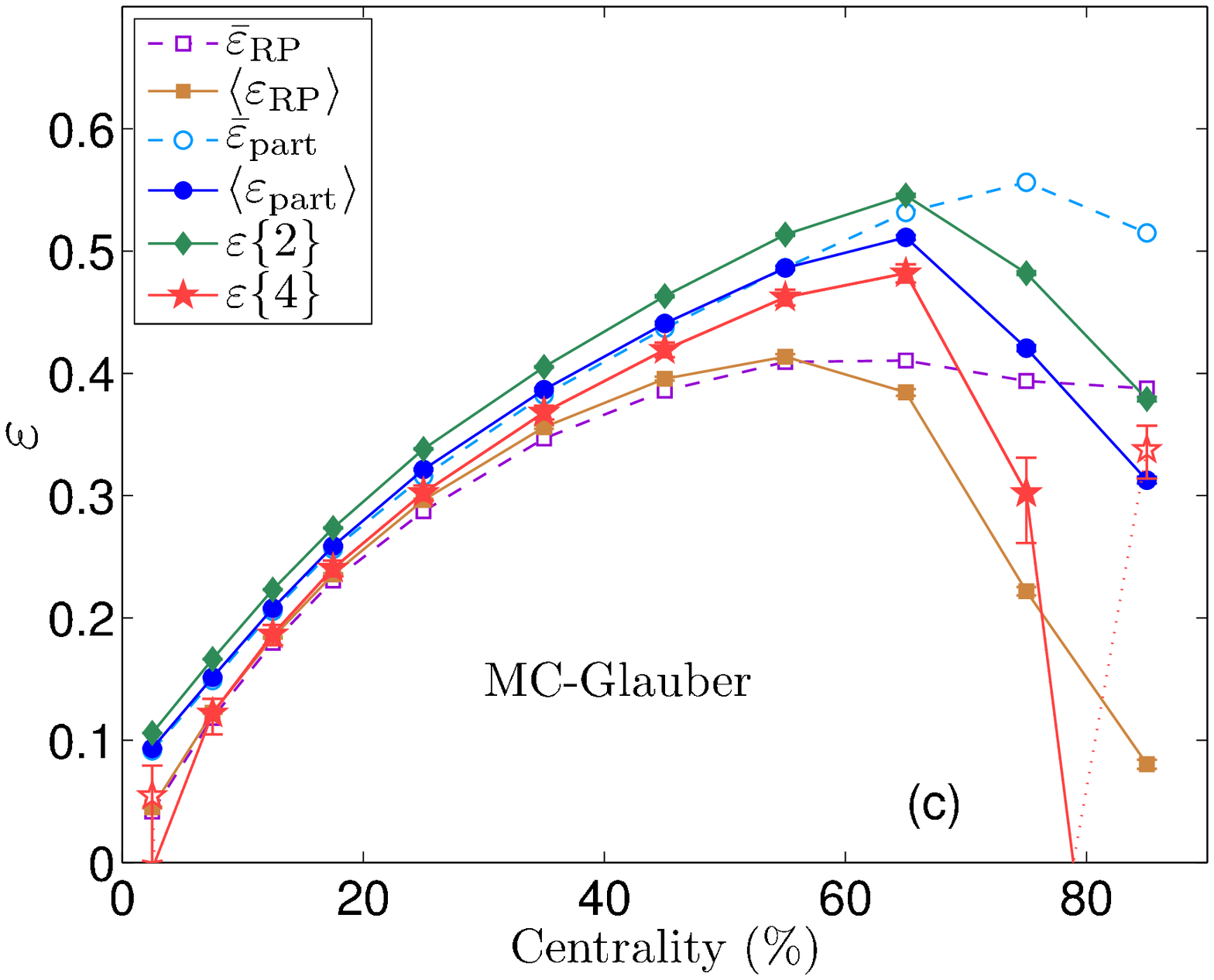}
 \includegraphics[width=0.42\linewidth]{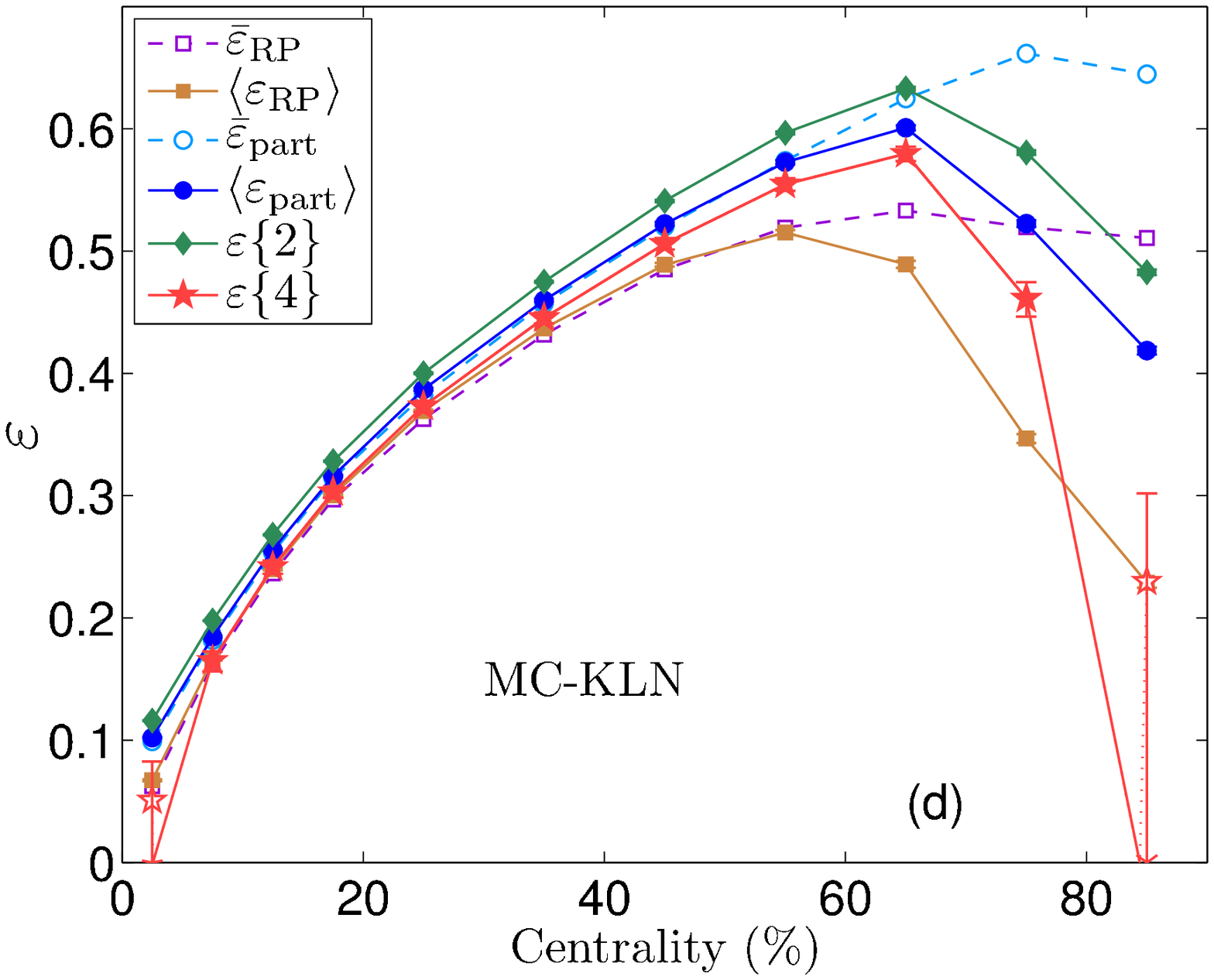}
\caption{(Color online) Different ellipticities as a function of
   impact parameter (top row) or collision centrality (bottom
   row), for the MC-Glauber (panels (a) and (c)) and the MC-KLN
   model (panels (b) and (d)). Panels (a,b) show $e$-weighted, panels
   (c,d) show $s$-weighted ellipticities. (See Figs.~\ref{F3}, \ref{F4} 
   below for a direct comparison between $e$- and $s$-weighted eccentricities.)
   Open stars indicate negative values for $\ecc\{4\}^4$.
   \label{F1}
}
\end{center}
\end{figure*}
%

In this work we compare results obtained from an event-by-event 
hydrodynamical evolution of fluctuating initial conditions with
the traditional method of ``single-shot'' hydrodynamic evolution,
where one first averages over many fluctuating initial profiles to obtain 
a smooth average profile, and then evolves this smooth profile 
hydrodynamically. The question addressed in this comparison is to what 
extent the average harmonic flow coefficients from event-by-event 
hydrodynamics can be faithfully represented by the harmonic flow 
coefficients extracted (at much lower numerical expense) from the 
hydrodynamic evolution of an ``average event''. 

Taking the initial density profiles from the Monte Carlo generator and 
superimposing them directly without additional manipulations generates a 
``reaction plane averaged'' profile with ellipticity 
$\bar{\ecc}_\mathrm{RP}$ (Eq.~(\ref{eq7})). After recentering 
each event to the origin of the $x$-$y$-plane, we can compute event
by event the reaction and participant plane ellipticities (Eqs.~(\ref{eq6}) 
and (\ref{eq2},\ref{eq3})) and evaluate their ensemble averages (\ref{eq8}) 
and (\ref{eq4}), respectively. To generate a smooth average profile with
ellipticity $\bar{\ecc}_\mathrm{part}$ (Eq.~\ref{eq5}) we rotate each
recentered event by the angle $\psi_2^\mathrm{PP}(e)$ ($\psi_2^\mathrm{PP}(s)$)
if we want to determine the eccentricity of the average energy (entropy)
density. For the calculation of entropy-weighted average eccentricities 
we perform any ensemble average first and convert the result to energy 
density later; in this case all events are rotated by their 
$\psi_2^\mathrm{PP}(s)$ angles. For energy-weighted ensemble averages we 
convert $s$ to $e$ in each event 
first, rotate by $\psi_2^\mathrm{PP}(e)$ and perform the ensemble average 
last. Other sequences or mixtures of these steps are technically possible 
but physically not meaningful. Note that the processes of computing the 
energy density from the entropy density via the EOS and of averaging the 
event profiles do not commute: The energy density obtained via the EOS 
from the ensemble-averaged entropy density profile is not the same as the 
ensemble-averaged energy density where the EOS is used in each event to 
convert $s$ to $e$. 

\vspace*{-3mm}
\section{Eccentricities}
\label{sec3}
\vspace*{-3mm}

\subsection{Centrality dependence of different ellipticities}
\label{sec3a}
\vspace*{-3mm}

Fig.~\ref{F1} shows a comparison between the different ellipticities 
defined in Sec.~\ref{sec2a}, as functions of the impact parameter $b$
in panels (a) and (b) and as functions of collision centrality (as defined
in Sec~\ref{sec2d}) in panels (c) and (d). For panels (a) and (b) we 
generated 10,000 initial profiles for each impact parameter (except for 
$b\eq0,\,1,$ and 2\,fm for which we generated 30,000 events each); for 
panels (c) and (d) we averaged over 10,000 profiles for each centrality 
bin. Within the centrality bins, the impact parameters were sampled between 
$b_\mathrm{min}$ and $b_\mathrm{max}$ with $b\,db$ weight. Compared to 
panels (a) and (b), this leads to additional ellipticity fluctuations 
related to the fluctuating impact parameter, whereas in Fig.~\ref{F1}a,b 
only $N_\mathrm{part}$ fluctuations at fixed $b$ contribute.

As discussed in Sec.~\ref{sec2a}, Eq.~(\ref{eq10}), $\ecc\{4\}^4$ can 
become negative when fluctuations grow large. Whenever this happens, we 
replace $\ecc\{4\}$ by $\sqrt[4]{\left|\ecc\{4\}^4\right|}$ and indicate 
this by an open star in Fig.~\ref{F1} (connected by dotted lines to other
points in the graph). One sees that $\ecc\{4\}^4$ has a tendency
to turn negative in the most peripheral collisions. In very central
collisions $\ecc\{4\}^4$ becomes very small, with central values that
can have either sign depending on whether we keep the impact parameter
fixed (Figs.~\ref{F1}a,b) or average over events with different impact 
parameters in a given centrality bin (see the $0{-}5\%$ centrality values
in Figs.~\ref{F1}c,d). Statistical errors are large, however, and within 
errors $\ecc\{4\}^4$ is compatible with zero for impact parameters 
$b<3$\,fm, i.e. in the most central ($0{-}5\%$ centrality) collisions.
We indicate this by open-ended error bars for 
$\sqrt[4]{\left|\ecc\{4\}^4\right|}$, pointing from its upper limit all 
the way to zero. 

%
\begin{figure}[b!]
 \includegraphics[width=0.9\linewidth]{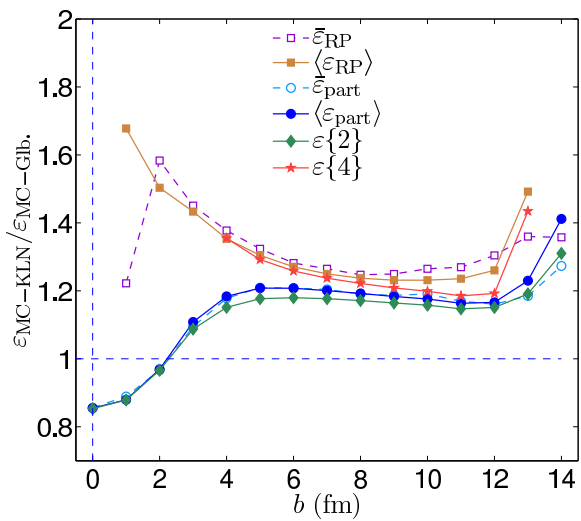}
\caption{(Color online) Impact parameter dependence of the ratio of 
   ellipticities obtained from the MC-KLN and MC-Glauber models 
   as shown in Figs.~\ref{F1}a,b.
   \label{F2}
}
\end{figure}
%

Comparing panels (a,c) for the MC-Glauber model with panels (b,d) for
the MC-KLN model we see great similarities in shape, but systematic
differences in magnitude of the ellipticities. The ratio of the MC-KLN
and MC-Glauber ellipticities is shown in Fig.~\ref{F2}. Except for the 
most central and most peripheral collisions, the MC-KLN ellipticities
exceed the MC-Glauber ones by an approximately constant factor of around
1.2. Please note the difference in the ratios for the reaction plane and 
participant eccentricities at small $b$. (The point for 
$\bar{\ecc}_\mathrm{RP}$ at $b\eq1$\,fm is obtained from a ratio of very 
small numbers and probably not statistically robust -- we had only 30,000 
events to determine the ensemble-averaged density profile.) For the 
$\ecc\{4\}$ ratio we dropped all points where the statistical error 
for $\ecc\{4\}^4$ extended into the region of negative values for either
the MC-Glauber or MC-KLN model.

Figure~\ref{F1} shows that, for central and mid-peripheral collisions,
the ensemble-averaged participant and reaction plane eccentricities 
$\la\ecc_\mathrm{part}\ra$ and $\la\ecc_\mathrm{RP}\ra$ agree very well
with the mean eccentricities $\bar{\ecc}_\mathrm{part}$ and 
$\bar{\ecc}_\mathrm{RP}$ of the corresponding ensemble-averaged profiles.
For strongly peripheral collisions ($b\agt10$\,fm), however, the average
of the ratio (Eqs.~(\ref{eq3},\ref{eq4},\ref{eq8})) differs strongly from 
the ratio of averages (Eqs.~(\ref{eq5},\ref{eq7})), indicating strong 
event-by-event fluctuations. We note that in very peripheral collisions the 
average event ellipticity drops quickly with increasing $b$ while the
ellipticity of the average profile remains large; single-shot hydrodynamic
calculations based on a smooth average initial profile thus overestimate
the effective initial source ellipticity and produce more elliptic flow 
than expected from event-by-event hydrodynamic evolution of individual
peripheral events. Still, as first emphasized in \cite{Song:2011hk}, the 
calculated $v_2$ from single-shot hydrodynamics decreases steeply at large 
collision centralities \cite{Luzum:2010ag,Hirano:2010jg,Song:2011qa,%
Shen:2011eg}, due to the decreasing fireball lifetime, which contrasts 
with the initially reported experimentally observed behaviour 
\cite{:2008ed,Aamodt:2010pa}, but agrees qualitatively with a recent 
reanalysis \cite{Collaboration:2011vk} where non-flow effects have been
largely eliminated and/or corrected for. We do point out that our 
Monte-Carlo simulations do not include fluctuations in the amount of 
entropy generated per nucleon-nucleon collision \cite{Qin:2010pf}; these 
could have important effects on the ellipticities in very peripheral 
collisions. 

\begin{figure}[h!]
 \includegraphics[width=0.9\linewidth]{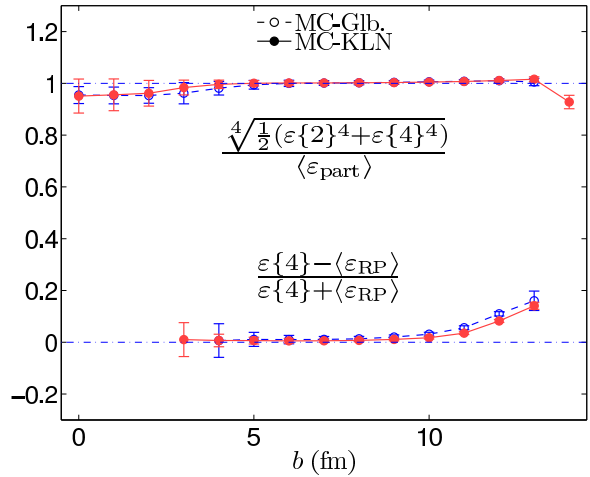}
 \caption{(Color online) Checks of the assumption of Gaussian
    and Bessel-Gaussian fluctuations for $\epart$ (see text for 
    discussion).  
 \label{F3}
 }
\end{figure}

Comparing the curves for $\la\ecc_\mathrm{part}\ra$, $\ecc\{2\}$ and
$\ecc\{4\}$ in Fig.~\ref{F1} we see that (as is manifest in the Gaussian 
model analysis in Eq.~(\ref{eq11a})) $\ecc\{2\}$ receives a positive and 
$\ecc\{4\}$ receives a negative contribution from event-by-event 
ellipticity fluctuations. In Fig.~\ref{F3} we check, as a function of 
impact parameter, the validity of the identities (\ref{eq11}) and 
%
%
(\ref{eq11b}) which follow from Bessel-Gaussian and Gaussian $\epart$ 
distributions, respectively. We see that both hold with good accuracy in 
the mid-centrality range ($b{\,\alt\,}10$\,fm for Eq.~(\ref{eq11}),
$5{\,\alt\,}b{\,\alt\,}11$\,fm for Eq.~(\ref{eq11b})) but break down in 
the most peripheral collisions. Both the Gaussian and Bessel-Gaussian 
hypotheses work slightly better for the MC-KLN than for the MC-Glauber 
model. Consistent with the analysis in Ref.~\cite{Voloshin:2007pc}, the 
Gaussian fluctuation hypothesis for $\epart$ breaks down at small impact 
parameters whereas (as theoretically expected \cite{Voloshin:2007pc}) the 
Bessel-Gaussian hypothesis appears to continue to hold, although we are 
unable to make this statement with statistical confidence. (For the ratio 
$(\ecc\{4\}{-}\la\ecc_\mathrm{RP})/(\ecc\{4\}{+}\la\ecc_\mathrm{RP})$
we again dropped all points for which the error band for $\ecc\{4\}^4$
reaches into negative territory.)

The assumption of Gaussian fluctuations of the real and imaginary parts 
of the complex ellipticity (\ref{eq1}) is often used to argue that the 
average reaction-plane ellipticity $\la\ecc_\mathrm{RP}\ra$ can serve as 
a proxy for $\ecc\{4\}$ (see Eq.~(\ref{eq11})), and that therefore 
reaction-plane averaged initial density profiles can be used in 
single-shot hydrodynamics (which ignores event-by-event fluctuations) 
to simulate the experimentally measured $v_2\{4\}$ values. Fig.~\ref{F1} 
and the bottom curves in Fig.~\ref{F3} show that $v_2\{4\}$ values obtained 
from single-shot hydrodynamic simulations with reaction-plane averaged 
initial conditions \cite{Hirano:2010jg,Song:2011qa} should not be trusted 
quantitatively for centralities $>40\%$.

To summarize this subsection, all the simplifying assumptions that allow
to focus attention on the three quantities $\la\ecc_\mathrm{part}\ra$,
$\ecc\{2\}$ and $\ecc\{4\}$ only (by substituting $\la\ecc_\mathrm{part}\ra$
for $\bar{\ecc}_\mathrm{part}$ and $\ecc\{4\}$ for $\bar{\ecc}_\mathrm{RP}$
or $\la\ecc_\mathrm{RP}\ra$) hold well for central to mid-central collisions
(${\,\leq\,}40\%$ centrality) but break down for peripheral collisions.
For ${\,>\,}40\%$ centrality there exists no substitute for event-by-event 
hydrodynamics if one aims for quantitative precision in the comparison with 
experimental data, since the latter are strongly affected by non-Gaussian 
event-by-event fluctuations at those centralities. 

\subsection{Ellipticities with different weight functions}
\label{sec3b}

%
\begin{figure}[htb]
 \includegraphics[width=\linewidth]{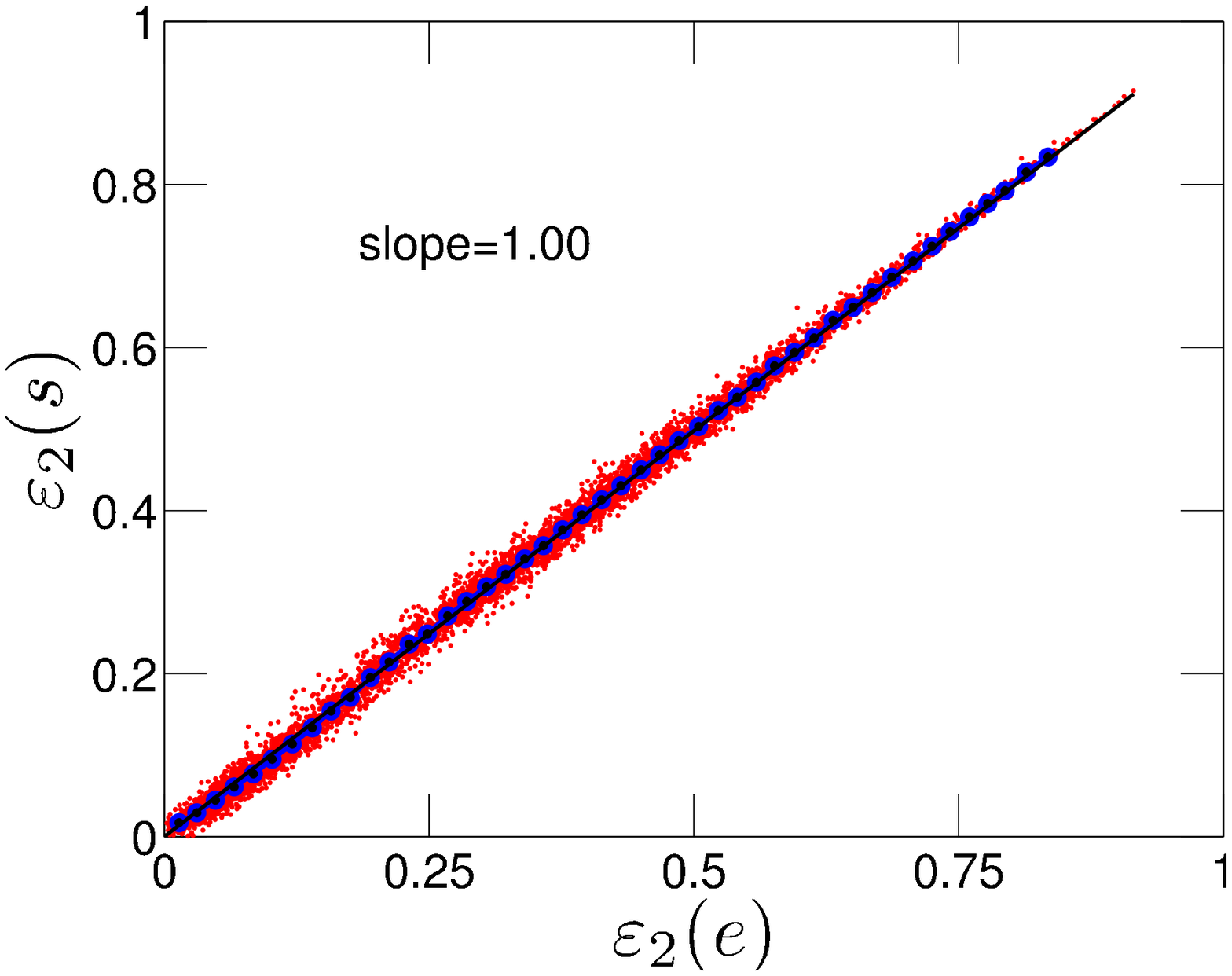}
 \caption{(Color online) $\ecc_\mathrm{part}(e)$ vs. $\ecc_\mathrm{part}(s)$ 
      for 6000 initial profiles from the MC-KLN model (1000 each for $b\eq0$,
      $0{-}5\%$, $15{-}20\%$, $20{-}30\%$, $30{-}40\%$, and $50{-}60\%$ 
      centrality).
 \label{F4}
 }
\end{figure}
%

Figure~\ref{F4} shows a comparison between the energy- and entropy-weighted
ellipticities of the initial profiles generated with the MC-KLN model, on 
an event-by-event basis. The scatter plot is based on 6000 events, 1000 
each for $b=0$ and for the following finite-width centrality bins: 
$0{-}5\%$, $15{-}20\%$, $20{-}30\%$, $30{-}40\%$, and $50{-}60\%$.
This is not a realistic mix in the experimental sense, but permits us to 
explore the full range from very small to very large event ellipticities.
The blue dots in Fig.~\ref{F4} represent bin averages, and the solid black
line is a linear fit through the origin. The fitted slope is 1.00, the 
scatter plot is seen to be tightly clustered around this fitted line, and 
only at small ellipticities $\ecc_2{\,<\,}20\%$ the $e$-weighted values are 
seen to be slightly larger on average than their $s$-weighted counterparts
(see also Fig.~\ref{F5}a below).  

\subsection{Higher order harmonics}
\label{sec3c}

%
\begin{figure*}[b!]
 \includegraphics[width=0.49\linewidth]{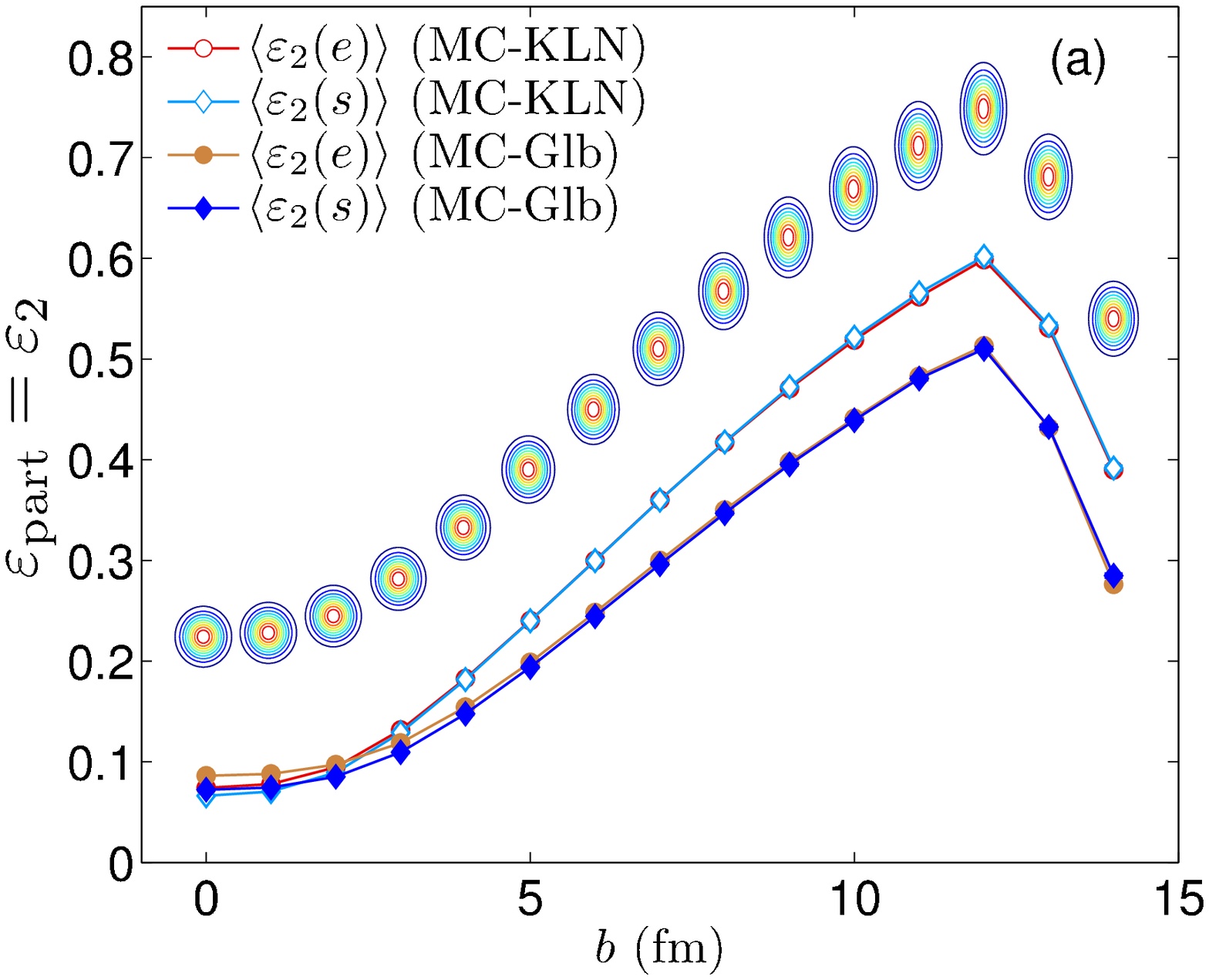}
 \includegraphics[width=0.49\linewidth]{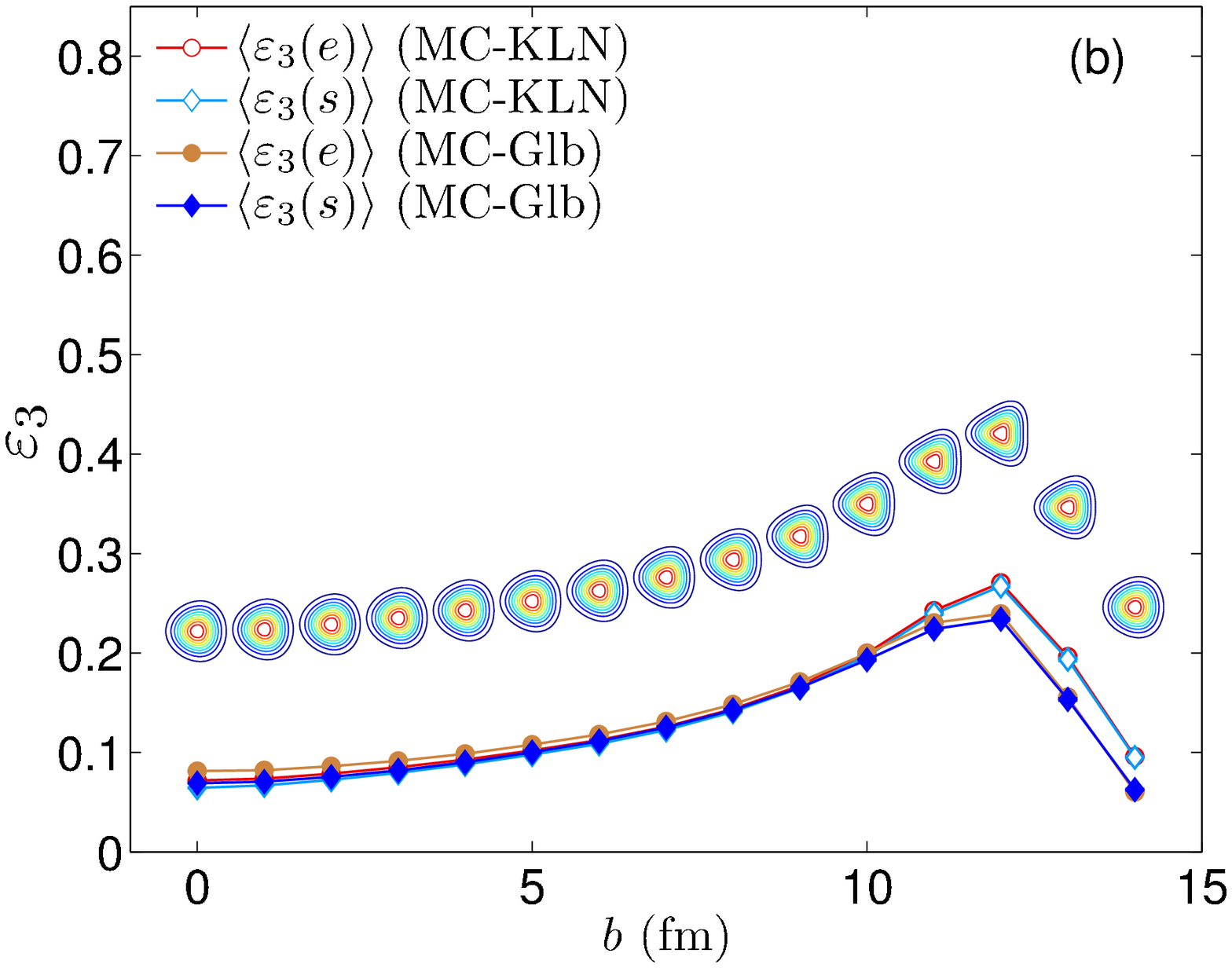}\\
 \includegraphics[width=0.49\linewidth]{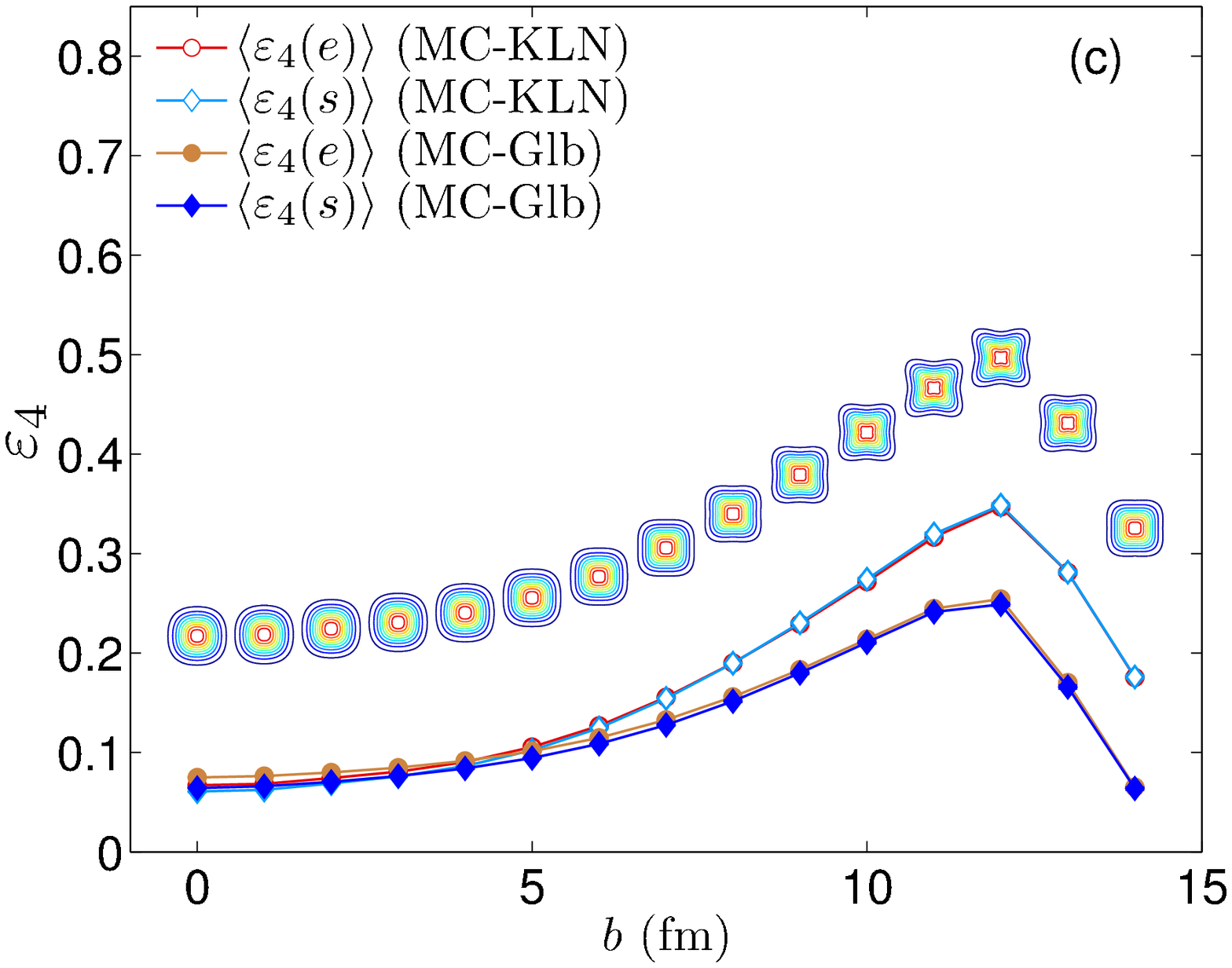}
 \includegraphics[width=0.49\linewidth]{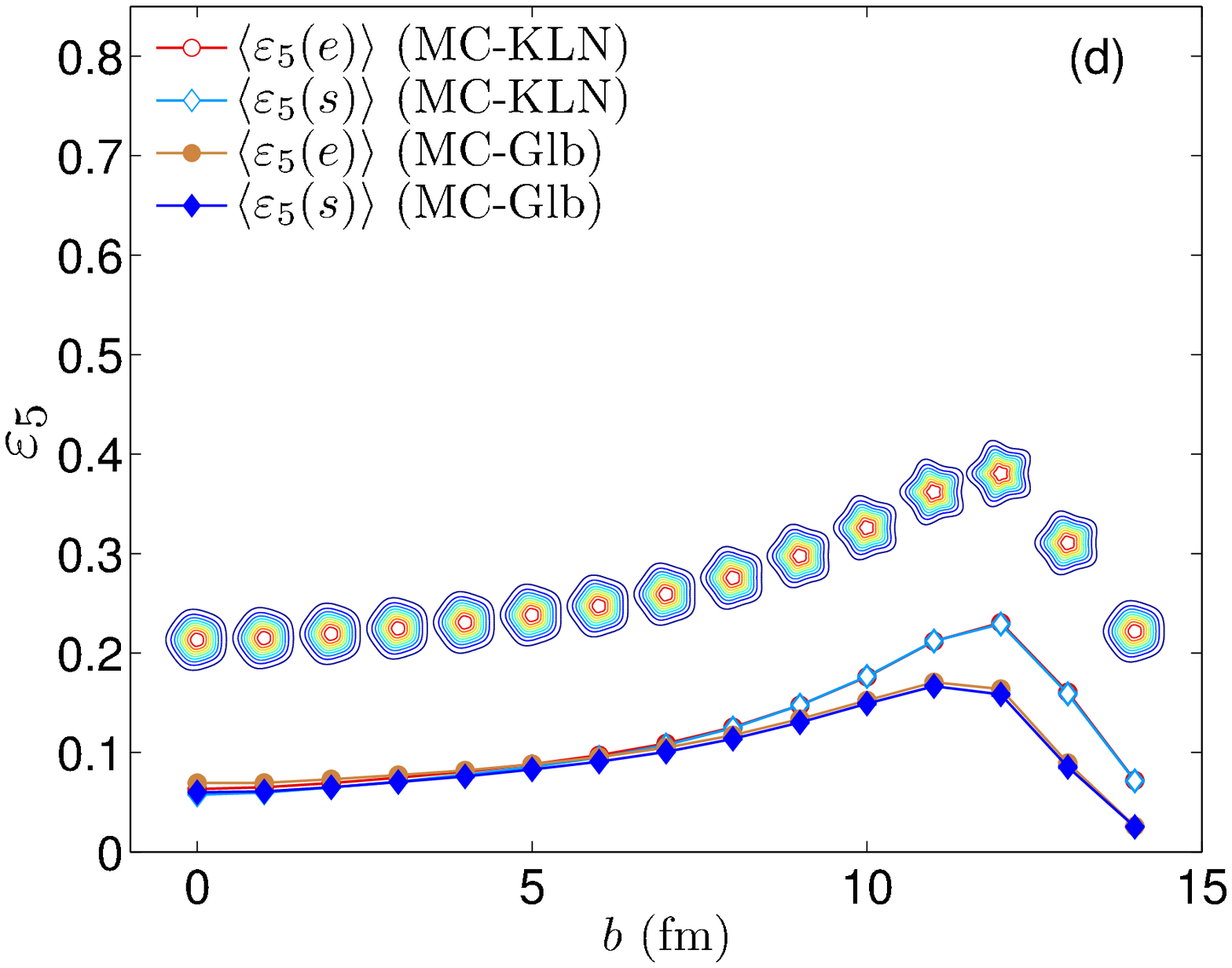}
 \caption{(Color online) Harmonic eccentricity coefficients 
          $\ecc_2\eq\ecc_\mathrm{part}$ (a), $\ecc_3$ (b),
          $\ecc_4$ (c) and $\ecc_4$ (d) as functions of impact parameter,
          calculated from the MC-Glauber (filled symbols, solid lines) and 
          MC-KLN models (open symbols, dashed lines), using the energy 
          density (circles) or entropy density (triangles) as weight function.
          The contour plots illustrate deformed Gaussian profiles 
          $e(r,\phi)\eq{e}_0\,\exp\left[-\frac{r^2}{2\rho^2}
          \bigl(1{+}\ecc_n\cos(n\phi)\bigr)\right]$, with
          eccentricity $\ecc_n(e)$ taken from the MC-KLN model at the
          corrsponding impact parameter.
 \label{F5}
 }
\end{figure*}
%

In Figs.~\ref{F5}a-d we compare the centrality dependences of the 
ensemble-averaged second to fifth harmonic eccentricity coefficients 
(energy- and entropy-weighted) from the MC-Glauber and MC-KLN models. 
The contour plots give a visual impression of the degree of deformation 
corresponding to the (larger) MC-KLN eccentricities, assuming (for 
illustration) the absence of any other eccentricity coefficients than 
the one shown in the particular panel.

First, one observes very little difference between the eccentricities of 
the entropy and energy density profiles, except for very central collisions 
($b{\,\alt\,}5$\,fm for the MC-Glauber, $b{\,\alt\,}3$\,fm for the MC-KLN 
model) where the energy-weighted eccentricities lie systematically somewhat 
above the entropy-weighted ones (for all orders $n$ studied here). The 
difference between $s$- and $e$-weighted eccentricities at small $b$ is 
bigger in the MC-Glauber than in the MC-KLN model.

%
\begin{figure*}
 \includegraphics[width=0.32\linewidth]{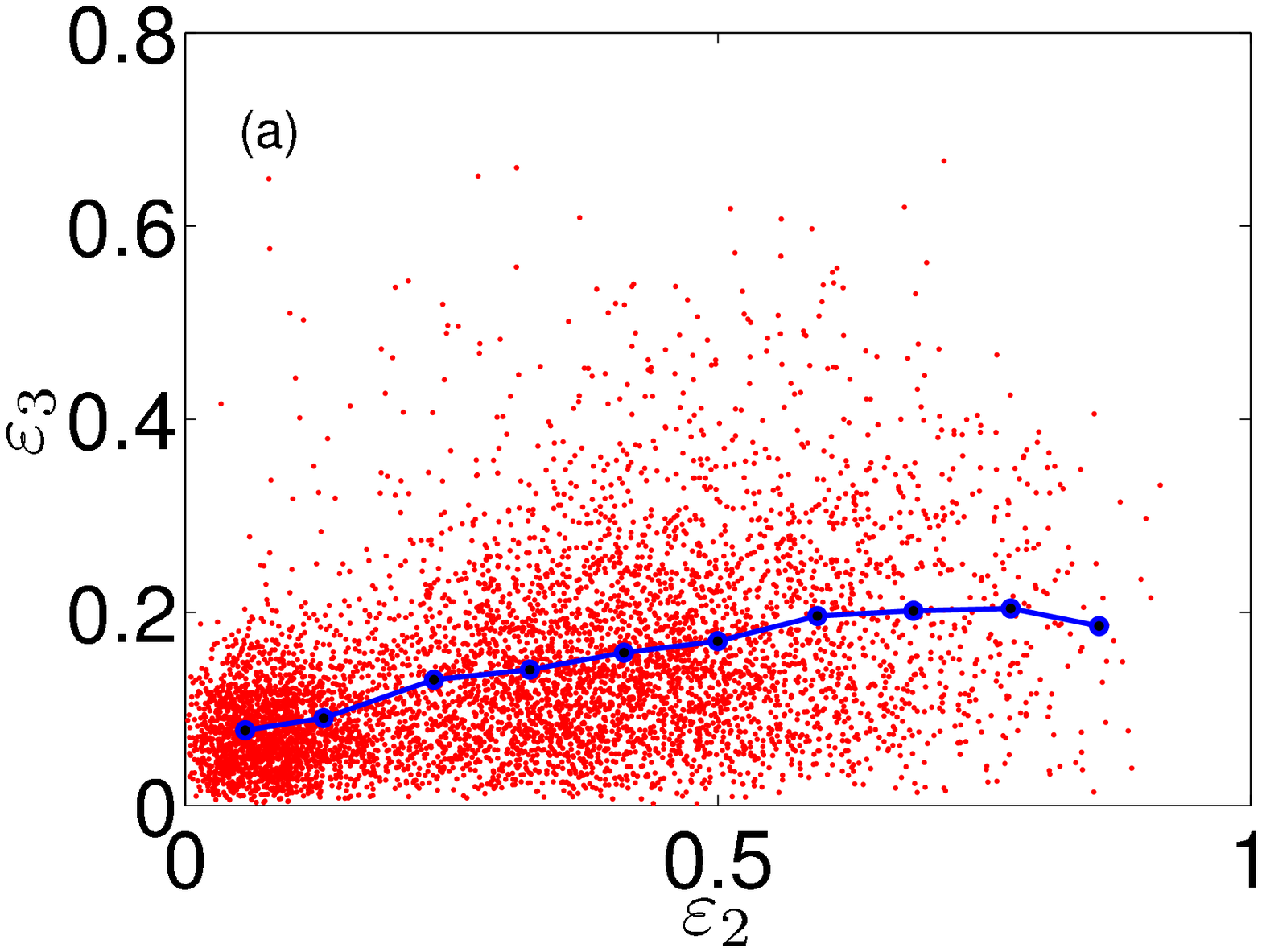}
 \includegraphics[width=0.32\linewidth]{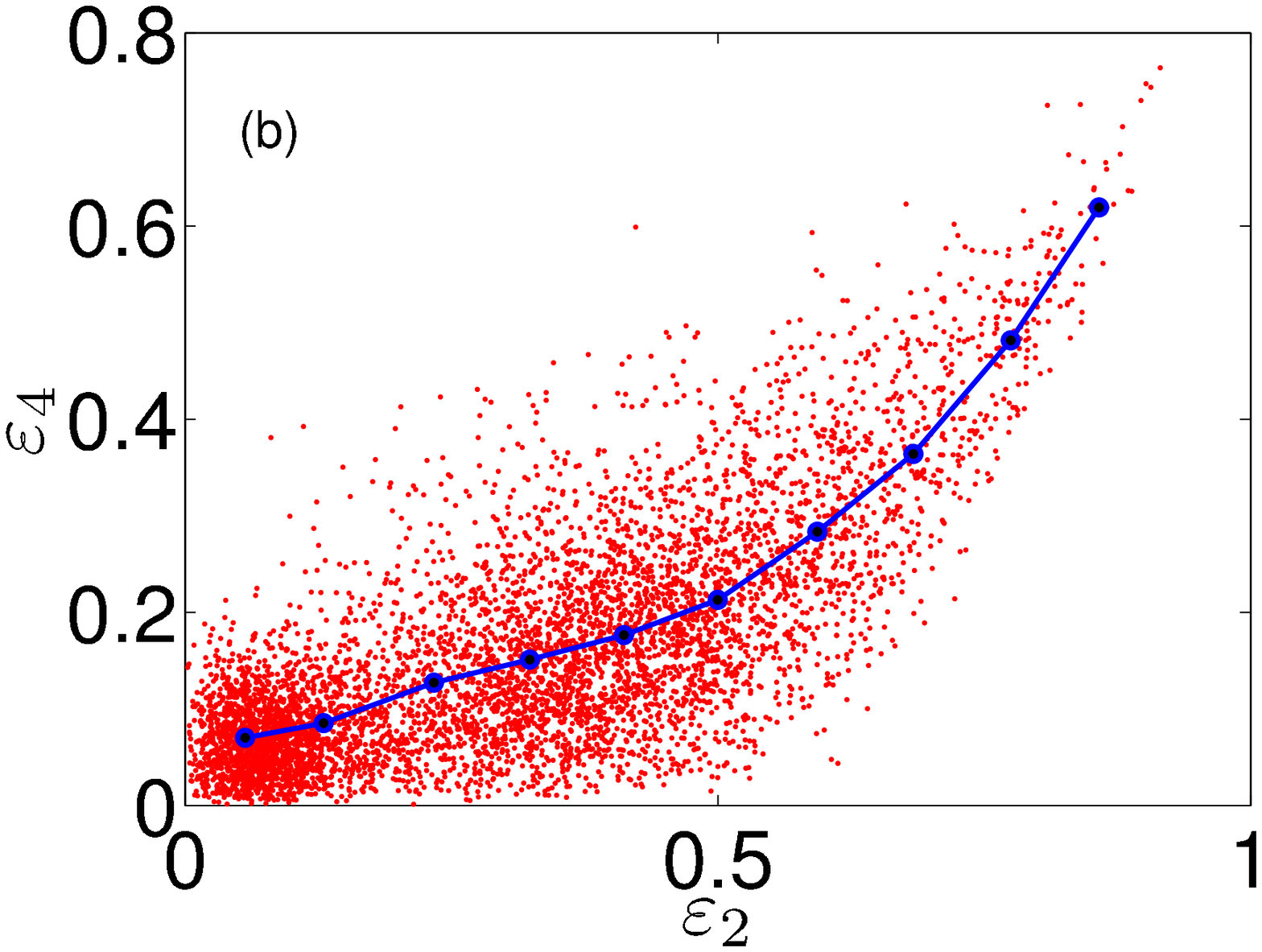}
 \includegraphics[width=0.32\linewidth]{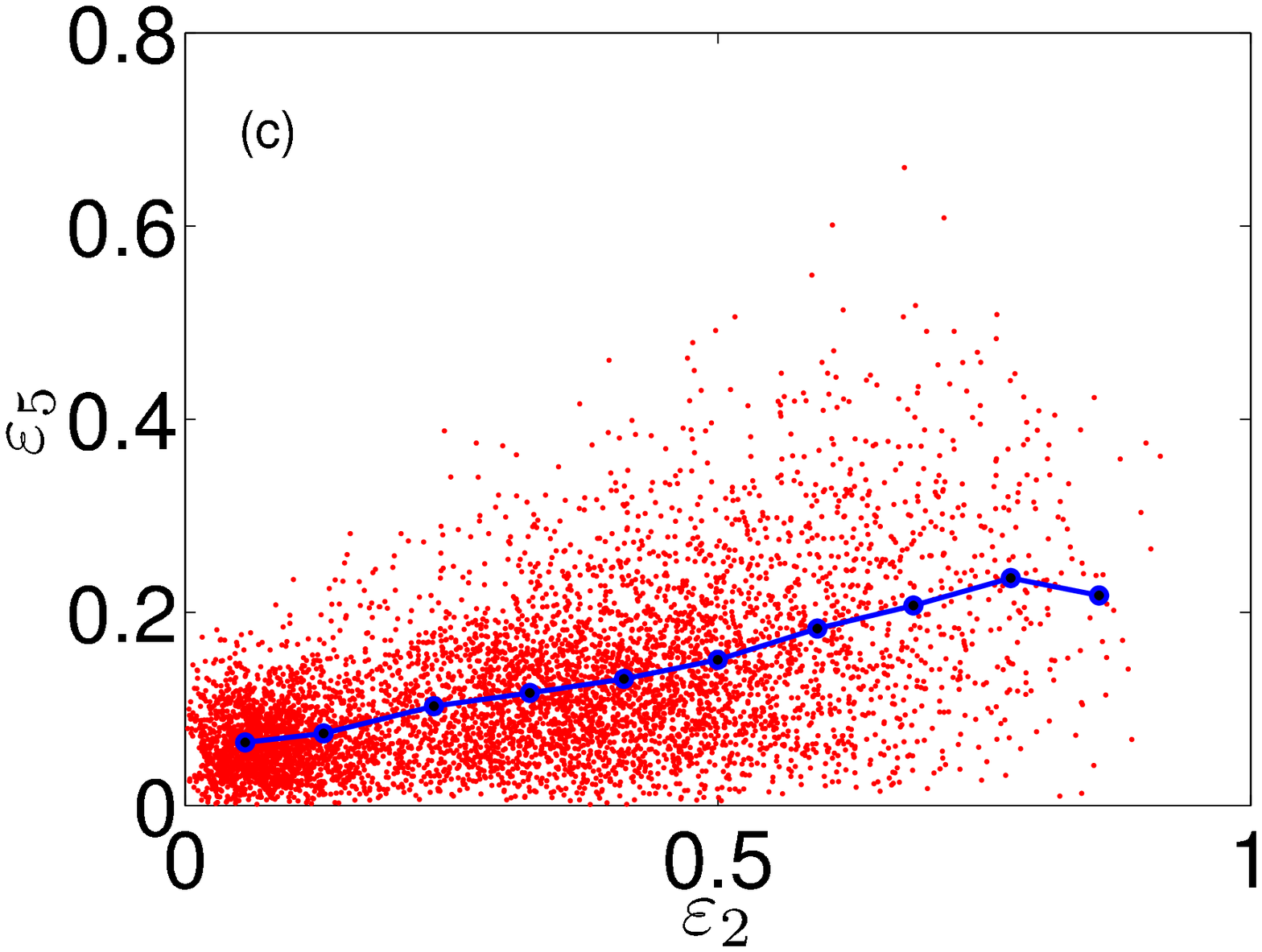}
 \caption{(Color online) Scatter plots illustrating the event-by-event
   correlation of $\ecc_3$ (a), $\ecc_4$ (b), and $\ecc_5$ (c) with the 
   fireball ellipticity $\ecc_2\eq\ecc_\mathrm{part}$, for the same set
   of 6000 event profiles analyzed in Fig.~\ref{F4}. The blue dots
   connected by a line indicate $\ecc_2$-bin averages, to guide the eye.
 \label{F6}
 }
\end{figure*}
%

Next, one notes the significantly larger ellipticities and quadrangularities
of the MC-KLN distributions compared to those from the MC-Glauber model,
for all but the most central collisions. These are driven by geometry, i.e.
by the almond-shaped deformation of the nuclear overlap zone in non-central
collisions, which in the KLN-model is more eccentric than in the Glauber
model. The third and fifth order harmonics, which are entirely due to
fluctuations (and whose associated angles $\psi_n^\mathrm{PP}$ are therefore
completely uncorrelated to the reaction plane -- see Ref.~\cite{Qin:2010pf} 
and discussion below), show remarkably similar eccentricity values in the 
two initialization models, except for the most peripheral events. 
Comparing the viscous suppression of elliptic and triangular flow thus
should allow to distinguish experimentally between the MC-Glauber and
MC-KLN models \cite{new3}.
 
Third, in central collisions all four eccentricity coefficients are
roughly of the same size. In peripheral collisions, the fluctuation-dominated
eccentricity coefficients ($\ecc_3$ and $\ecc_5$) are generically smaller 
than the geometry-dominated ones ($\ecc_2$, but also to some extent 
$\ecc_4$).\footnote{We checked that the centrality dependences
  of the ratios $\ecc_n/\ecc_3$ agree qualitatively, but not
  quantitatively with Fig.~3 in Ref.~\cite{Lacey:2010hw}. We suspect
  that the differences, which are larger for the MC-Glauber than the
  MC-KLN model, are due to somewhat different Woods-Saxon and (in the 
  MC-Glauber case) fluctuation size parameters used in 
  Ref.~\cite{Lacey:2010hw}.} 
This is less obvious when one defines the higher order eccentricities
with $r^n$ instead of $r^2$ weight \cite{Qin:2010pf}, which tends to
increase the values of the higher harmonics in peripheral collisions.

Even with ``only'' an $r^2$ weight, $\ecc_4$ and $\ecc_5$ are seen to 
become large enough around $b\sim10-13$\,fm that, if collective acceleration
happens predominantly in the directions of steepest descent of the 
density profile, one has to expect cross-currents in the developing 
anisotropic flow patterns. These can lead to destructive interference
and a correspondingly reduced efficiency of converting $n^\mathrm{th}$-order
eccentricities $\ecc_n$ into $n^\mathrm{th}$-order harmonic flows $v_n$
\cite{Alver:2010dn}. In realistic situations this issue is exacerbated
by the simultaneous presence of {\em several} large eccentricity components
$\ecc_n$, which is expected to lead to a strongly non-diagonal and
probably non-linear response matrix relating $v_n$ to $\ecc_n$ 
\cite{Qin:2010pf}. This will be discussed in Sec.~\ref{sec4}. 
  
\subsection{Eccentricity correlations}
\label{sec3d}

It is reasonable to ask whether and how the different harmonic 
eccentricity coefficients $\ecc_n$ are correlated with each other.
Figure~\ref{F6} shows scatter plots of the correlations between 
$\ecc_{3,4,5}$ and the ellipticity $\ecc_2$ which, for large $\ecc_2$
values, is dominated by geometric overlap effects. We note that,
according to the definition (\ref{eq15}), all eccentricity coefficients
are positive definite, $\ecc_n{\,\geq\,}0$. Keeping this in mind, 
Figs.~\ref{F6}a,c show that $\ecc_3$ and $\ecc_5$ are uncorrelated
with the fireball ellipticity; the slight growth of $\la\ecc_{3,5}\ra$
with increasing $\ecc_2$ is related to the growth of the variances of
their distributions in more peripheral collisions. 

\begin{figure*}
 \includegraphics[width=0.32\linewidth]{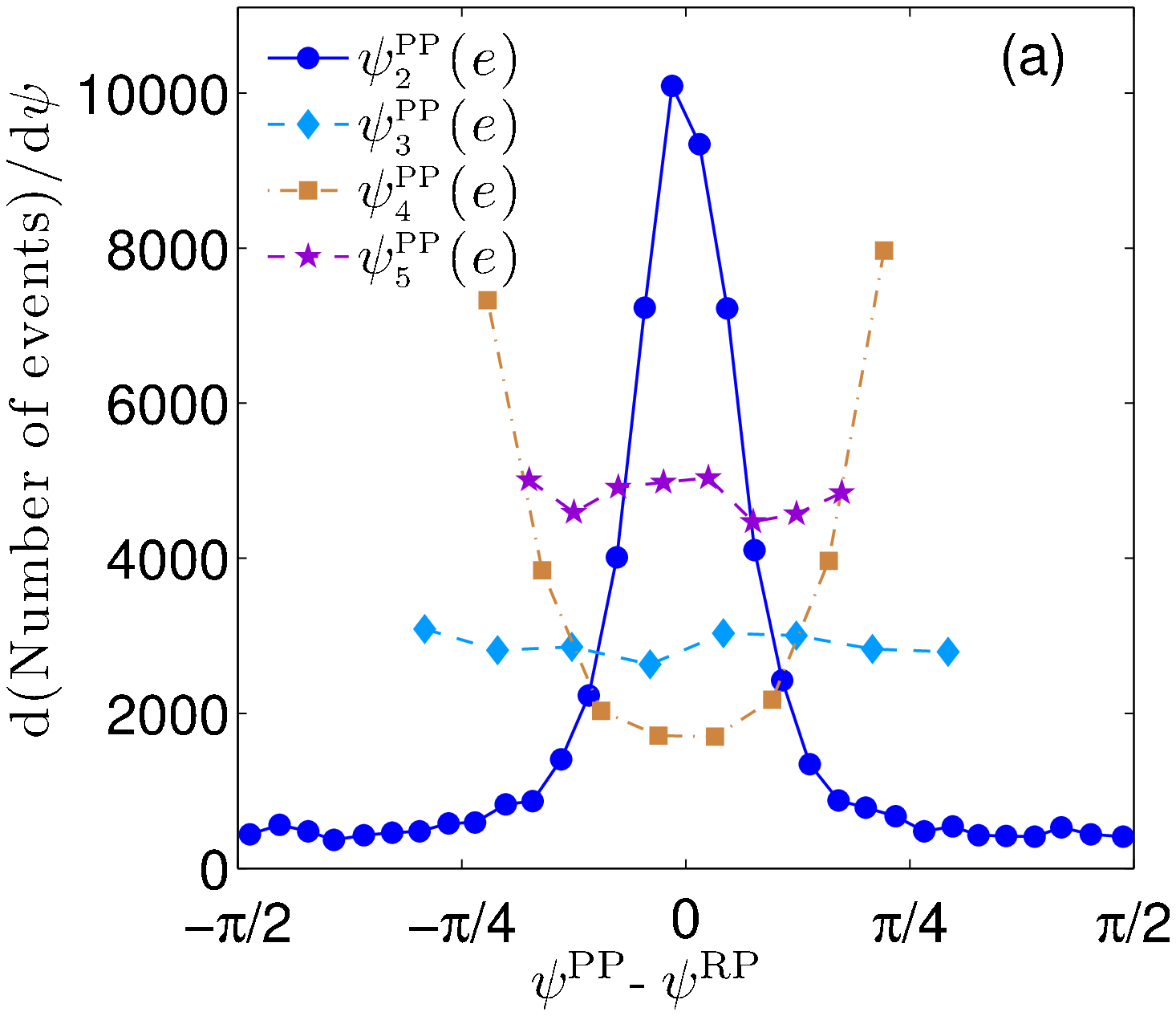}
 \includegraphics[width=0.32\linewidth]{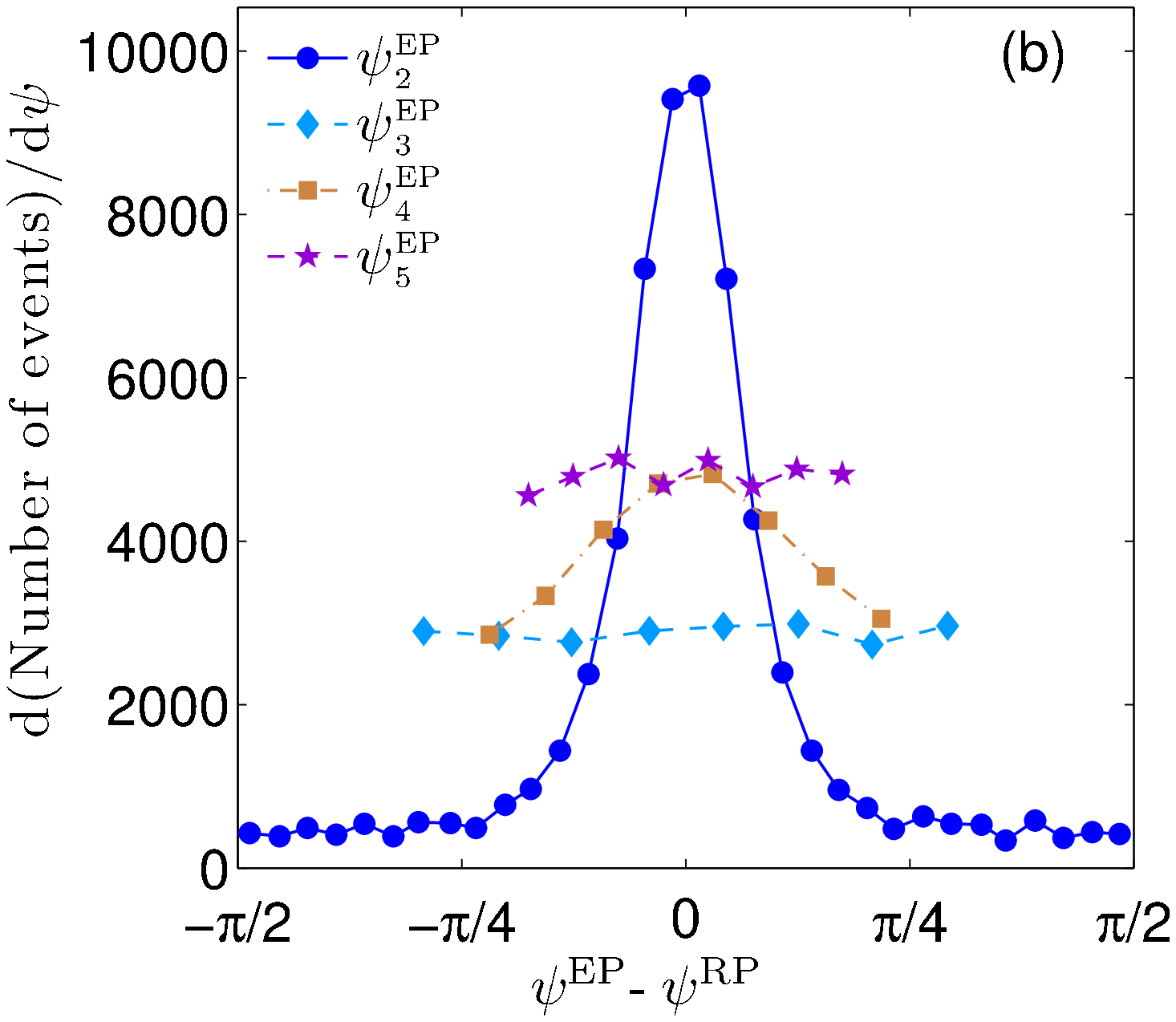}
 \includegraphics[width=0.32\linewidth,height=5.08cm]{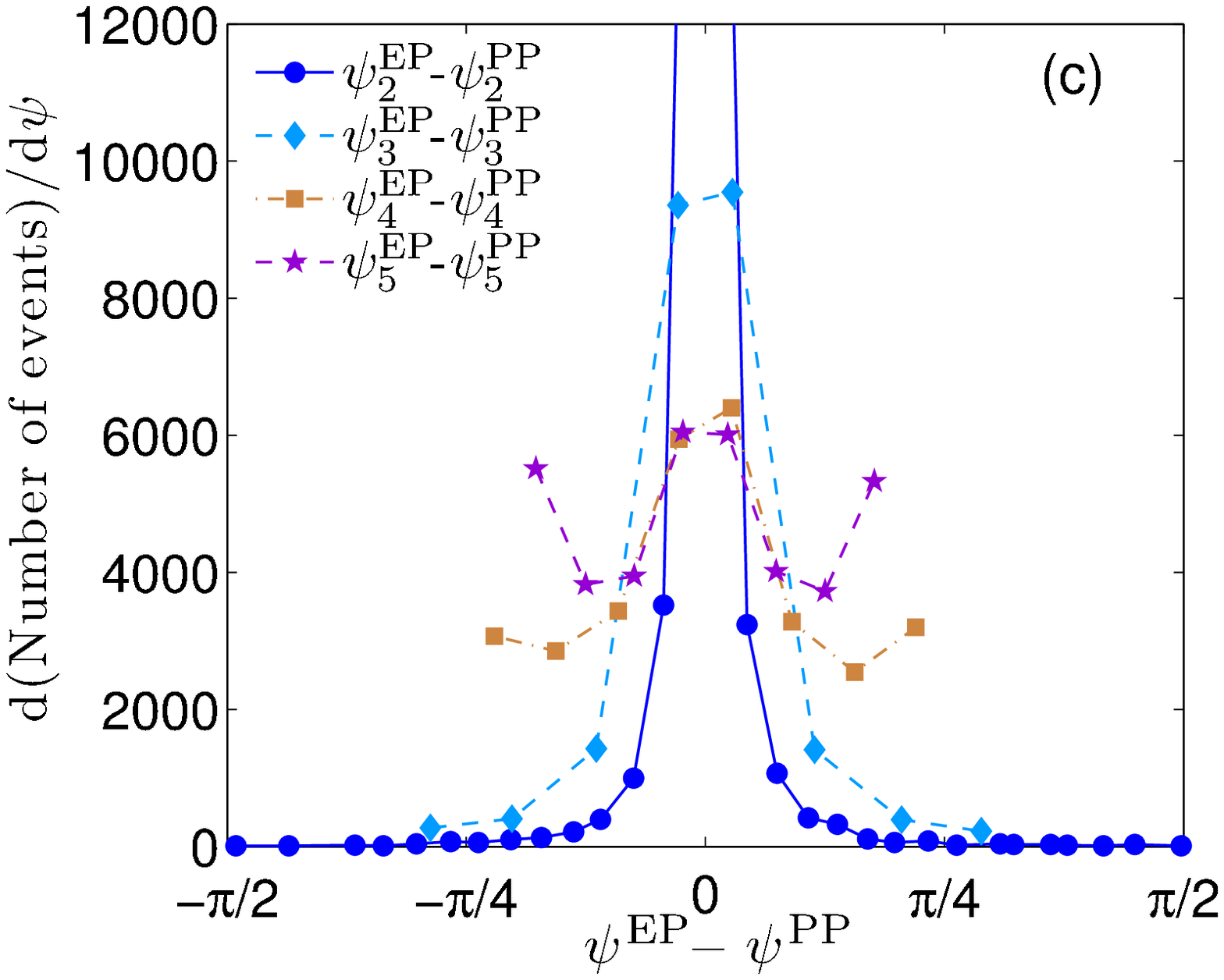}
 \caption{(Color online) Event-by-event correlation of the participant 
    plane (PP,(a)) and event plane (EP,(b)) angles with the reaction plane 
    (RP), as well as the correlation between participant and event plane 
    angles (c), for different harmonic eccentricity and flow coefficients.
    The same 6000 events as in Fig.~\ref{F4} were analyzed.
 \label{F7}
 }
\end{figure*}

In contrast, the quadrangularity $\ecc_4$ shows a clear positive 
correlation with the ellipticity, see Fig.~\ref{F6}b. It is of geometrical 
origin: it reflects the football or almond shape of the overlap zone in 
non-central collisions which is a little sharper than a pure $\cos(2\phi)$ 
deformation. This is corroborated by the angle $\psi_4^\mathrm{PP}$ shown 
in Fig.~\ref{F7}a which, on average, points $45^\circ$ relative to 
$\psi_2^\mathrm{PP}$ (which again points in $x$-direction). This means 
that the quadrangular component of the initial fireball definition is 
oriented like a diamond, with its corners on the $x$ and $y$ axes. 
Superimposing it on a pure $\cos(2\phi)$ deformation leads to a somewhat 
sharper shape of the density distribution.

\section{Event-by-event hydrodynamics and flow fluctuations}
\label{sec4}

In this section we analyze the results from ideal fluid event-by-event
hydrodynamic evolution of the fluctuating initial profiles studied in
the previous section. We focus on the anisotropic flow coefficients $v_n$,
their relationship to the initial eccentricity coefficients $\ecc_n$,
and the correlation between the $n^\mathrm{th}$-order flow angles 
$\psi_n^\mathrm{EP}$ and the corresponding $n^\mathrm{th}$-order 
participant-plane angles $\psi_n^\mathrm{PP}$ associated with $\ecc_n$.

%
%

\subsection{Correlations between participant plane, event plane, and
            reaction plane}
\label{sec4a}

One of the key characteristics of fluid dynamics is its ability
to transform initial geometric deformation into a deformation of
the final momentum distribution, via collective flow. This happens
through spatially anisotropic hydrodynamic forces (i.e. pressure
gradients) which cause anisotropic acceleration of the fluid. As a 
result, correlations between participant and event planes are expected:
The angle $\psi_n^\mathrm{PP}$ points in the direction of the largest
pressure gradient associated with the $n^\mathrm{th}$ harmonic component
of the spatial deformation of the initial density distribution, while
$\psi_n^\mathrm{EP}$ points into the direction where the $n^\mathrm{th}$ 
harmonic component of the final collective flow is largest. Without
interference between harmonics of different order, we would thus expect
$\psi_n^\mathrm{PP}$ and $\psi_n^\mathrm{EP}$ to point, on average
and up to event-by-event fluctuations, in the same direction.

In Figs.~\ref{F7}a,b we show the distribution of participant and event plane 
angles, associated with the $n^\mathrm{th}$-order eccentricities and 
harmonic flows, relative to the $x$-$z$ reaction plane. The analysis
uses the same 6000 events as before. In panel (a) we see that 
$\psi_{3,5}^\mathrm{PP}$ are completely uncorrelated with the reaction 
plane \cite{Qin:2010pf}, as expected from the fact that the corresponding 
eccentricities are entirely fluctuation-driven, without contribution from 
the collision geometry. Panel (b) shows that the same holds true
for $\psi_{3,5}^\mathrm{EP}$, which is (at least superficially) consistent
with the expectation that $v_3$ is mostly or entirely driven by $\ecc_3$,
and $v_5$ by $\ecc_5$. We will revisit this below. $\psi_2^\mathrm{PP}$
and $\psi_2^\mathrm{EP}$ are strongly correlated with the reaction plane,
at least for this mixed-centrality set of events. This is expected since,
for non-central collisions, $\ecc_2$ is mostly controlled by the almond-shaped
overlap geometry, and $v_2$ is mostly a collective flow response to this
geometric deformation; event-by-event fluctuations contribute to $\ecc_2$
(and thus $v_2$), but in general do not dominate them.

The behavior of $\psi_4^\mathrm{PP}$ in Fig.~\ref{F7}a is interesting 
because it is on average strongly ``anti-correlated'' with the reaction 
plane, in the sense that it points (on average) at $45^\circ$ relative to 
the $x$-axis. The geometric reason for this has already been discussed above 
in subsection \ref{sec3d}. On the other hand, Fig.~\ref{F7}b shows that
the angle $\psi_4^\mathrm{EP}$ points on average {\em into} the reaction
plane. This correlation of  $\psi_4^\mathrm{EP}$ with the reaction plane 
is somewhat weaker than the anti-correlation of $\psi_4^\mathrm{PP}$ with
that plane seen in panel (a). Still, it suggests that quadrangular flow 
$v_4$ does not, on average, develop predominantly in the direction of the 
steepest pressure gradient associated with $\ecc_4$, but in the direction 
of steepest $\ecc_2$-induced pressure gradient. This can be understood 
as follows: since $\ecc_2$ generates a second harmonic deformation of the 
flow velocity profile which elliptically deforms {\em the exponent} of the 
flow-boosted Boltzmann factor $\exp[-p\cdot{u}(x)/T(x)]$ describing the
local thermal momentum distribution of particles, it leads to harmonic 
contributions $v_{2k}$ of {\em all} even orders $n\eq2k$ in the momentum 
distributions of the finally emitted particles \cite{Borghini:2005kd}. 
Fig.~\ref{F7}b suggests that, on average, this effect wins over 
initial-state quadrangular deformation effects.  

Figure~\ref{F7}c, however, in which we analyze directly the correlation 
between the event and participant plane angles, paints a more subtle 
picture. It shows, surprisingly, a correlation peak at zero relative
angle between $\psi_4^\mathrm{EP}$ and $\psi_4^\mathrm{PP}$, whereas the
above discussion should have led us to expect a correlation peak at $45^\circ$.
The resolution of this paradox is presented in the next subsection:  
The relative importance of geometric and fluctuation-induced contributions
to $\ecc_n$, $v_n$, and their associated angles changes with collision 
centrality, with geometry playing a relatively larger role in peripheral 
collisions. One should therefore look at the angle correlations as a
function of collision centrality. One finds that the correlation function
peaks in Figs.~\ref{F7}a,b for the $4^\mathrm{th}$-order angles relative 
to the reaction plane are almost entirely due to geometric effects in 
peripheral collisions, while in central collisions both $\psi_4^\mathrm{PP}$ 
and $\psi_4^\mathrm{EP}$ are fluctuation-dominated and thus essentially
uncorrelated with the reaction plane. On the other hand, precisely because
in central collisions geometric effects such as geometrically driven elliptic
flow do not dominate the hydrodynamic response to the fluctuation-driven 
higher-order eccentricities, $\psi_4^\mathrm{EP}$ and $\psi_4^\mathrm{PP}$ 
remain relatively strongly correlated in near-central collisions. This 
is the reason for the peak at $0^\circ$ for $n\eq4$ in Fig.~\ref{F7}c. 
(A hint of the ``anti-correlation'' at $45^\circ$ is still visible in 
Fig.~\ref{F7}c, and it would be stronger if we had not (for unrelated 
reasons) strongly oversampled central collisions in our mixed-centrality 
sample.)

%
\begin{figure*}
 \includegraphics[width=0.45\linewidth,height=6.87cm]{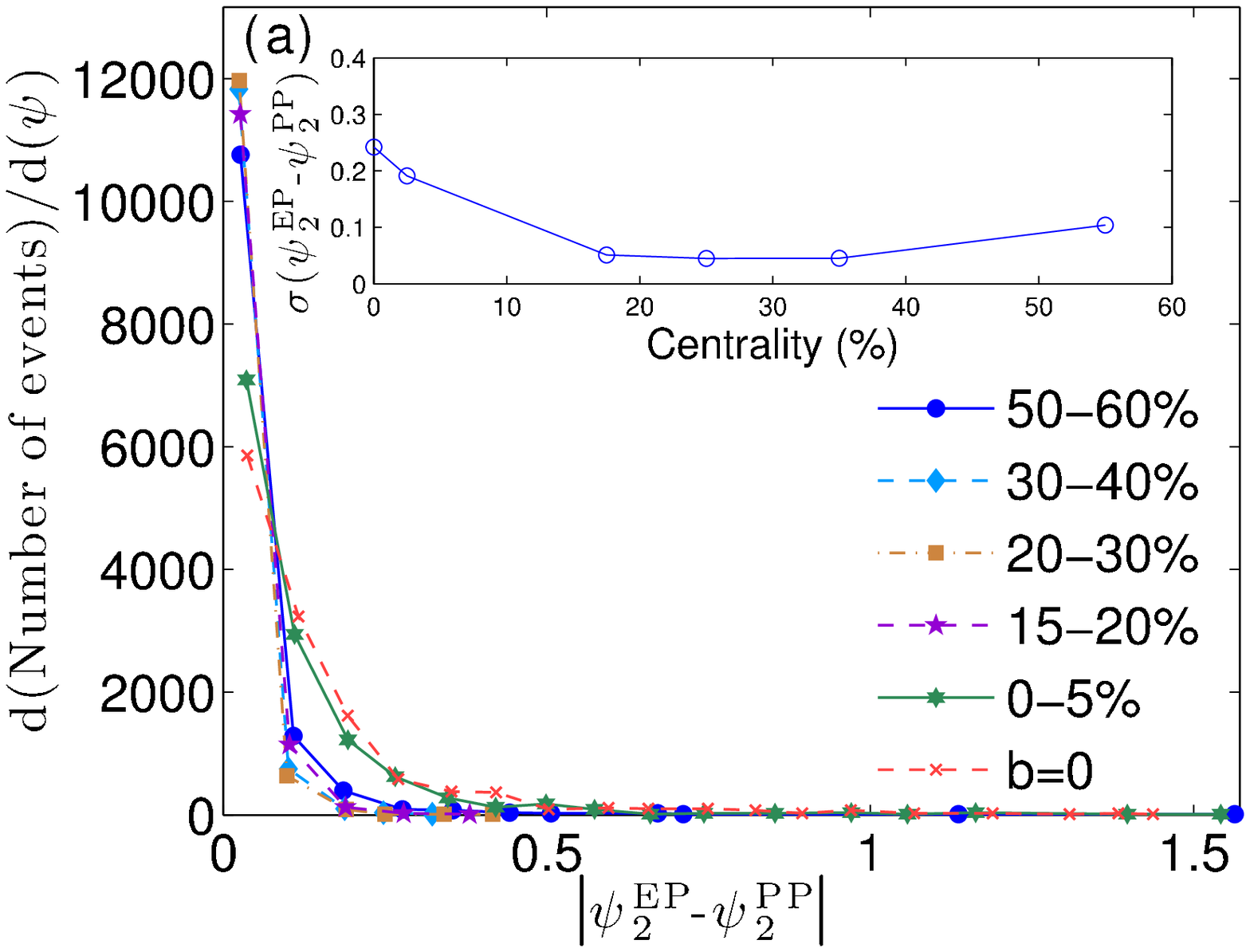}
 \includegraphics[width=0.45\linewidth]{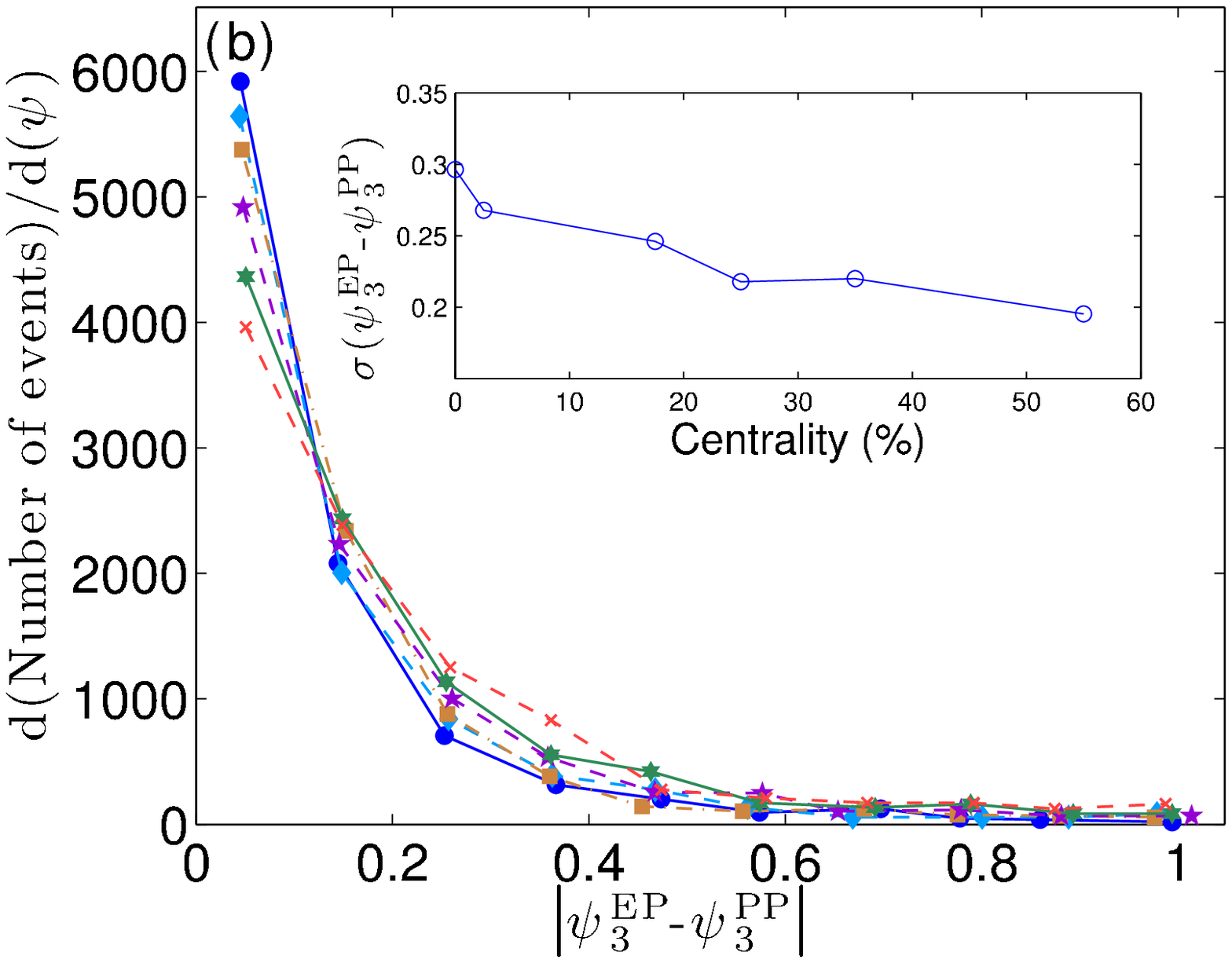}\\
 \includegraphics[width=0.45\linewidth]{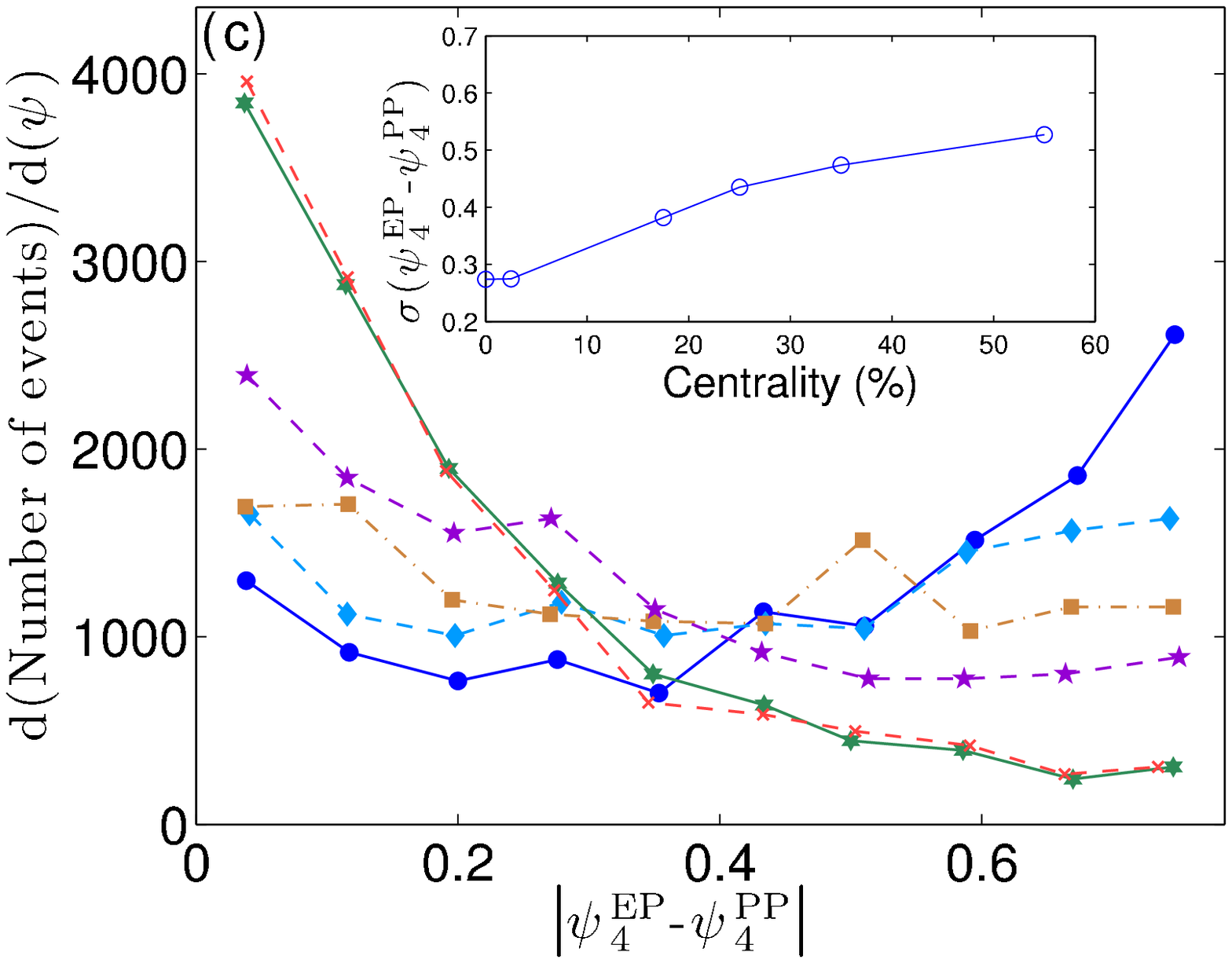}
 \includegraphics[width=0.45\linewidth]{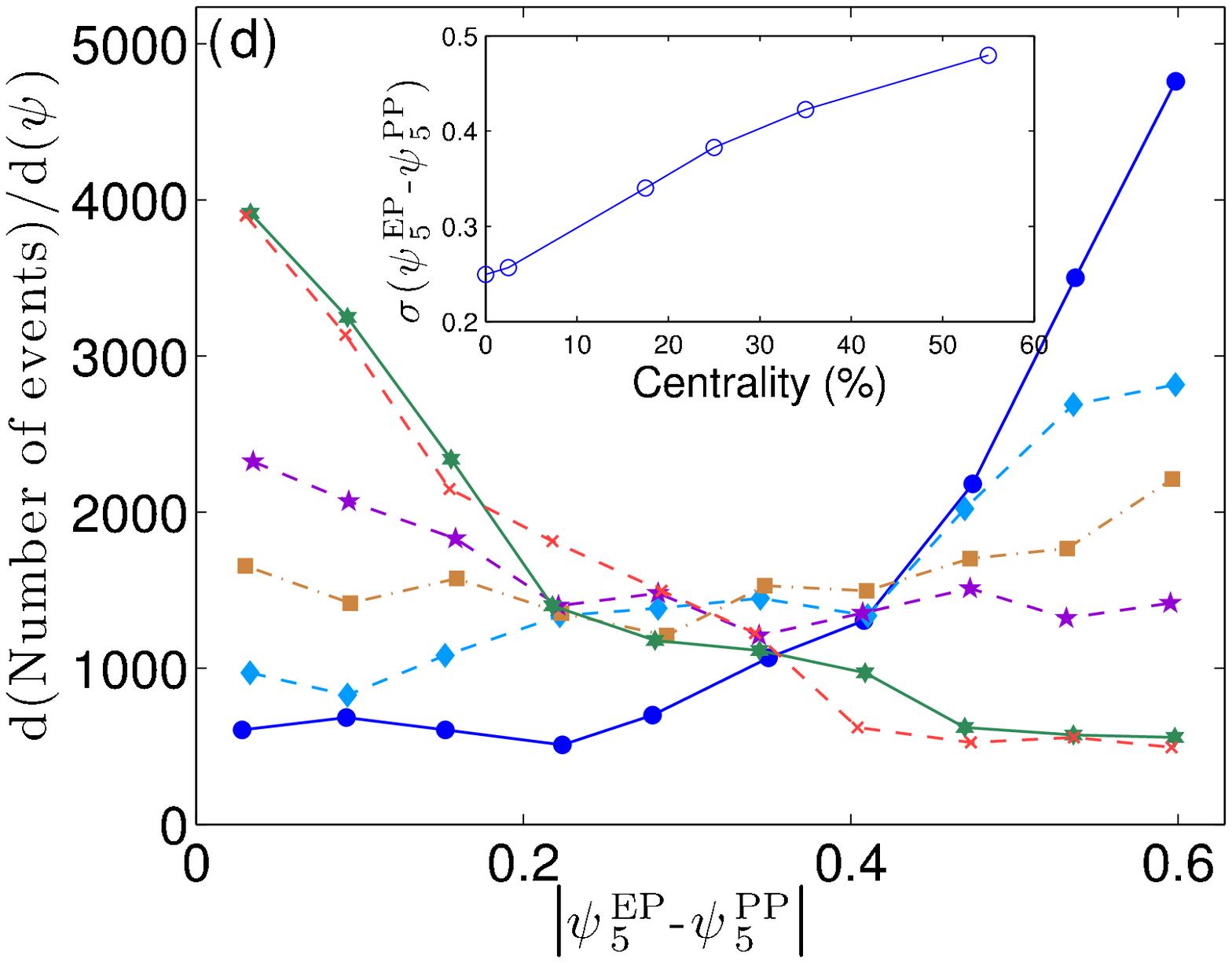}
 \caption{(color online) Event-by-event correlation between the absolute 
   values of the event plane (EP) and participant plane (PP) angles for
   the harmonics of order $2-5$ (panels (a)-(d)), for events in different 
   centrality classes as indicated in the legend. (The same set of MC-KLN 
   events as in Fig.~\ref{F4} was used.) The insets show the centrality 
   dependence of the widths of these correlations around zero.
   \label{F8}
 }
\end{figure*}
%

We close this discussion with the following additional observations
about Fig.~\ref{F7}c: (i) The second-order participant and event planes
are much more strongly correlated with each other than either one of them
is with the reaction plane. This shows that even in very central collisions, 
where the source ellipticity is mostly fluctuation-driven and its angle 
therefore only weakly correlated with the reaction plane, elliptic flow 
develops event-by-event in the direction of the short axis of the 
ellipsoid. (ii) Even though the angles associated with $\ecc_3$ and $v_3$
are uncorrelated with the reaction plane (Figs.~\ref{F7}a,b), they
are strongly correlated with each other. This indicates that $v_3$
is mostly driven by $\ecc_3$, especially in the more central collisions, 
with relatively little interference from other harmonics. (iii) The  
$5^\mathrm{th}$-order event and participant plane angles show correlation 
peaks both at $0$ and $\pi/5$. As we will see in the following subsection, 
the former results from central and the latter from peripheral collisions.
The peak at $\pi/5$ indicates significant cross-feeding between modes 
with $n\eq2,\,3$, and 5.

\subsection{Centrality dependence of event and participant plane 
            correlations}
\label{sec4b}

Figure~\ref{F8} looks at the correlation between the $n^\mathrm{th}$-order 
EP and PP angles at different collision centralities. This generalizes a 
similar analysis for $n\eq2$ in Ref.~\cite{Holopainen:2010gz} to higher 
harmonics. Plotted are the distributions of the absolute value of the 
difference between the two angles in the main graph and the rms of this 
distribution (i.e. the width around zero of the correlation) in the inset, 
as a function of collision centrality. Panel (a) shows that the 
second-order participant and event planes are strongly correlated at all 
collision centralities. This demonstrates that elliptic flow is 
generated almost exclusively by the source ellipticity. The variance of 
the correlation is $\sim0.05$\,rad in the mid-central range (15-40\% 
centrality) and increases in very central and very peripheral collisions 
due to growing ellipticity fluctuations.

A similar correlation exists for the $3^\mathrm{rd}$-order participant
and event planes, at all collision centralities, but with a larger
variance of order $0.2-0.3$\,rad (depending on centrality). The relatively
strong correlation suggests that $\ecc_3$ is the dominant driver for $v_3$ 
\cite{Alver:2010gr}.

%
\begin{figure*}
\begin{center}
 \includegraphics[width=0.45\linewidth]{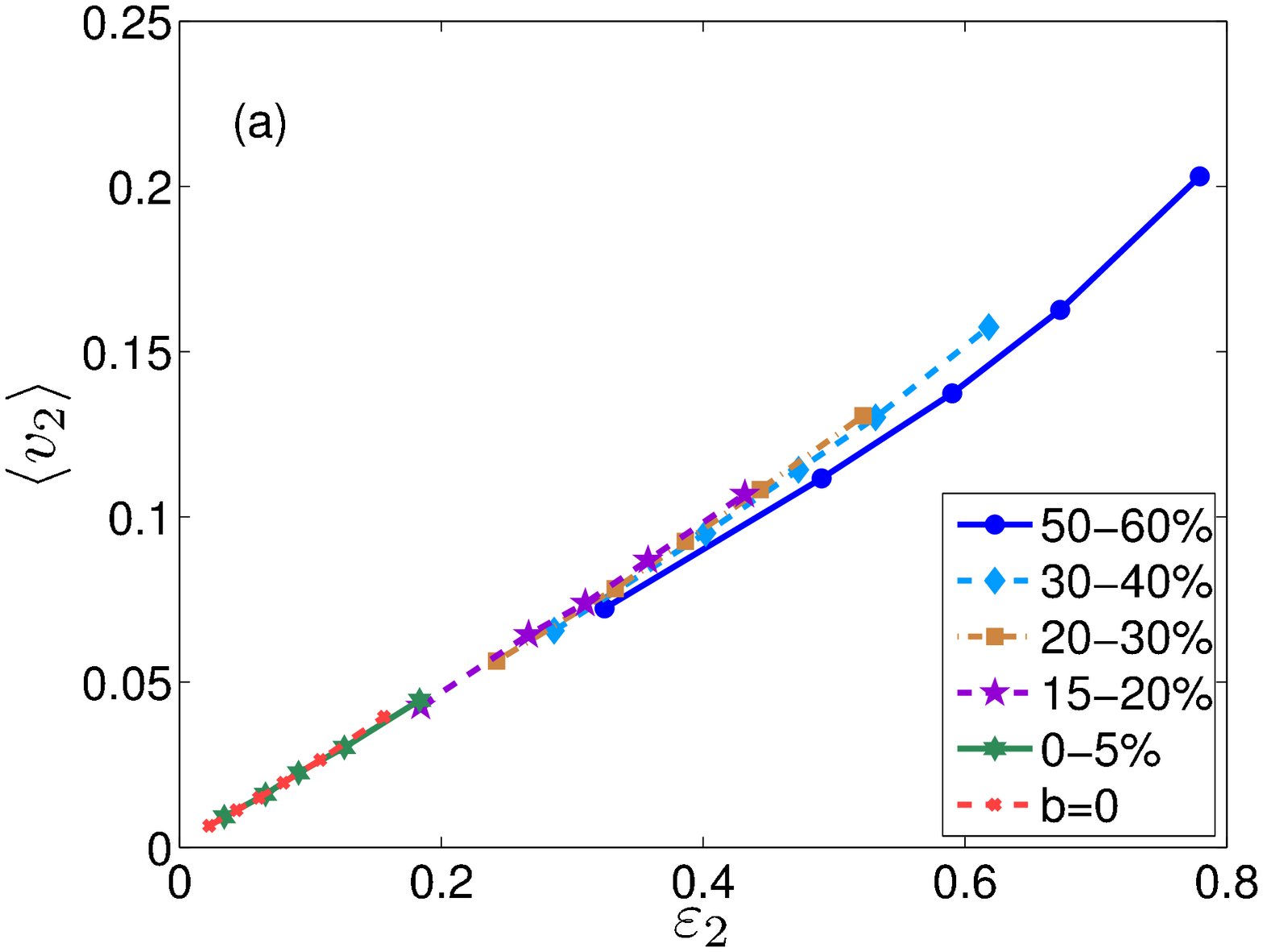}
 \includegraphics[width=0.45\linewidth]{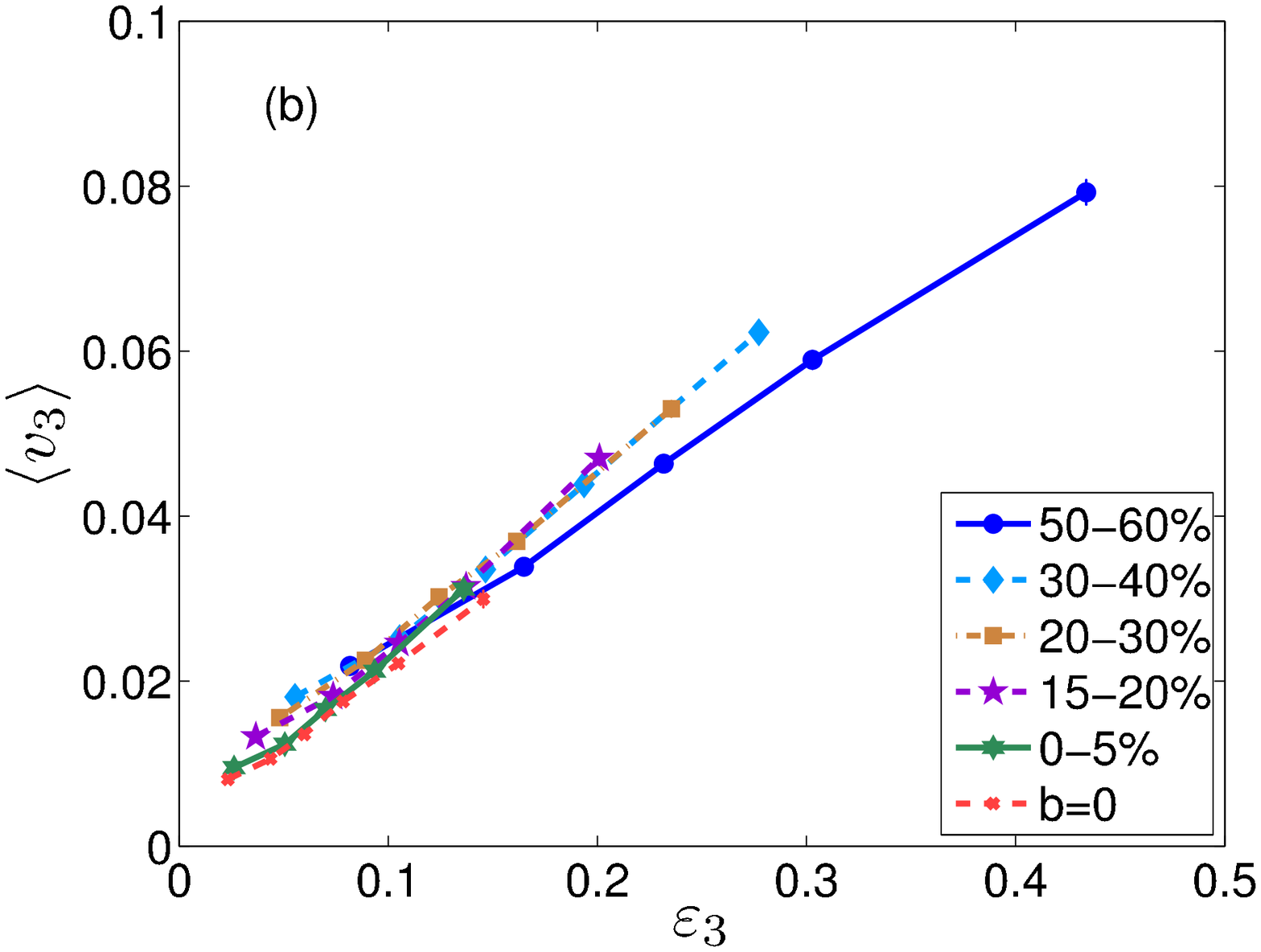}\\
 \includegraphics[width=0.45\linewidth]{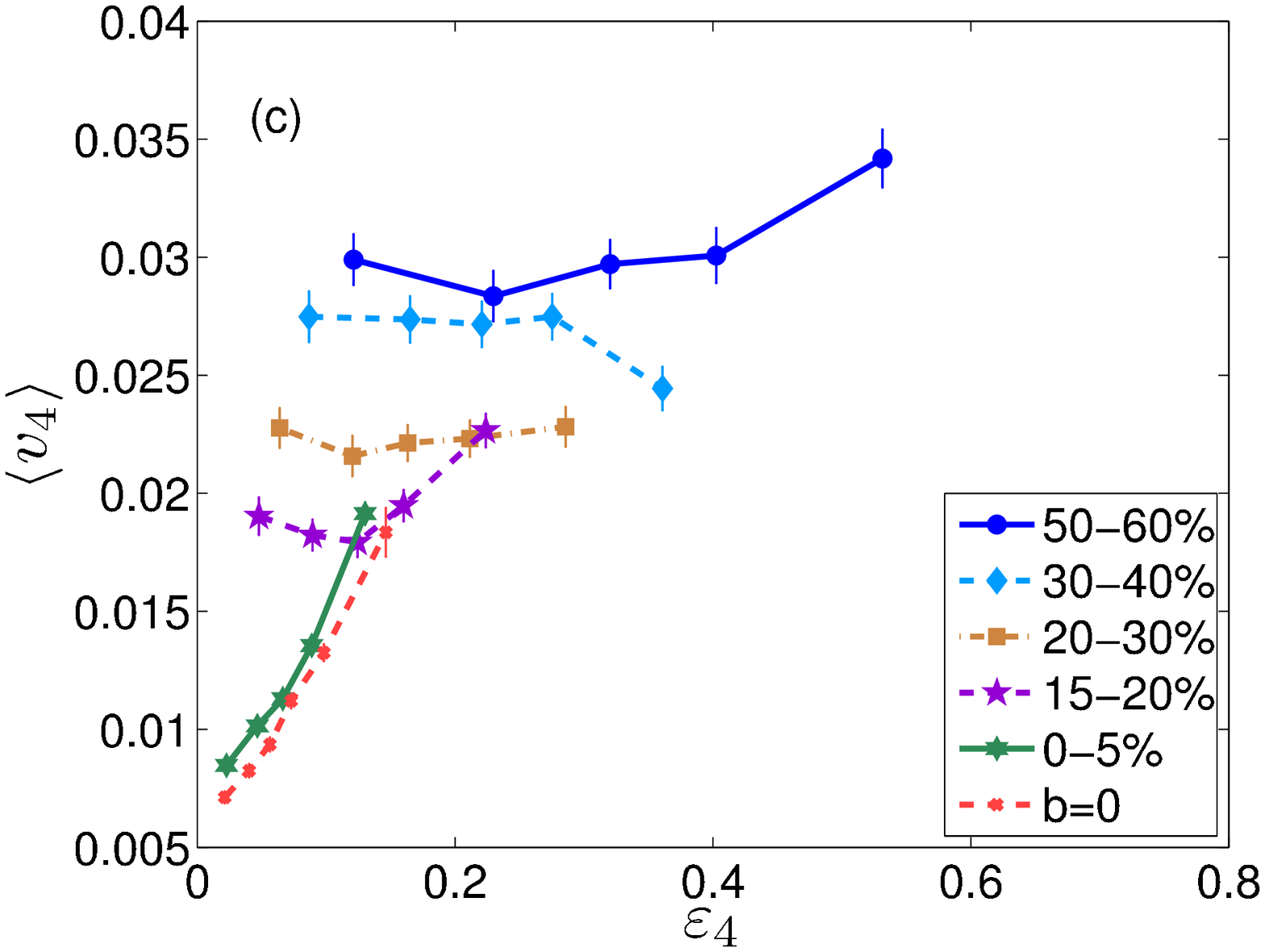}
 \includegraphics[width=0.45\linewidth]{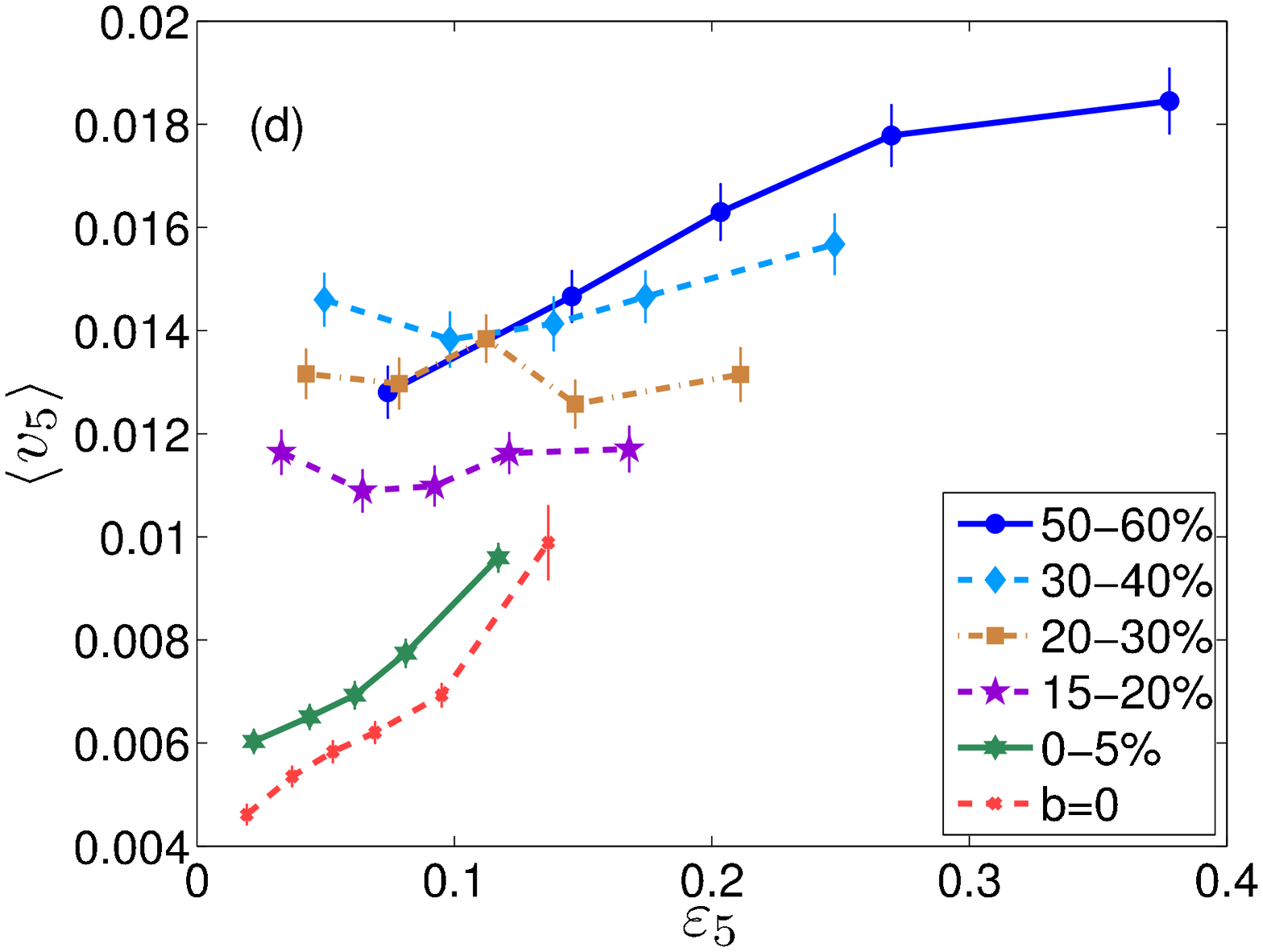}
 \caption{(Color online) $\la v_n\ra(\ecc_n)$ for $n\eq2,3,4,5$ 
   (panels (a)-(d)). As in Fig.~\ref{F4}, each centrality 
   class contains 1000 MC-KLN events, grouped in eccentricity bins 
   of sufficient width to have reasonable statistics in each bin.
 \label{F9}
 }
\end{center}
\end{figure*}
%

For the $4^\mathrm{th}$- and $5^\mathrm{th}$-order participant and event
planes the situation is complicated, as seen in panels (c) and (d). The
planes are correlated with each other (i.e. the distributions peak
at zero difference angle) in central collisions, become essentially 
uncorrelated in mid-central collisions and anti-correlated (i.e.
peaked at a difference angle of $\pi/n$, $n\eq4,5$) in peripheral collisions.
The anti-correlation in peripheral collisions indicates strong mode-mixing,
driven by the large ellipticity $\ecc_2$ and strong elliptic flow $v_2$
at large impact parameters which generates $v_4$ and $v_5$ contributions
by coupling to lower harmonics, as described in the previous subsection.
For $v_4$ in particular, a strong $\cos(2\phi)$ component in the collective
flow velocity generates a $v_4$ of the final momentum distribution, without
any need for nonzero $\ecc_4$. At large impact parameters, $\ecc_2$-induced
quadrupolar flow from the initial elliptic deformation of the overlap 
region thus dominates over any contribution from initial quadrangular 
deformation. In near-central collisions, on the other hand, where all 
$\ecc_n$ stem mostly from shape fluctuations, $v_{4,5}$ are dominantly 
driven by $\ecc_{4,5}$.

\vspace*{-3mm}
\subsection{Harmonic flows and their corresponding initial 
eccentricities: nonlinear hydrodynamic response}
\label{sec4c}
\vspace*{-3mm}

As discussed in the Introduction, it is often assumed that the harmonic
flows $v_n$ respond linearly to the eccentricities $\ecc_n$, at least as
long as the latter are small. This assumption receives support from
hydrodynamic simulations \cite{Alver:2010dn} as long as one probes 
deformed initial profiles with only a single non-vanishing harmonic 
eccentricity coefficient. In Fig.~\ref{F9} we investigate the validity of 
this assumption with fluctuating MC-KLN events which feature nonzero 
$\ecc_n$ values for all $n$. 
  
Figure~\ref{F9}a generally provides support for the assumption of a
linear dependence of the elliptic flow $v_2$ on initial ellipticity 
$\ecc_2$, with two important caveats: 
\begin{itemize}
\item[(i)]
At small and large ellipticities,
$v_2$ deviates upward from a best-fit line through the origin, indicating
additional contributors to the elliptic flow. Indeed, for zero ellipticity
$\ecc_2\eq0$ we find a nonzero average $\la v_2\ra$. These are events 
with typically large nonzero values for eccentricities of higher harmonic
order which generate elliptic flow through mode-mixing (e.g. between $\ecc_3$
and $\ecc_5$). We see that this happens at all centralities, even for
$b\eq0$, due to event-by-event fluctuations of the eccentricity coefficients.
\item[(ii)]
The slope of the curve $\la v_2\ra(\ecc_2)$ decreases in very peripheral
collisions, indicating destructive interference via mode-mixing from other
harmonics in the hydrodynamic evolution of the small and highly fluctuating
fireballs created at large impact parameters.
\end{itemize} 

%
\begin{figure*}
\begin{center}
 \includegraphics[width=0.44\linewidth]{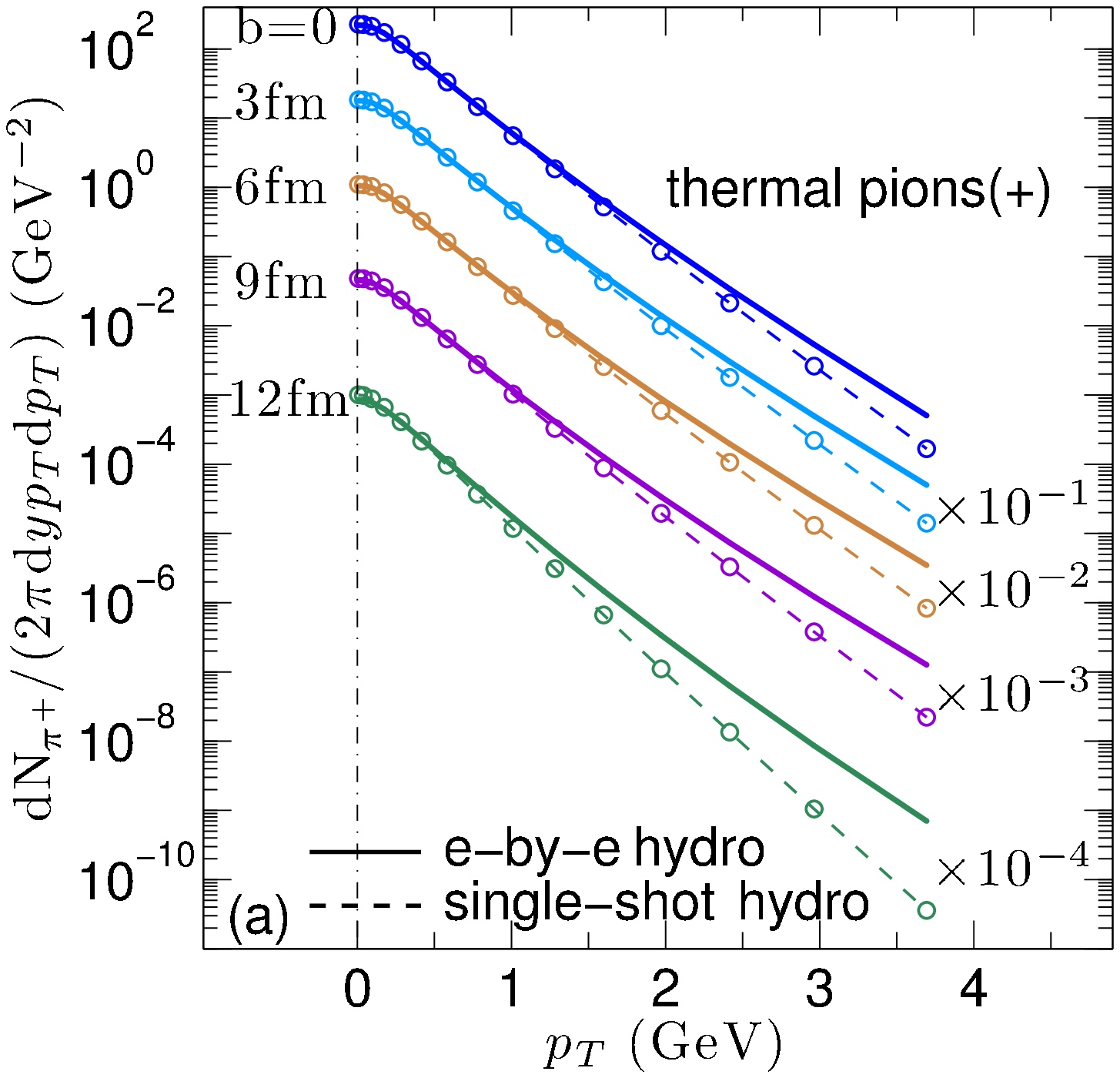}
 \includegraphics[width=0.45\linewidth]{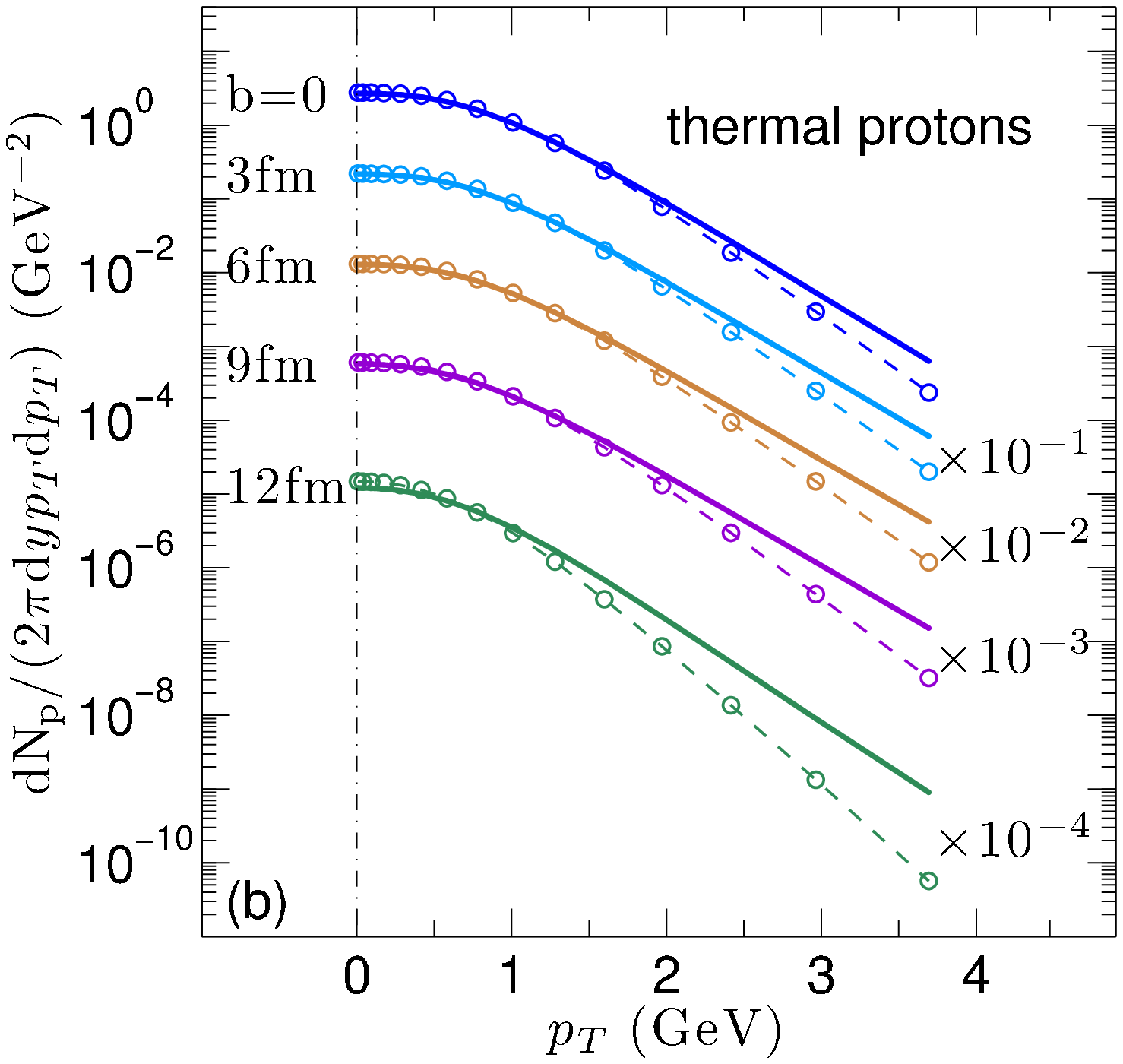}
 \caption{(Color online) Transverse momenrum spectra for directly
 emitted $\pi^+$ (a) and protons (b) from event-by-event (solid lines)
 and single-shot hydrodynamics (dashed lines), for $200\,A$\,GeV Au+Au 
 collisions at five selected impact parameters. 
 \label{F10}
 }
\end{center}
\end{figure*}
%

The $\ecc_3$-dependence of triangular flow $\la v_3\ra$, shown in 
Fig.~\ref{F9}b, shows a qualitatively similar story, but the deviations
from linear response are stronger, with significant non-zero triangular
flow in events with zero initial triangularity, especially for larger
impact parameters. 

For $\la v_4\ra$ and $\la v_5\ra$, shown in Figs.~\ref{F9}c and \ref{F9}d, 
mode-mixing effects are very strong, and a linear response of $v_n$ to
$\ecc_n$ ($n=4,5$) can no longer be claimed. This is quite different from
the results in \cite{Alver:2010dn} where $v_4$ was studied for a source
that had only $\ecc_4$ deformation: in this case $v_4(\ecc_4)$ was found to
be approximately linear for small $\ecc_4$, with a downward bend at larger
$\ecc_4$ values due to negative interference from cross-currents for
sources with large quadrangularities. (This approximately linear dependence
survived in the $p_T$-integrated $v_4$ even though it was noticed in a 
related study \cite{Luzum:2010ae} that, for mid-central collisions, the 
differential quadrangular flow $v_4(p_T)$ appears at high $p_T$ to be 
mostly determined by the elliptic deformation of the hydrodynamic flow 
profile generated by $\ecc_2$.) Our present study 
shows that it is unlikely that the anisotropic flow resulting from highly 
inhomogeneous initial profiles with nonzero eccentricity coefficients of 
all harmonic orders can be obtained by some sort of linear superposition 
of flows generated from sources with only a single nonzero harmonic 
eccentricity coefficient, as suggested \cite{Teaney:2010vd}. The 
hydrodynamic response $\{v_n\}$ to a set of initial eccentricity 
coefficients $\{\ecc_n\}$ is not only non-diagonal, but also (via 
mode-mixing) non-linear, and there is no suitable single-shot substitute 
for event-by-event hydrodynamic evolution of fluctuating initial 
conditions.

We note, however, that non-linear mode-mixing effects appear to be 
minimal for the elliptic and triangular flow (Figs.~\ref{F9}a,b). $v_2$
and $v_3$ remain therefore the best candidates for an extraction of the 
fluid's viscosity, by studying (with quantitative precision) the fluid's 
efficiency in converting initial spatial deformations into final momentum 
anisotropies and anisotropic flows. We will further elaborate on this 
theme in the next section. 

\section{Single-shot versus event-by-event hydrodynamics}
\label{sec5}

We now discuss the effects of event-by-event initial-state fluctuations 
on the finally observed pion and proton $p_T$-spectra and anisotropic flow, 
comparing traditional single-shot hydrodynamic evolution of an 
appropriately constructed smooth average initial profile with
event-by-event evolution of fluctuating initial conditions (with an 
ensemble average taken at the end). Since the calculation of resonance 
decay feeddown corrections is computationally expensive but not expected 
to cause qualitative changes, we here concentrate on directly emitted 
(``thermal'') pions and protons. For the graphs shown in this section, 
we generated for each impact parameter 1000 fluctuating events and
propagated them either individually (``event-by-event hydrodynamics'') 
or in a single hydrodynamic run after rotating and averaging their profiles
(``single-shot hydrodynamics'') down to a decoupling temperature
of 120\,MeV.

\subsection{Transverse momentum spectra}
\label{sec5a}

%
\begin{figure*}
 \includegraphics[width=0.45\linewidth]{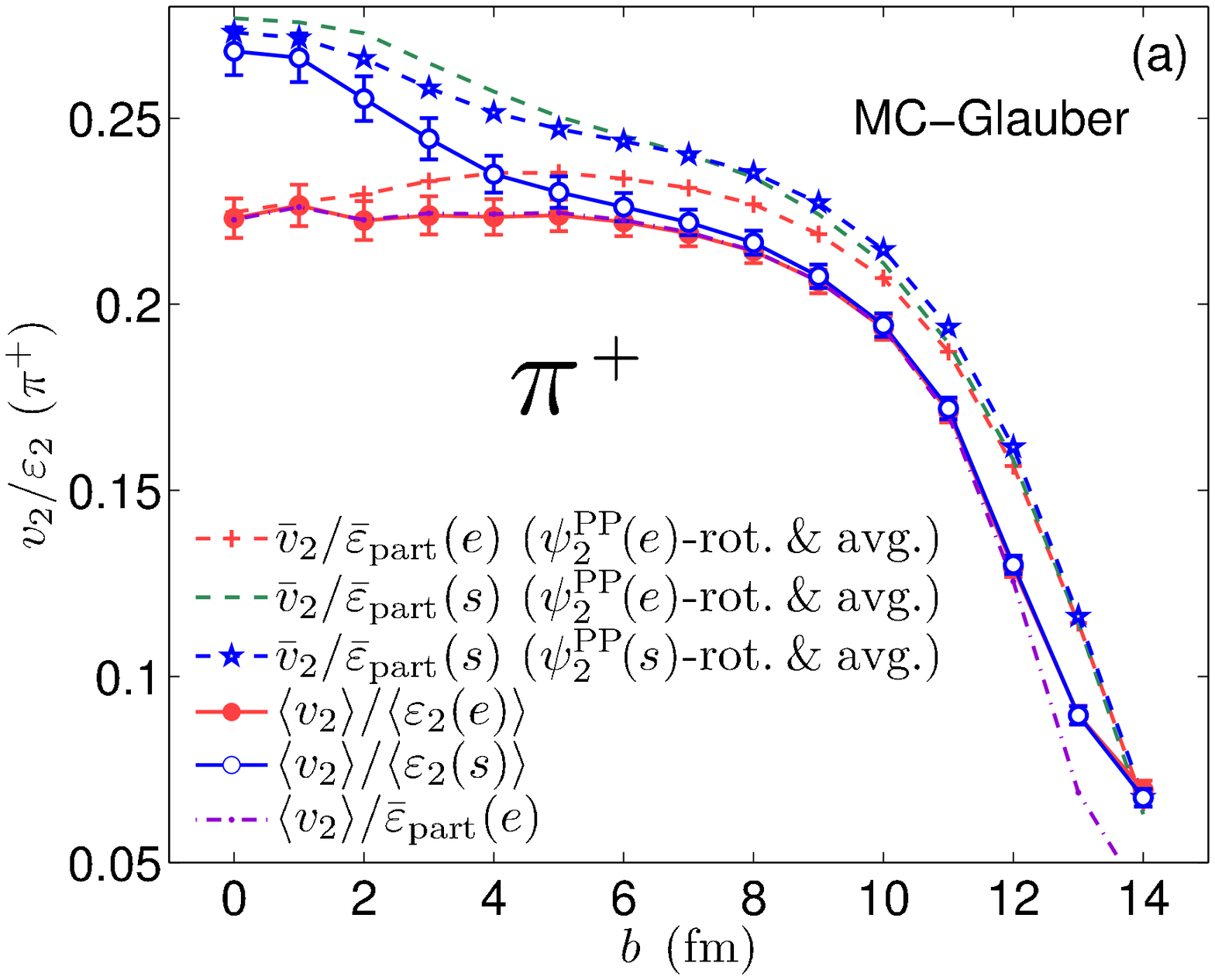}
 \includegraphics[width=0.45\linewidth]{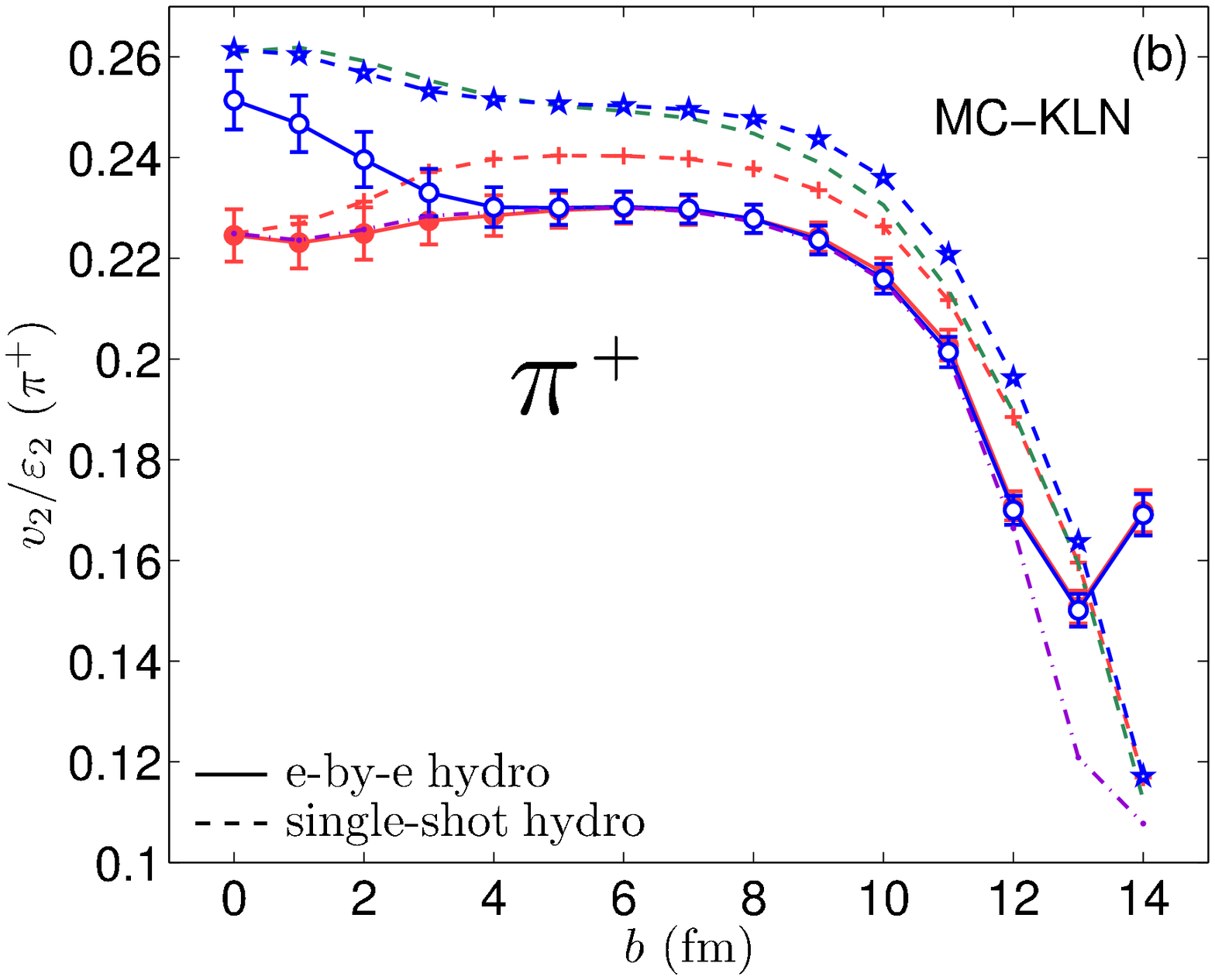}\\
 \includegraphics[width=0.45\linewidth]{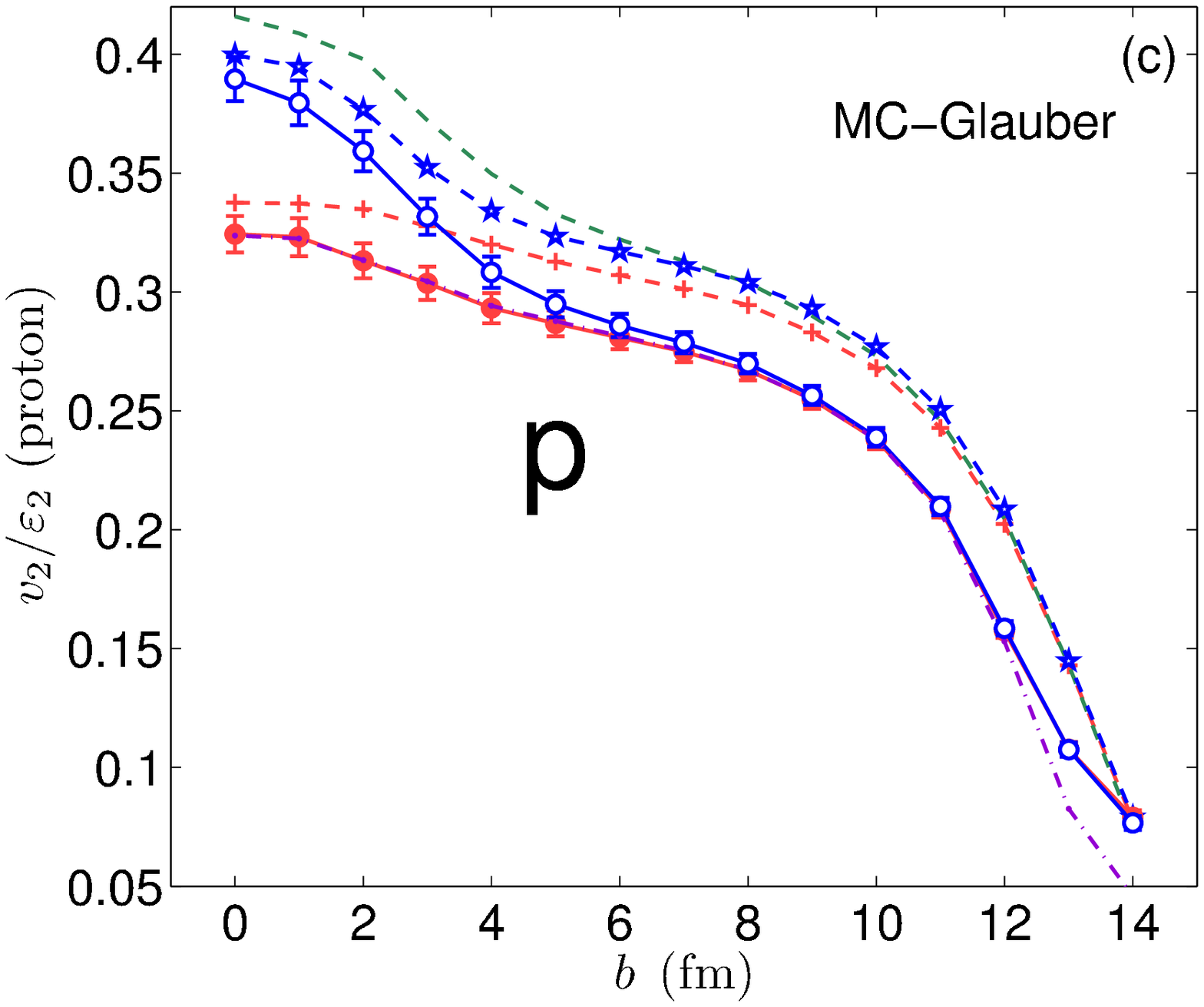}
 \includegraphics[width=0.45\linewidth]{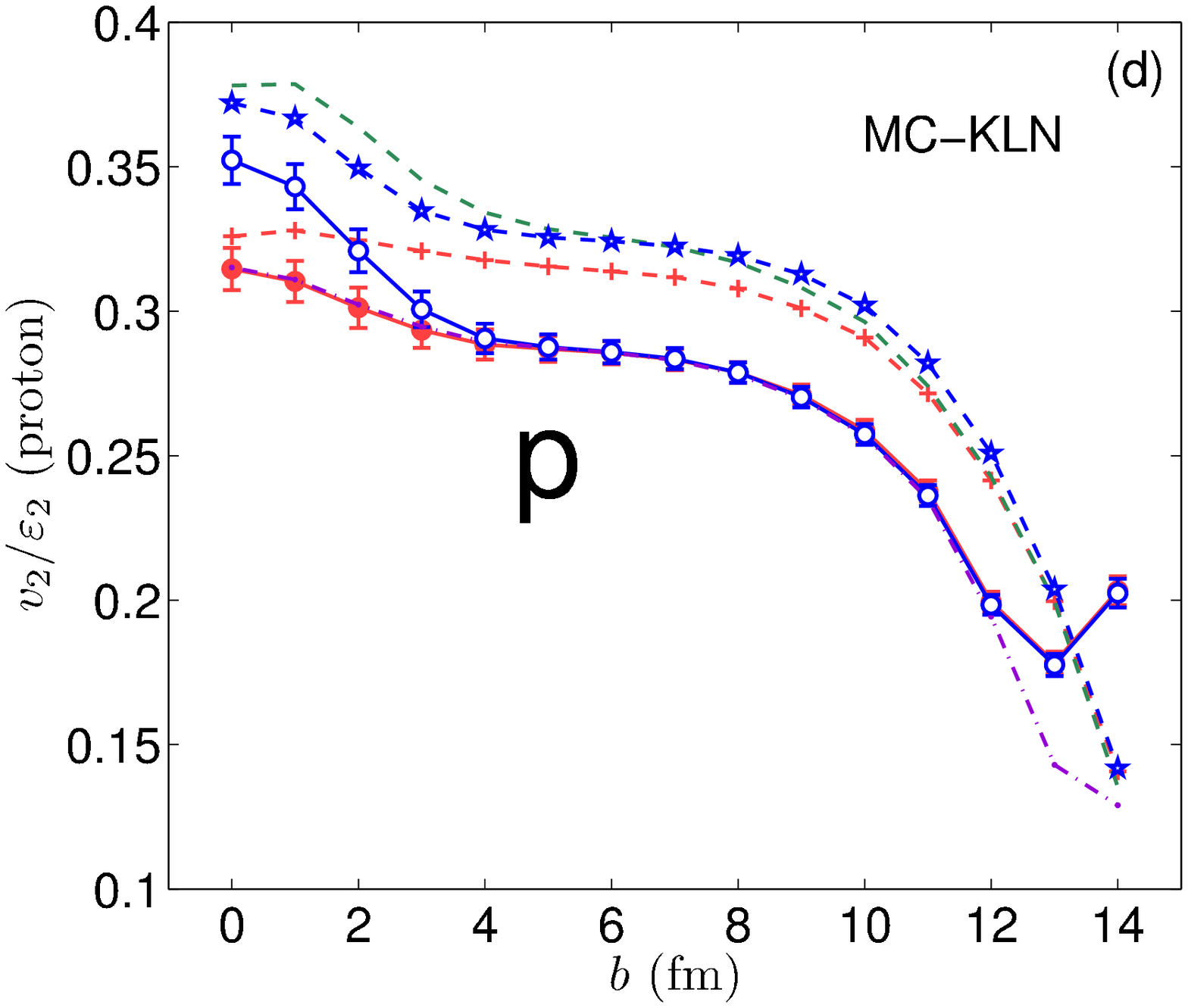}
 \caption{(Color online) Eccentricity-scaled elliptic flow $v_2/\ecc_2$ as 
  function of impact parameter, for pions (panels (a,b)) and protons 
  (panels (c,d)), from the ideal fluid dynamic evolution of initial
  MC-Glauber (a,c) and MC-KLN (b,d) density profiles. Solid (dashed)
  lines correspond to event-by-event (single-shot) hydrodynamics.
  See text for discussion. 
 \label{F11}
 }
\end{figure*}
%

In a very interesting recent paper \cite{Chatterjee:2011dw} Chatterjee 
{\it et al.} showed that thermal photon spectra from exploding heavy-ion
collision fireballs with fluctuating initial conditions which were 
hydrodynamically evolved event-by-event are significantly harder than
those obtained from single-shot hydrodynamic evolution of the corresponding 
ensemble-averaged (and therefore much smoother) initial profiles. The 
authors of \cite{Chatterjee:2011dw} attributed this effect to the existence 
of ``hot spots'' in the fluctuating initial conditions that radiate photons 
at a higher than average temperature. Figure~\ref{F10} shows that the same
hardening occurs in the pion and proton spectra even though these strongly
interacting hadrons are emitted only at freeze-out, with the same decoupling 
temperature assumed in both types of evolution.\footnote{A similar 
  effect was also seen in \cite{Holopainen:2010gz} whose authors further 
  pointed out that the strength of this ``hardening effect'' depends on 
  the fluctuation size parameter in the initial conditions (i.e. the area 
  over which the entropy produced in a nucleon-nucleon collision is 
  distributed).}
This proves that the effect 
is due to stronger radial flow in the event-by-event evolved fluctuating 
fireballs, driven by the stronger than average pressure gradients associated 
with the ``hot spots'' (i.e. over-dense regions) in the initial profile. The
importance of initial-state fluctuation effects on the final $p_T$-spectra
becomes stronger in peripheral collisions where the initial fireballs are 
smaller and ``hot spots'' have a relatively larger influence. If stronger
radial flow is the explanation of the fluctuation-driven hardening of the
pion and proton spectra observed in Fig.~\ref{F10}, it is probably also a 
dominant contributor to the hardening of the photon spectra noted in 
Ref.~\cite{Chatterjee:2011dw}, at least for low $p_T$ (i.e. in
the hydrodynamic regime). This could be checked by comparing the
photon radiation from the late hadronic stage in event-by-event and
single-shot hydrodynamics which, if our interpretation is correct, should 
show the same fluctuation-driven, flow-induced hardening as the total 
photon spectra.  

%
\begin{figure*}
 \includegraphics[width=0.45\linewidth]{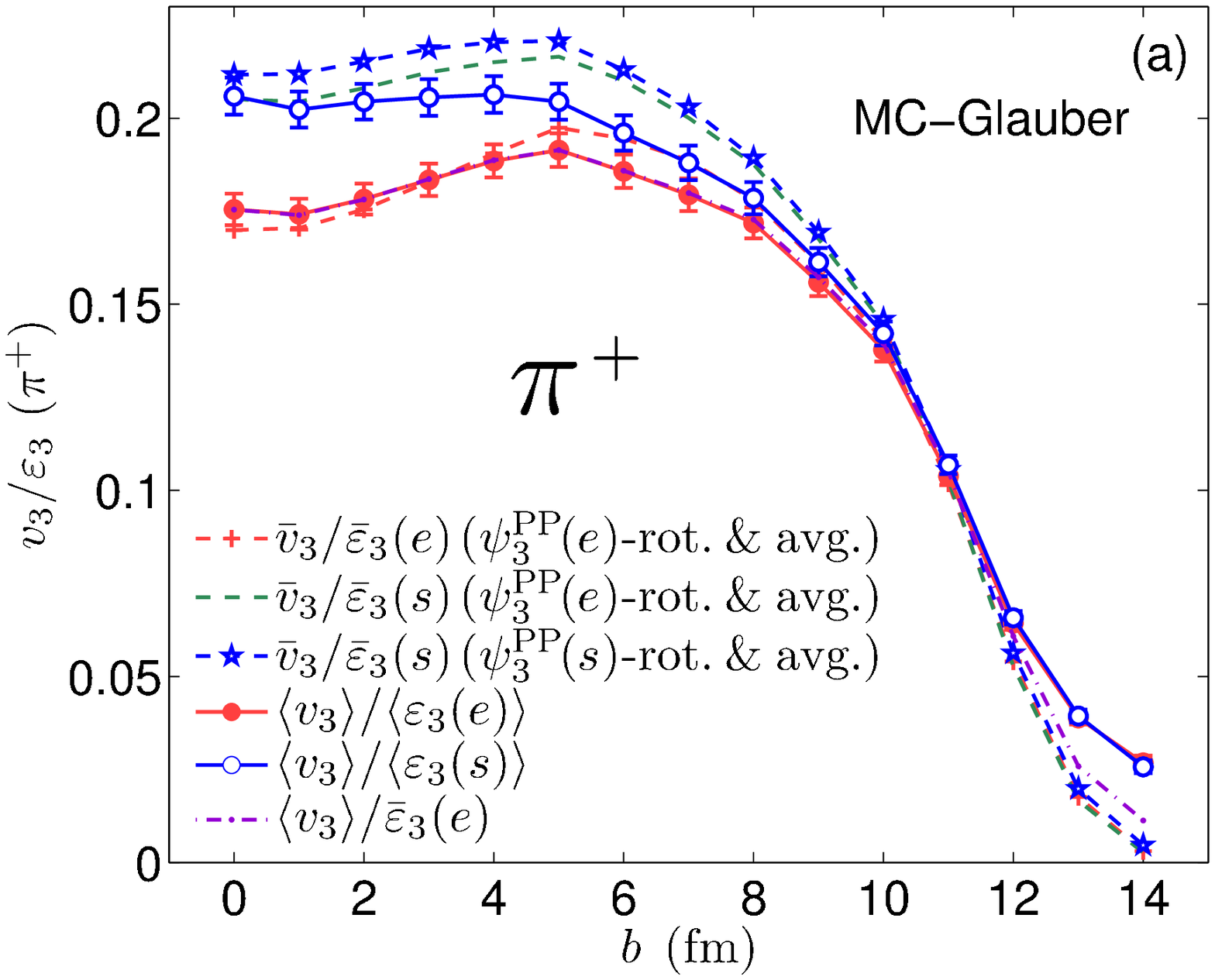}
 \includegraphics[width=0.45\linewidth]{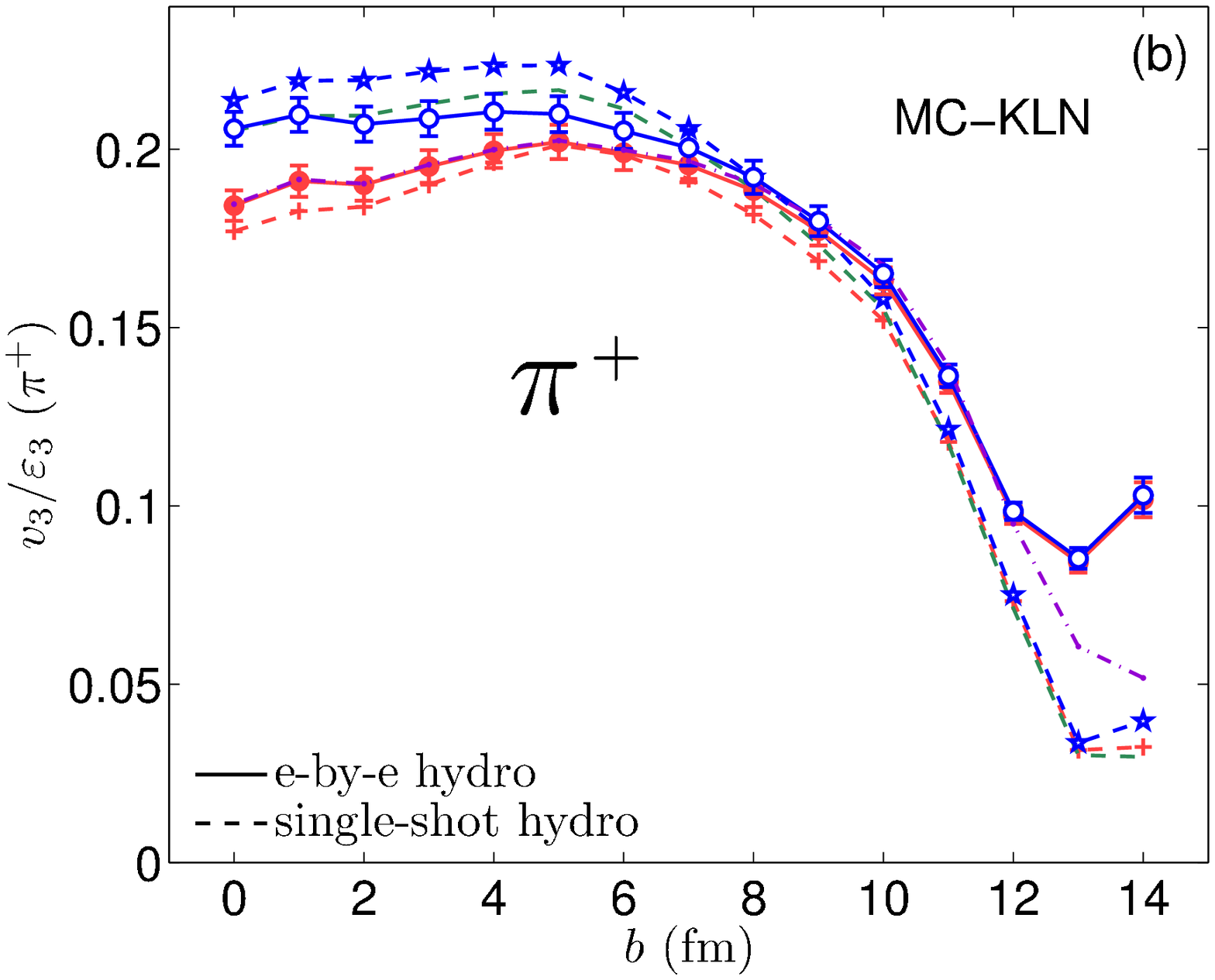}\\
 \includegraphics[width=0.45\linewidth]{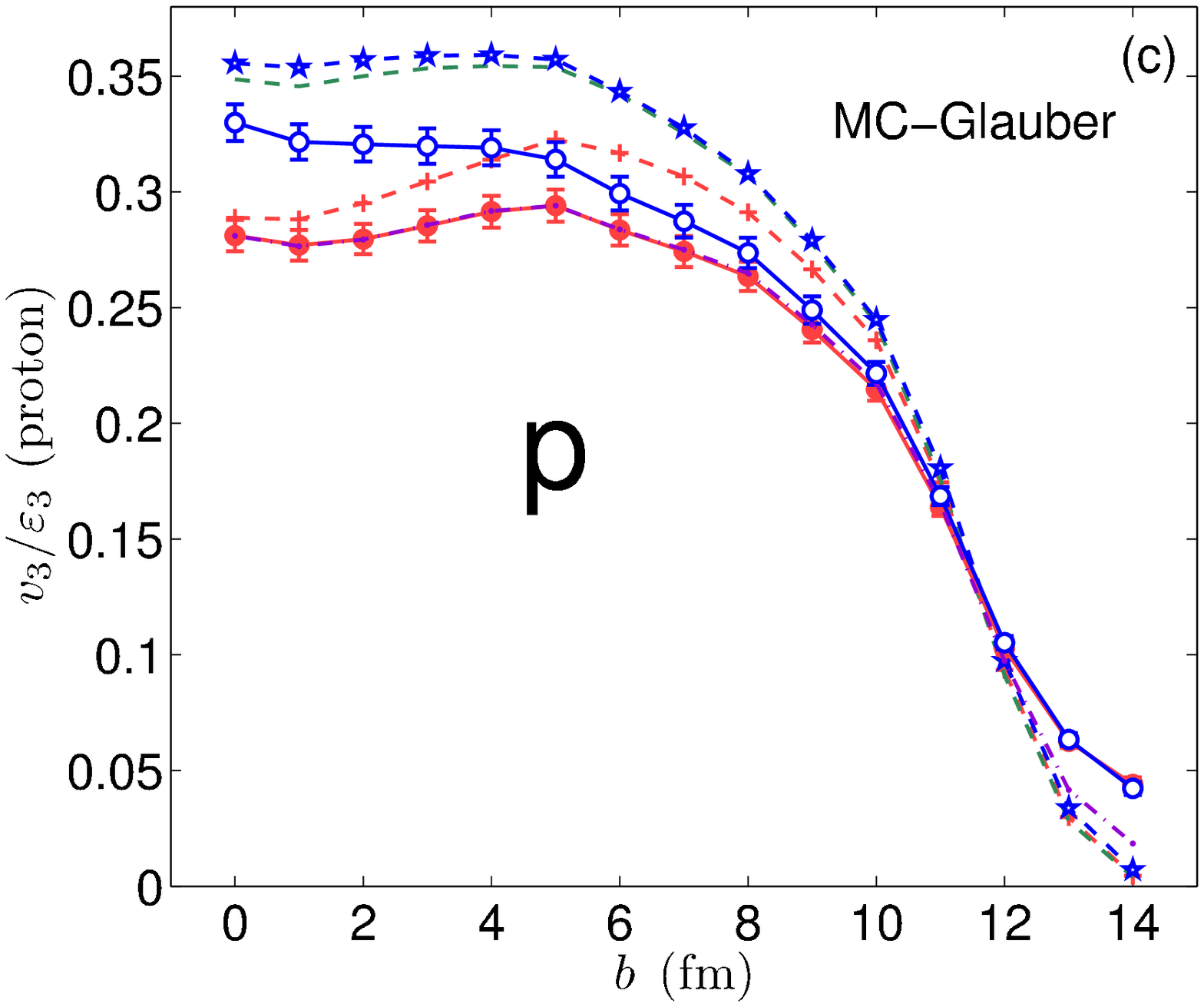}
 \includegraphics[width=0.45\linewidth]{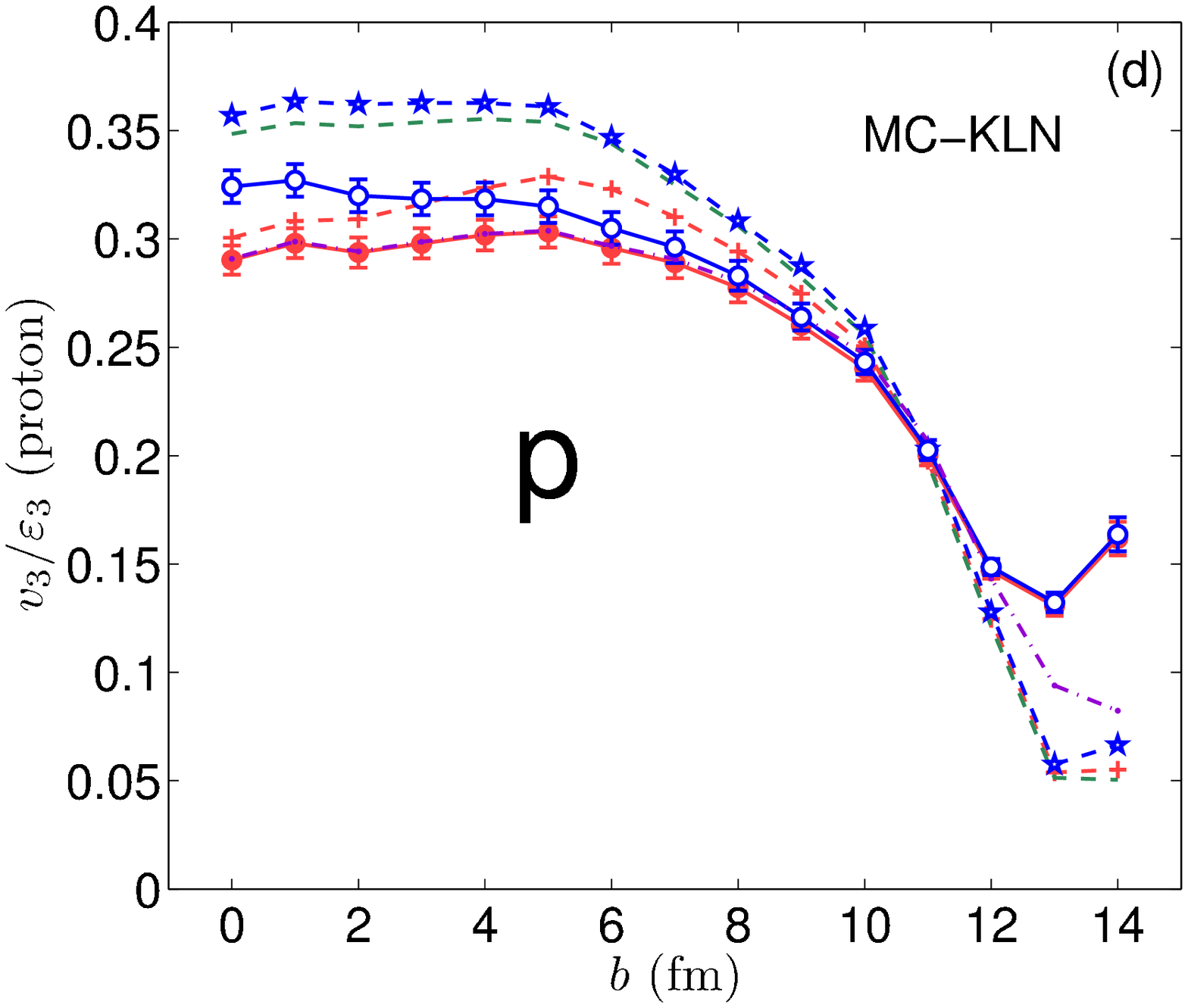}
 \caption{(Color online) Same as Fig.~\ref{F11}, but for the
 eccentricity-scaled triangular flow $v_3/\ecc_3$.
 \label{F12}
 }
\end{figure*}
%

\subsection{Elliptic and triangular flow}
\label{sec5b}

In Figures~\ref{F11} and \ref{F12}, we compare the eccentricity-scaled 
elliptic and triangular flows, $v_2/\ecc_2$ and $v_3/\ecc_3$, for pions 
and protons as a function of impact parameter, from single-shot (dashed
lines) and event-by-event hydrodynamics (solid lines). These ratios 
represent the efficiency of the fluid for converting initial spatial 
deformations into final-state momentum anisotropies. This conversion
efficiency is affected (i.e. reduced) by shear viscosity, so these 
ratios form the basis of many analyses that aim to extract this 
transport coefficient from experimental heavy-ion data. 

For event-by-event hydrodynamics we show two curves, using either the 
entropy density (blue open circles) or the energy density weighted (red 
solid circles) average eccentricities to normalize the average final 
flow $\la v_n\ra$. For the ellipticity (Fig.~\ref{F11}) this choice is seen 
to make a difference only in rather central collisions ($b{\,<\,}4$\,fm), 
but for the triangularity the differences are significant out to average 
impact parameters probed in minimum bias samples, $b{\,\alt\,}8$\,fm. As 
stated earlier, we prefer the energy density weighted eccentricities 
(solid circles) as deformation measures because energy density and 
pressure are closely related through the EOS, and it is the pressure 
gradients (and their anisotropies) that drive the collective flow (and 
its anisotropies).

For the single-shot hydrodynamic simulations, a question arises as to
how exactly one should construct the ensemble-averaged smooth initial 
profile which is then evolved hydrodynamically. We have explored three 
reasonable procedures (variations of which have been used in the 
literature) and show them as dashed lines in Figs.~\ref{F11} and \ref{F12}. 
For the {\em lines labeled by stars}, we rotate the entropy density for 
each fluctuating event by the corresponding entropy-weighted 
participant-plane angle $\psi_n^\mathrm{PP}(s)$ ($n\eq2,3$, see 
Eq.~(\ref{eq17})),\footnote{Note that for computation of $\bar{v}_3$ 
  we rotate the events by a different angle before averaging than
  for $\bar{v}_2$, i.e. $\bar{v}_3$ and $\bar{v}_2$ are obtained from
  two different single-shot hydrodynamic runs, starting from different
  averaged initial energy density profiles.}
then average the rotated entropy profiles, compute the eccentricity 
$\bar{\ecc}_n(s)$ of the resulting average entropy density profile and 
convert it to energy density using the EOS for input into the 
hydrodynamic code. For the {\em lines labeled by crosses}, we rotate the 
energy density for each fluctuating event (obtained from the EOS) by the 
corresponding energy-weighted participant-plane angle $\psi_n^\mathrm{PP}(e)$
(see Eqs.~(\ref{eq1},\ref{eq2})), compute the averaged rotated energy 
density profile and its eccentricity $\bar{\ecc}_n(e)$, and use it 
directly as hydrodynamic input. For the {\em dashed lines without symbols}, 
finally, the averaged initial energy density (and therefore the final 
$\bar{v}_n$) are exactly the same as for the lines with crosses, but the 
final $\bar{v}_n$ is scaled by the entropy-weighted (rather than 
energy-weighted) eccentricity of the averaged initial profile, where the 
entropy density is obtained from the smooth averaged energy density via 
the EOS. 

%
\begin{figure*}
 \includegraphics[width=0.48\linewidth]{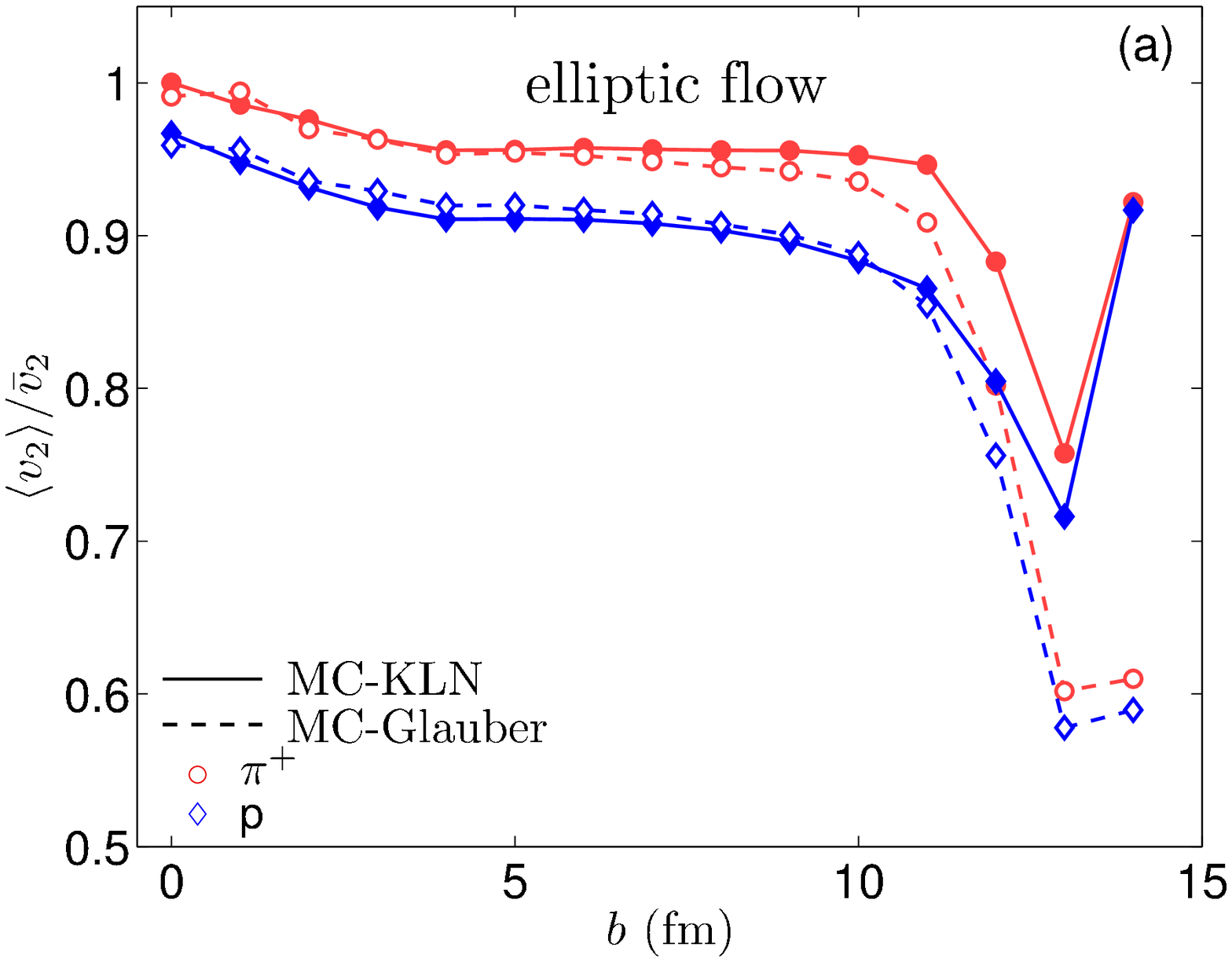}
 \includegraphics[width=0.48\linewidth]{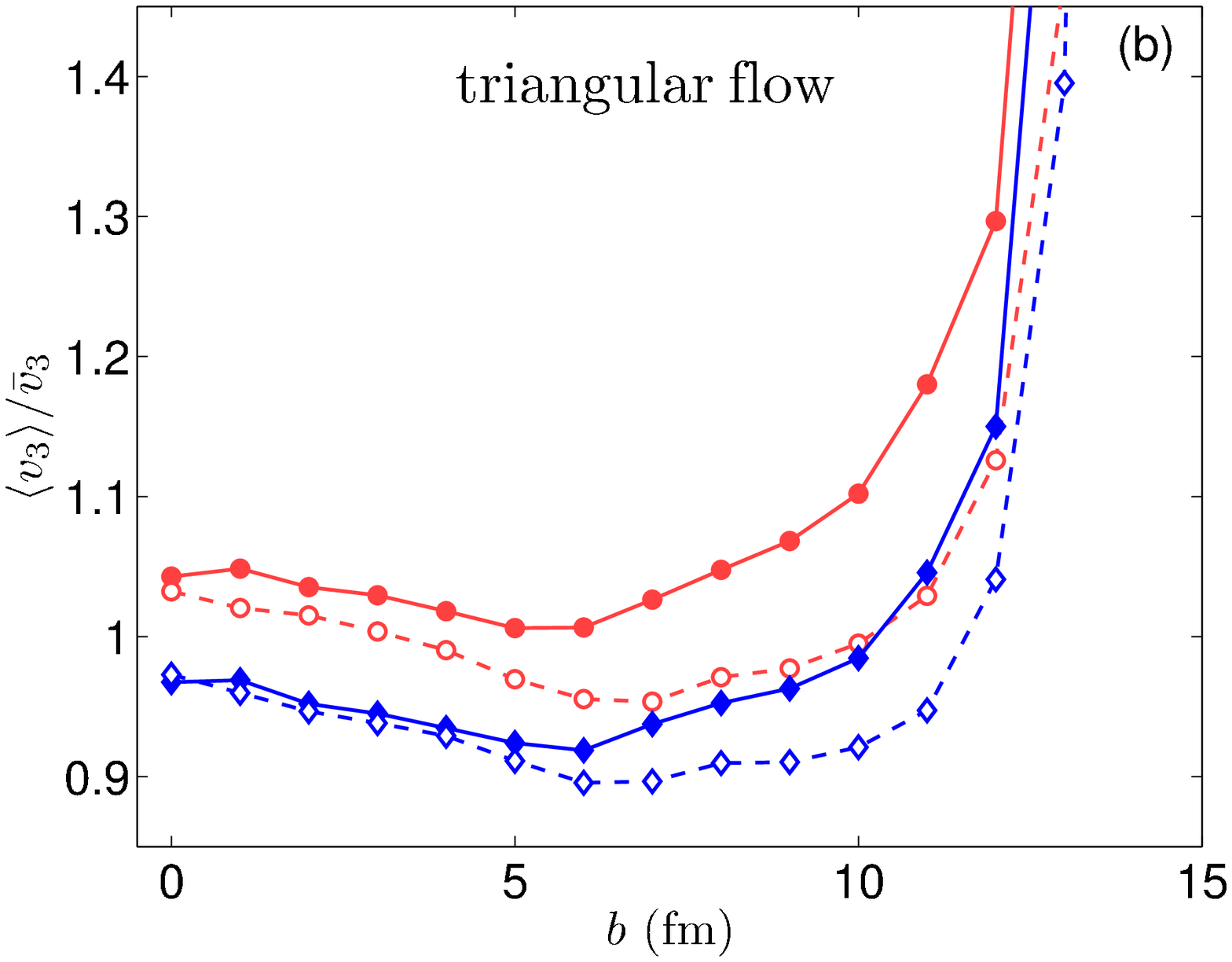}
 \caption{(Color online) Ratio of the average flow coefficient
  $\la v_n\ra$ from event-by-event hydrodynamics and the corresponding
  mean $\bar{v}_n$ from single-shot hydrodynamics, as a function of
  impact parameter in $200\,A$\,GeV Au+Au collisions, for $n\eq2$ (a)
  and $n\eq3$ (b). Shown are the ratios for directly emitted pions 
  (circles) and protons (diamonds) from fluctuating events using the 
  MC-Glauber (dashed lines) and MC-KLN models (solid lines). Average 
  events for computing $\bar{v}_n$ using single-shot hydrodyamics
  were obtained by rotating the energy density of each event by 
  $\psi_n^\mathrm{PP}(e)$ before superimposing them.
 \label{F13}
 }
\end{figure*}
%

The differences between the different dashed lines illustrate the 
uncertainties associated with the choice of averaging procedure for the 
initial state. Keeping in mind that a 20\% reduction in $v_2/\ecc_2$ 
corresponds (very roughly) to an increase of $\eta/s$ by $1/4\pi$ 
\cite{Song:2010mg}, one sees that these differences are not negligible 
if one aims for quantitative precision in the extraction of the specific 
shear viscosity. Comparing the three dashed lines we see that it doesn't 
make much difference whether we use the $s$-weighted or $e$-weighted
participant-plane angles to rotate the events before superimposing them
(the dashed lines without symbols and with stars are all very close to 
each other), but that in the more central collisions we obtain significantly
different values for the conversion efficiencies $\bar{v}_n/\bar{\ecc}_n$ 
if we normalize by $e$- or $s$-weighted mean eccentricities. Even though 
they look similar in Fig.~\ref{F5}a, at small impact parameters 
$\bar{\ecc}_\mathrm{part}(e)$ and $\la\ecc_2(e)\ra$ are larger than 
$\bar{\ecc}_\mathrm{part}(s)$ and $\la\ecc_2(s)\ra$, respectively, and 
this is the main reason why the red and blue lines in Fig.~\ref{F11} 
diverge at small $b$, for both event-by-event (solid lines) and 
single-shot hydrodynamics (dashed lines). 

An apples-to-apples comparison between event-by-event and single-shot 
hydrodynamics (and between theory and experimental data) therefore must 
ensure that the same (or at least conceptually compatible) eccentricities 
are used to normalize the anisotropic flow coefficients that are to be 
compared. In Figs.~\ref{F11},\,\ref{F12} we should therefore compare
blue solid with blue dashed, or red solid with red dashed lines, but
not curves of different colors. 

Even this is not good enough if one wants to accurately assess the relative 
space-to-momentum anisotropy conversion efficiency in single-shot and 
event-by-event hydrodynamics: in the single-shot hydro curves we use 
$\bar{\ecc}_\mathrm{part}$ to normalize the final elliptic flow, whereas 
the event-by-event hydro results were normalized with 
$\la\ecc_2\ra{\,\equiv\,}\la\ecc_\mathrm{part}\ra$. While each of these 
eccentricity measures makes perfect sense in its own context, they differ 
at large impact parameters, $\bar{\ecc}_\mathrm{part}$ being larger (see 
Figs.~\ref{F1}a,b). To avoid this problem we have added in Figs.~\ref{F11} 
and \ref{F12} an additional ``mixed ratio'' (dash-dotted purple line) which 
normalizes the ensemble-averaged anisotropic flow $\la v_n\ra$ ($n\eq2,3$) 
from event-by-event hydrodynamics (used in the ratio $\la v_n\ra/\la\ecc_n\ra$ 
denoted by solid lines with solid red circles) by the mean $e$-weighted
eccentricity $\bar{\ecc}_n$ from single-shot hydrodynamics (used in the 
ratio $\bar{v}_n/\bar{\ecc}_n$ denoted by dashed lines with crosses). This
dot-dashed purple line agrees almost perfectly with the solid red line
with circles over most of the impact parameter range, except for peripheral
collisions with $b{\,\agt\,}10$\,fm where $\bar{\ecc}_n$ and $\la\ecc_n\ra$
begin to diverge. The red dashed lines with crosses and purple dash-dotted 
lines show the anisotropic flows from single-shot and event-by-event 
hydrodynamics normalized by the {\em same} eccentricity measure 
characterizing the fluctuating event sample. Their comparison allows 
an unambiguous assessment of the different efficiencies of single-shot 
and event-by-event hydrodynamics in converting initial eccentricities 
to final momentum anisotropies. Their ratio is shown in Fig.~\ref{F13}.

From Fig.\,\ref{F13}a one concludes that, for ideal hydrodynamics,
event-by-event fluctuations on average reduce the efficiency of the 
fluid in converting initial source ellipticity into elliptic flow.
Over most of the centrality range this reduction is about 4\% for pions
and about twice as large for protons, and it is similar for MC-KLN
and MC-Glauber initial profiles. In very central collisions the ratio
of conversion efficiencies for event-by-event vs. single-shot hydrodynamics
is closer to 1, but it degrades strongly in very peripheral collisions
where event-by-event evolution generates on average $30{-}40\%$ less
elliptic flow than single-shot hydrodynamics. The generic tendency of 
event-by-event hydrodynamic evolution of fluctuating initial profiles 
to generate less elliptic flow than expected from hydrodynamic 
evolution of the corresponding smooth average profile has been observed 
before \cite{Andrade:2006yh,Schenke:2010rr}; our systematic study in 
Fig.~\ref{F13}a quantifies this effect over the full range of collision 
centralities. 
 
The situation with triangular flow, shown in Fig.~\ref{F13}b, is quite 
different: event-by-event propagation of initial-state flutuations can
lead to an increase or decrease of the triangular flow compared to
single-shot hydrodynamics, depending on particle mass (pions or protons), 
the nature of the fluctuations (MC-Glauber or MC-KLN), and collision 
centrality. Contrary to elliptic flow, in peripheral collisions 
event-by-event evolution leads to significantly {\em larger} average 
triangular flow than single-shot hydrodynamics. 

We expect that non-zero viscosity will dampen fluctuation effects and 
somewhat reduce the differences between event-by-event and single-shot
hydrodynamic evolution of elliptic and triangular flow shown in 
Fig.~\ref{F13}. Nevertheless it appears that, for quantitative studies
of the influence of viscosity on the generation of anisotropic collective 
flow, event-by-event hydrodynamic evolution is an essential and indispensable 
ingredient. 

\vspace*{-3mm}
\subsection{Elliptic flow fluctuations}
\label{sec5c}
\vspace*{-3mm}

%
\begin{figure*}
 \includegraphics[width=0.42\linewidth]{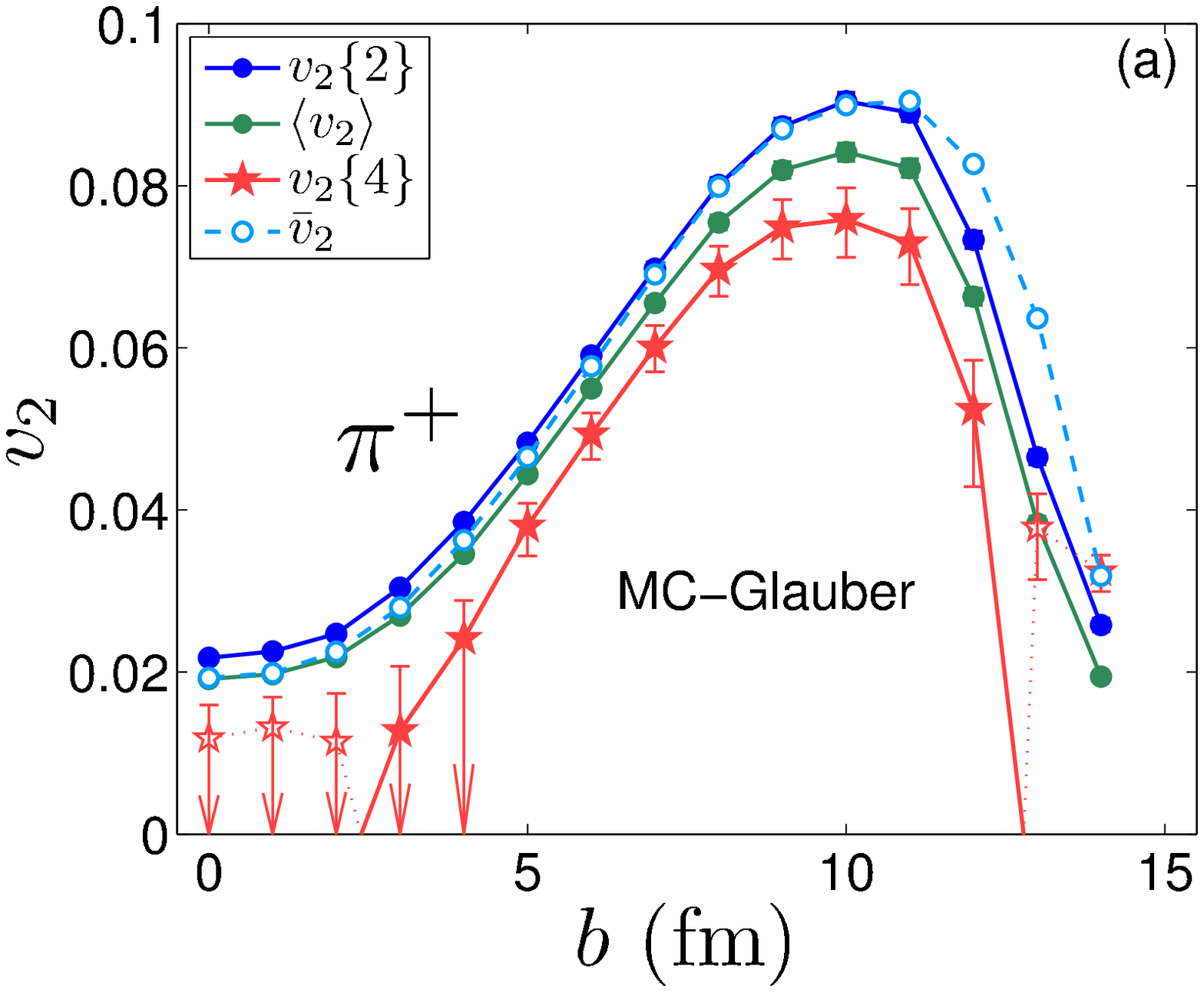}
 \includegraphics[width=0.42\linewidth]{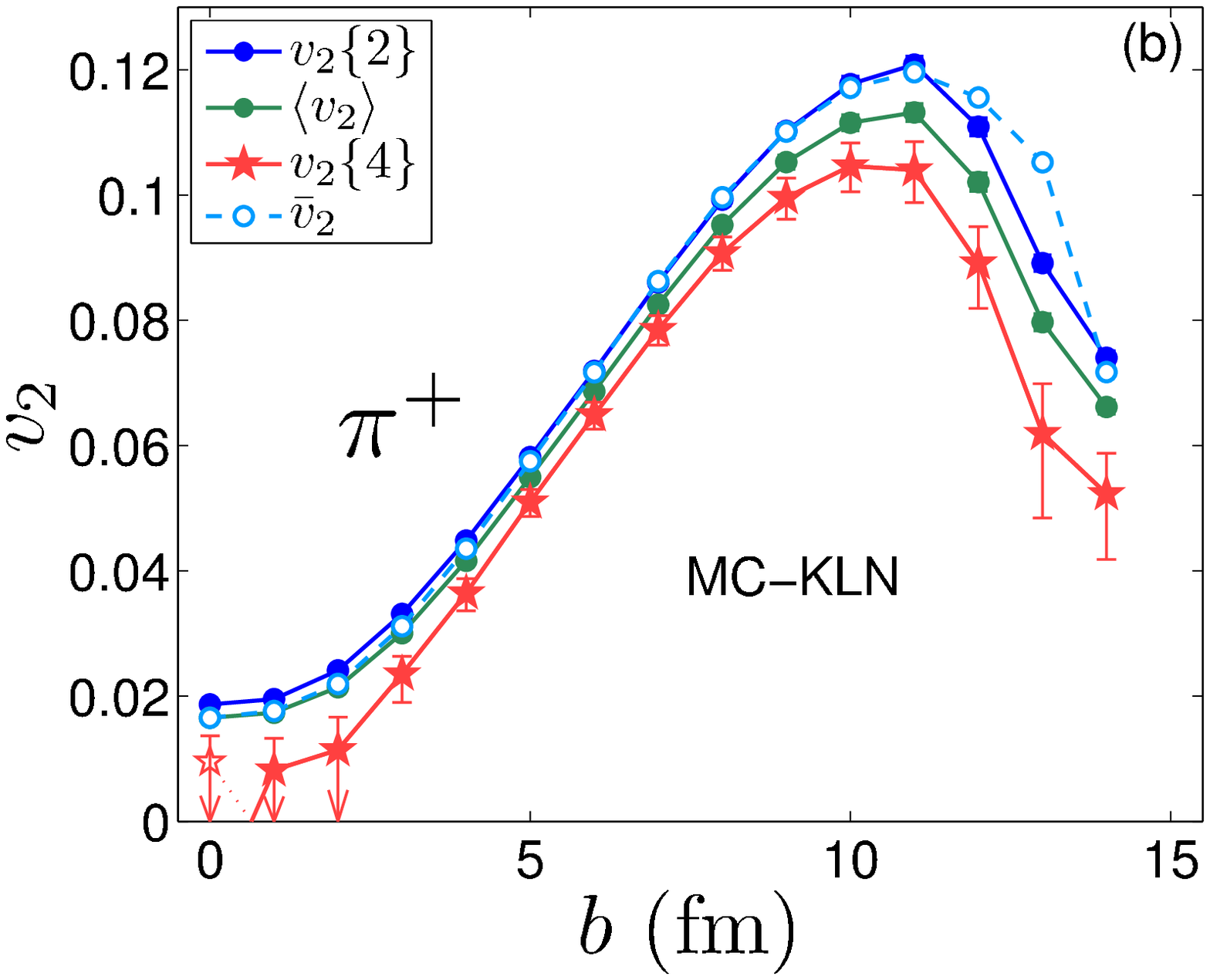}\\
 \includegraphics[width=0.42\linewidth]{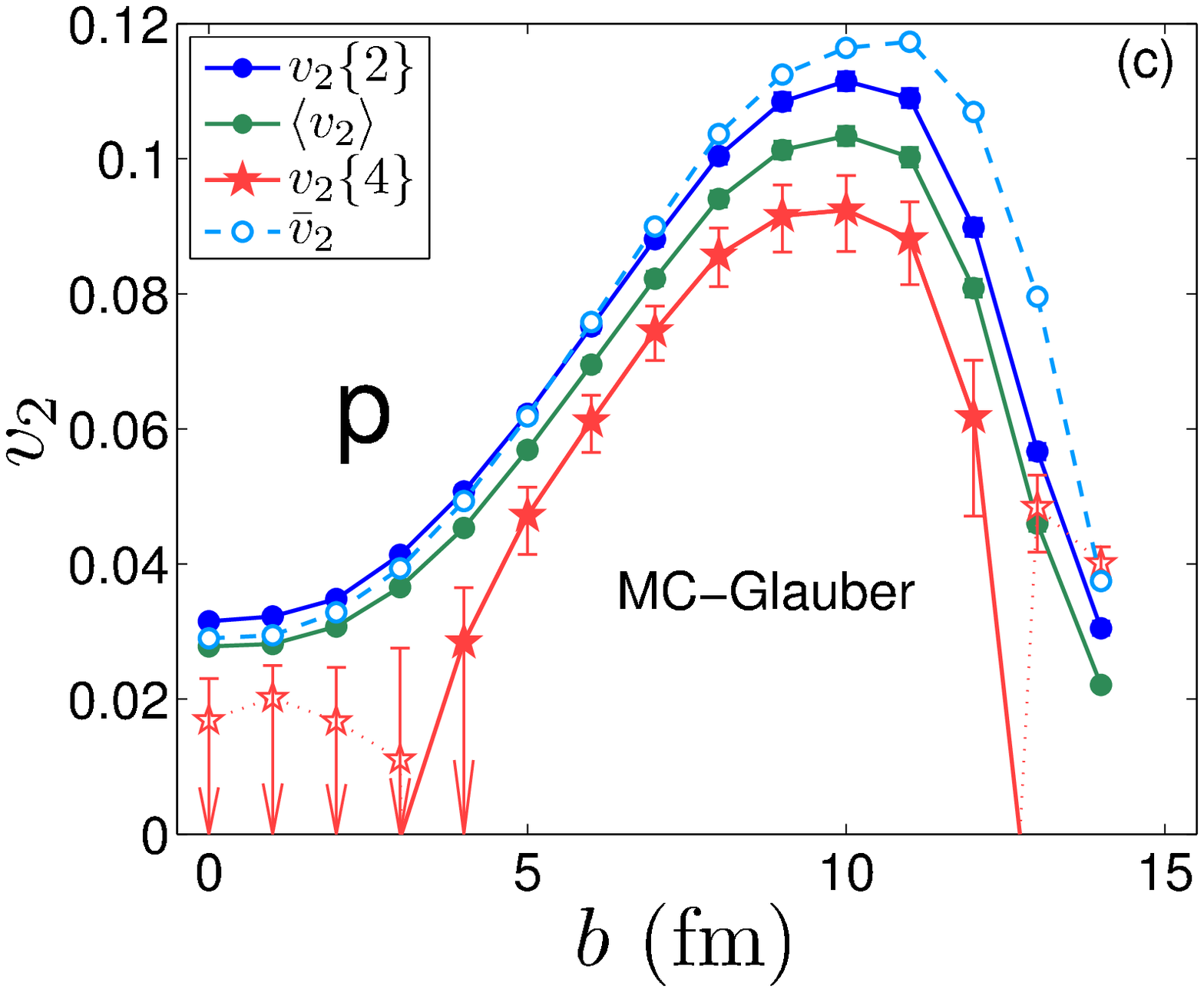}
 \includegraphics[width=0.42\linewidth]{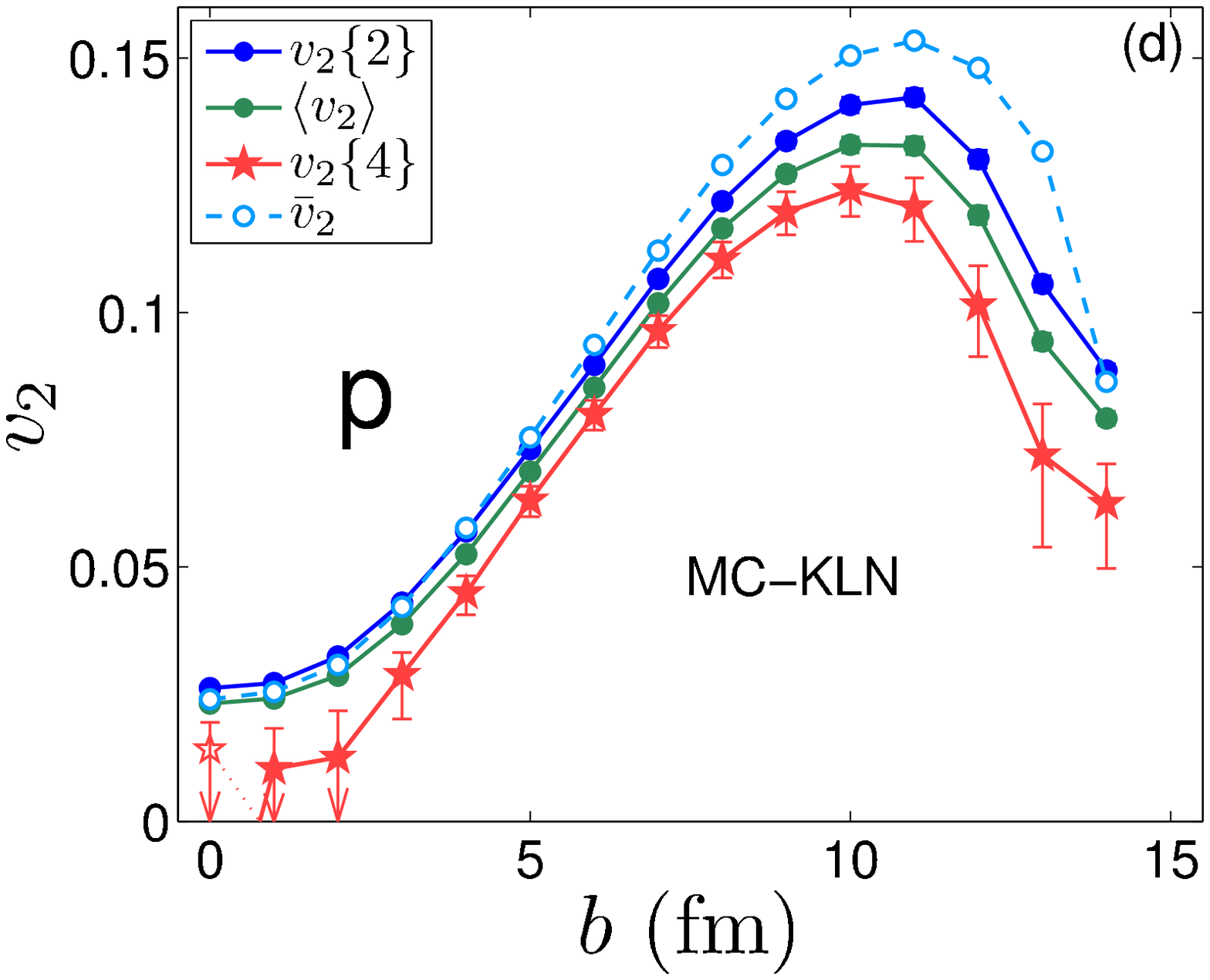}
 \caption{(Color online) Different measures for the final elliptic
    flow $v_2$ (similar to Fig.~\ref{F1}) for directly emitted pions 
    (a,b) and protons (c,d) as functions of impact parameter from 
    event-by-event ideal fluid dynamics, using MC-Glauber (a,c) and 
    MC-KLN (b,d) initial conditions for $200\,A$\,GeV Au+Au collisions.
 \label{F14}
 }
\end{figure*}
%

Similar to what is shown in Figs.~\ref{F1}a,b for the initial source 
ellipticities, Fig.~\ref{F14} shows the elliptic flow measures $\la v_2\ra$, 
$v_2\{2\}$ and $v_2\{4\}$ from event-by-event hydrodynamics, together with
$\bar{v}_2$ from single-shot hydrodynamic evolution of the corresponding
averaged initial profile, for pions and protons, using MC-Glauber and
MC-KLN initializations, respectively. $v_2\{2\}$ and $v_2\{4\}$ are defined
in analogy to Eqs.~(\ref{eq9},\ref{eq10}) by
\begin{eqnarray}
 \label{eq23}
 v_2\{2\}^2 &=& \la v_2^2 \ra,
\\
 \label{eq24}
 v_2\{4\}^4 &=& 2\la v_2^2\ra^2 - \la v_2^4\ra.
\end{eqnarray}
Here $v_2$ is calculated event-by-event via Eq.~(\ref{eq20}) from the 
Cooper-Frye spectrum at freeze-out (with zero statistical uncertainties
since it is determined with mathematical precision by the event-by-event 
hydrodynamic output). 

As in Fig.~\ref{F1}, open stars show the central values for 
$\sqrt[4]{|v_2\{4\}^4|}$ whenever $v_2\{4\}^4$ turns negative, and 
open-ended error bars indicate that the error band for $v_2\{4\}^4$ 
ranges from positive to negative values. Similar to the ellipticities
shown in Fig.~\ref{F1}, the latter happens at small impact parameters,
but for the MC-Glauber model the $b$-range over which this happens
for $v_2\{4\}$ (for both pions and protons) is somewhat larger than 
for $\ecc\{4\}$. Still, $v_2\{4\}$ is compatible with zero over this
entire range, and we do not find statistically significant negative 
values for $v_2\{4\}$ at small impact parameters. At large $b{\,>\,}12$\,fm
$v_2\{4\}^4$ turns negative for both pions and protons when we use 
MC-Glauber initial conditions whereas it remains positive for MC-KLN 
initial profiles.   

By comparing $\bar{v}_2$ (open circles in Fig.~\ref{F14}) with 
$\la v_2\ra$ (solid green circles) one sees that in mid-central to 
peripheral collisions the $v_2$-suppression from event-by-event 
hydrodynamic evolution is of the same order as or (especially for 
protons) even larger than the difference between $v_2\{2\}$ and 
$\la v_2\ra$ (solid blue vs. solid green circles) that arises from 
event-by-event flow fluctuations. As a result, $v_2\{2\}$ from 
event-by-event hydrodynamics lies in peripheral collisions even 
below $\bar{v}_2$ from single-shot hydrodynamics, in spite of its 
fluctuation-induced enhancement.

\begin{figure}[h!]
 \includegraphics[width=0.9\linewidth]{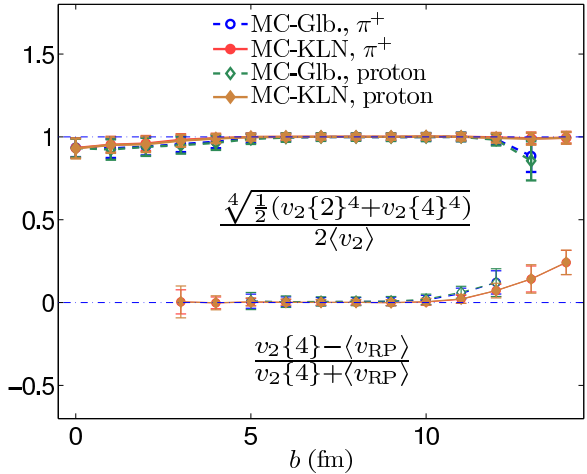}
 \caption{(Color online) Similar to Fig.~\ref{F3}, but for the
    elliptic flow $v_2$ of pions and protons. See text for discussion.
 \label{F15}
 }
\end{figure}

Similar to Eqs.~(\ref{eq11})-(\ref{eq11b}) we can test whether the 
$v_2$ fluctuations from event to event have Gaussian or Bessel-Gaussian
distributions. This is done in Fig.~\ref{F15}. The upper set of curves
(thick lines) test the $v_2$-analogue of relation (\ref{eq11b}) whereas
the lower set (thin lines) tests the validity of Eq.~(\ref{eq11}). (In 
the lower set of curves we dropped all $b$-values for which the error 
band for $v_2\{4\}^4$ extends ton negative values.) Just as we saw for 
the initial ellipticities in Fig.~\ref{F3}, both the Gaussian and
Bessel-Gaussian hypotheses for $v_2$-fluctuations are seen to hold quite 
well in mid-central ($4{\,\alt\,}b{\,\alt\,}10$\,fm) collisions. The 
Bessel-Gaussian hypothesis breaks down in peripheral collisions
($b{\,>\,}10$\,fm). Whether it holds (as expected \cite{Voloshin:2007pc})
in central collisions is a question that, with our present statistics of
1000 events per $b$-value, we can not reliably answer, but we see no
indications for the opposite. The assumption of Gaussian $v_2$-fluctuations
breaks down in central collisions ($b{\,<\,}5$\,fm), as expected.
For the MC-Glauber model it also breaks down in very peripheral collisions,
whereas for MC-KLN initial conditions the final elliptic flow exhibits
a nice Gaussian distribution all the way to the largest impact parameters.

Overall, a comparison of Figs.~\ref{F15} and \ref{F3} (as well as of
Figs.~\ref{F14} and \ref{F1}) shows that the statistical properties of 
$v_2$ fluctuations are qualitatively similar but quantitatively different 
from those of the initial ellipticity fluctuations. This is consistent 
%
%
with the fact that the main driver for elliptic flow is the initial
ellipticity, but that eccentricity coefficients of higher harmonic orders 
affect the evolution of $v_2$ weakly but measurably through non-linear 
mode-coupling effects.

\vspace*{-3mm}
\section{Summary and conclusions}
\label{sec6}
\vspace*{-2mm}

In this work we presented a comprehensive analysis of event-by-event
shape fluctuations in the initial state and flow fluctuations in the final
state of relativistic heavy-ion collisions, as quantified by the first
four non-trivial harmonic eccentricity and flow coefficients, $\ecc_n$ 
and $v_n$ ($n\eq2,3,4,5$). Using the MC-Glauber and MC-KLN models to
generate fluctuating initial entropy and energy density profiles, we
explored the centrality dependence of a number of different variants 
of these anisotropy measures that are being used by practitioners in the
field, and compared them with each other. Although they all exhibit similar 
qualitative behaviour, quantitative differences exist and must be carefully
taken into account in the theoretical analysis of experimental data. As 
far as we know, ours is the first comprehensive analysis quantifying these 
differences for both the Glauber and Color Glass Condensate models.

\bigskip

\noindent
We list a few key results:

\noindent
-- The average and mean ellipticities $\la \ecc_2\ra$ and $\bar{\ecc}_2$ 
agree with excellent accuracy over a wide range of impact parameters,
but diverge in very peripheral collisions ($\geq60\%$ centrality) where
$\bar{\ecc}_2{\,>\,}\la \ecc_2\ra$ (both for participant-plane and 
reaction-plane averaged profiles). 

\noindent
-- The average energy and entropy density weighted eccentricities agree 
with excellent accuracy over a wide range of impact parameters, except 
for central collisions ($b{\,\alt\,}4$\,fm) where 
$\la\ecc_n(e)\ra{\,>\,}\la\ecc_n(s)\ra$.

\noindent
-- Whether the fluctuating entropy density distributions for individual 
events are first converted to energy density and then rotated by 
$\psi_n^\mathrm{PP}(e)$ and averaged, or first rotated by 
$\psi_n^\mathrm{PP}(s)$ and averaged and then converted to energy density
has very little influence on the shape of the resulting smooth average 
initial energy density profile for single-shot hydrodynamics. We prefer
(and propose as standard procedure) the conversion to energy density as 
the first step, since in event-by-event hydrodynamics the energy density
gradients of each event generate (through the EOS) the pressure gradients 
that drive the evolution of collective flow.

\noindent
-- The shortcut of using reaction-plane averaging to generate a smooth
profile for single-shot hydrodynamics with ellipticity approximately equal 
to $\ecc\{4\}$ of the ensemble, in the hope of generating with a single 
hydrodynamic run an elliptic flow $\bar{v}_2$ that can be directly 
compared with $v_2\{4\}$ measurements, works only in the $0{-}40\%$ 
centrality range. For peripheral collisions this method cannot be trusted.   

\noindent
-- The assumption of Bessel-Gaussian fluctuations for initial source 
ellipticity and final elliptic flow work well for $b{\,\alt\,}10$\,fm 
but breaks down in more peripheral collisions. For more peripheral 
collisions the hypothesis that $\ecc_2$ and $v_2$ are Gaussian distributed
works better than the Bessel-Gaussian assumption, but it breaks down
for $b{\,<\,}5$\,fm. For MC-Glauber initial conditions, directly emitted 
pions and protons feature negative values of $v_2\{4\}^4$ in very 
peripheral collisions. The fluctuations of initial source ellipticities 
and final elliptic flow values have qualitatively similar but
quantitatively different statistical properties.

\noindent
-- Except for rather central collisions, the eccentricities $\la\ecc_2\ra$, 
$\la\ecc_4\ra$ and $\la\ecc_5\ra$ from the MC-KLN model are all significantly
larger than those from the MC-Glauber model. In contrast, $\la\ecc_3\ra$
is numerically very similar for the two models over most of the impact 
parameter range. The viscous suppression of triangular flow $v_3$ may thus 
allow for a determination of the QGP shear viscosity $(\eta/s)_\mathrm{QGP}$ 
that is free from the large model uncertainties that arise from the different 
MC-Glauber and MC-KLN ellipticities when using $v_2$ for such an extraction
\cite{new3}.

\noindent
-- The second and fourth order eccentricities $\ecc_2$ and $\ecc_4$
are strongly correlated by collision geometry, and $v_4$ receives strong
contributions even from a purely elliptical deformation of the final flow 
velocity distribution. These complications make $v_4$ a poor candidate for
systematic studies of viscous effects on the evolution of collective
flow. Similar comments apply to $v_5$ since it couples via mode-coupling
to triangularity from fluctuations and to ellipticity from collision 
geometry. This mixture of contributions from conceptually different 
origins complicates a systematic analysis. In general, flow coefficients
$v_n$ of high harmonic order ($n{\,>\,}3$) show poor correlation with the
eccentricity coefficients $\ecc_n$ of the same harmonic order, except
for very central collisions where all eccentricities are driven by
fluctuations alone (and not by overlap geometry).   
 
\noindent
-- In spite of non-linear mode-coupling effects, the basic response of
elliptic flow $v_2$ to ellipticity $\ecc_2$, and of triangular flow $v_3$
to triangularity $\ecc_3$, is approximately linear. These two observables 
thus remain prime candidates for systematic studies of viscous effects 
on collective hydrodynamic flow.

\noindent
-- Event-by-event hydrodynamics generates harder $p_T$-spectra for
the emitted hadrons than single-shot hydrodynamic evolution of the
corresponding averaged initial profile. This is due to additional
radial flow generated by large pressure gradients arising from ``hot
spots'' in the initial fluctuating density distribution. The hardening
effect is particularly strong in peripheral collisions which produce
small fireballs that fluctuate strongly.  

\noindent
-- Event-by-event hydrodynamic evolution of fluctuating initial conditions
leads to smaller average elliptic flow than obtained by evolving the
corresponding averaged initial condition in a single shot. This 
suppression depends somewhat on collision centrality, and for ideal fluids 
it is generically of order 4-5\% for pions and 8-10\% for protons. The 
effect is sufficiently large to lead to a significant over-estimate of 
the fluid's specific shear viscosity if one extracts it from elliptic 
flow measurements by comparing with single-shot hydrodynamic simulations. 
Even though we expect the discrepancy between event-by-event and 
single-shot hydrodynamics to decrease a bit in viscous fluid dynamics, 
we believe that a quantitatively precise experimental determination of 
$\eta/s$ from collective flow data will require comparison with 
event-by-event hydrodynamical calculations.

%
\begin{figure*}
 \includegraphics[width=0.32\linewidth]{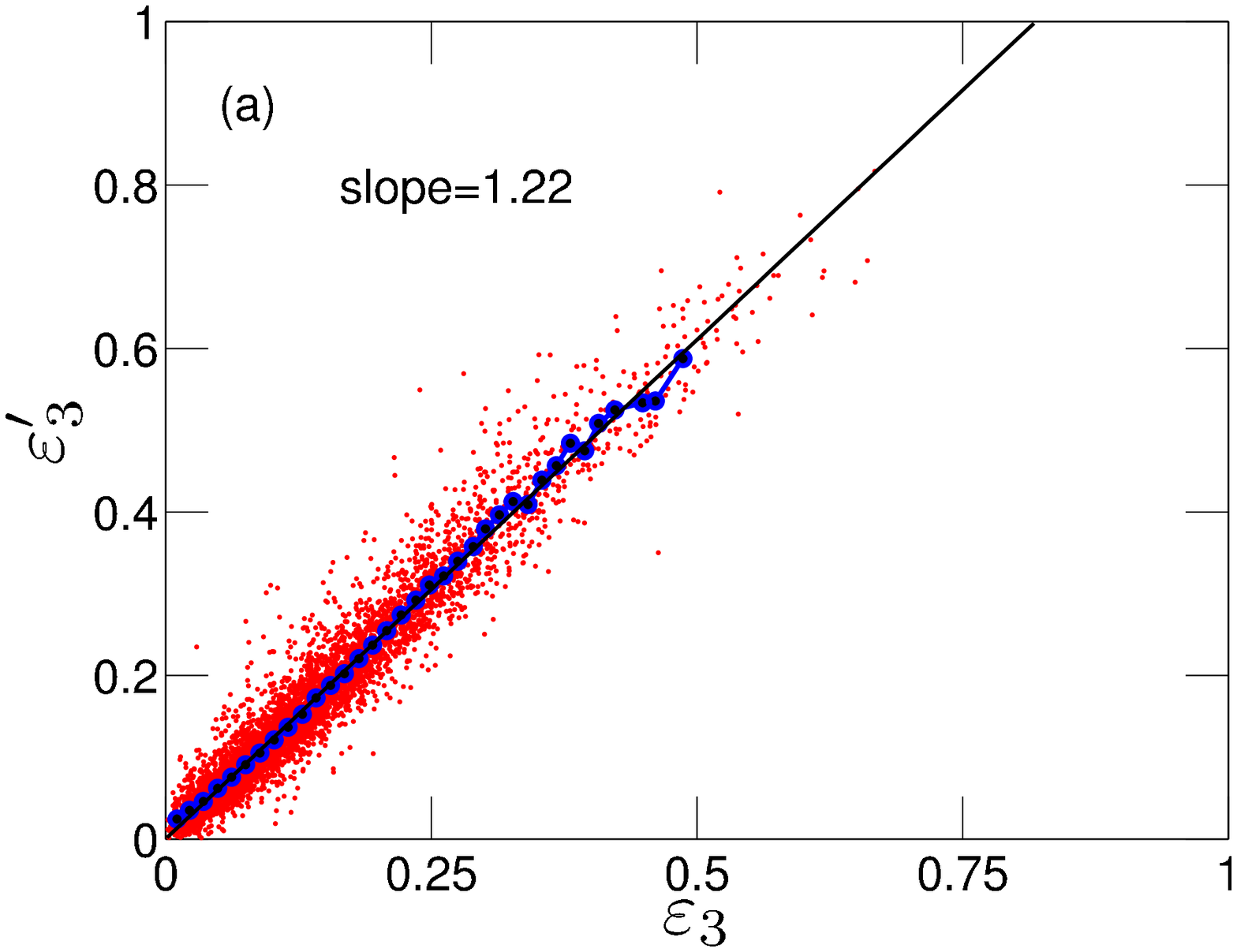}
 \includegraphics[width=0.32\linewidth]{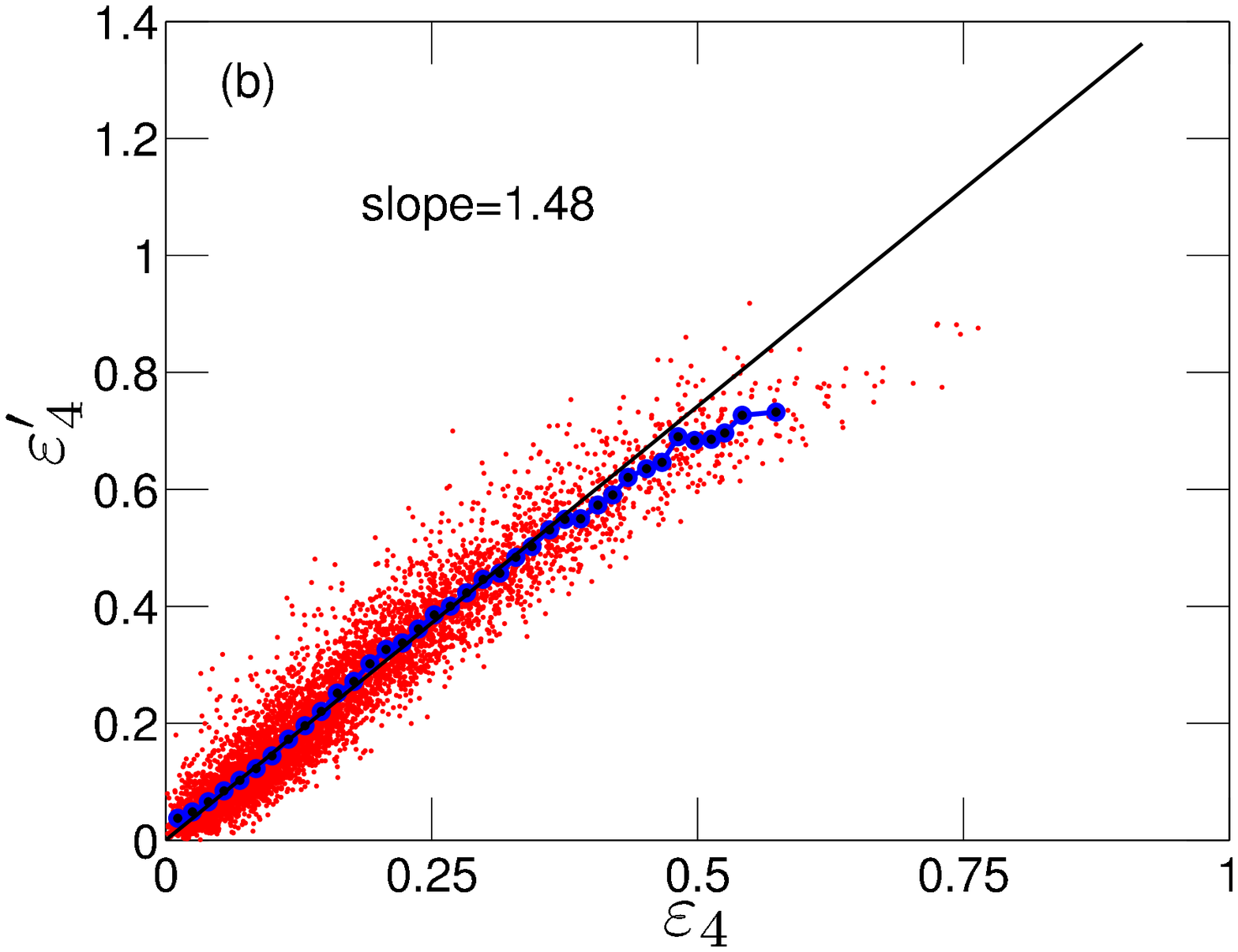}
 \includegraphics[width=0.32\linewidth]{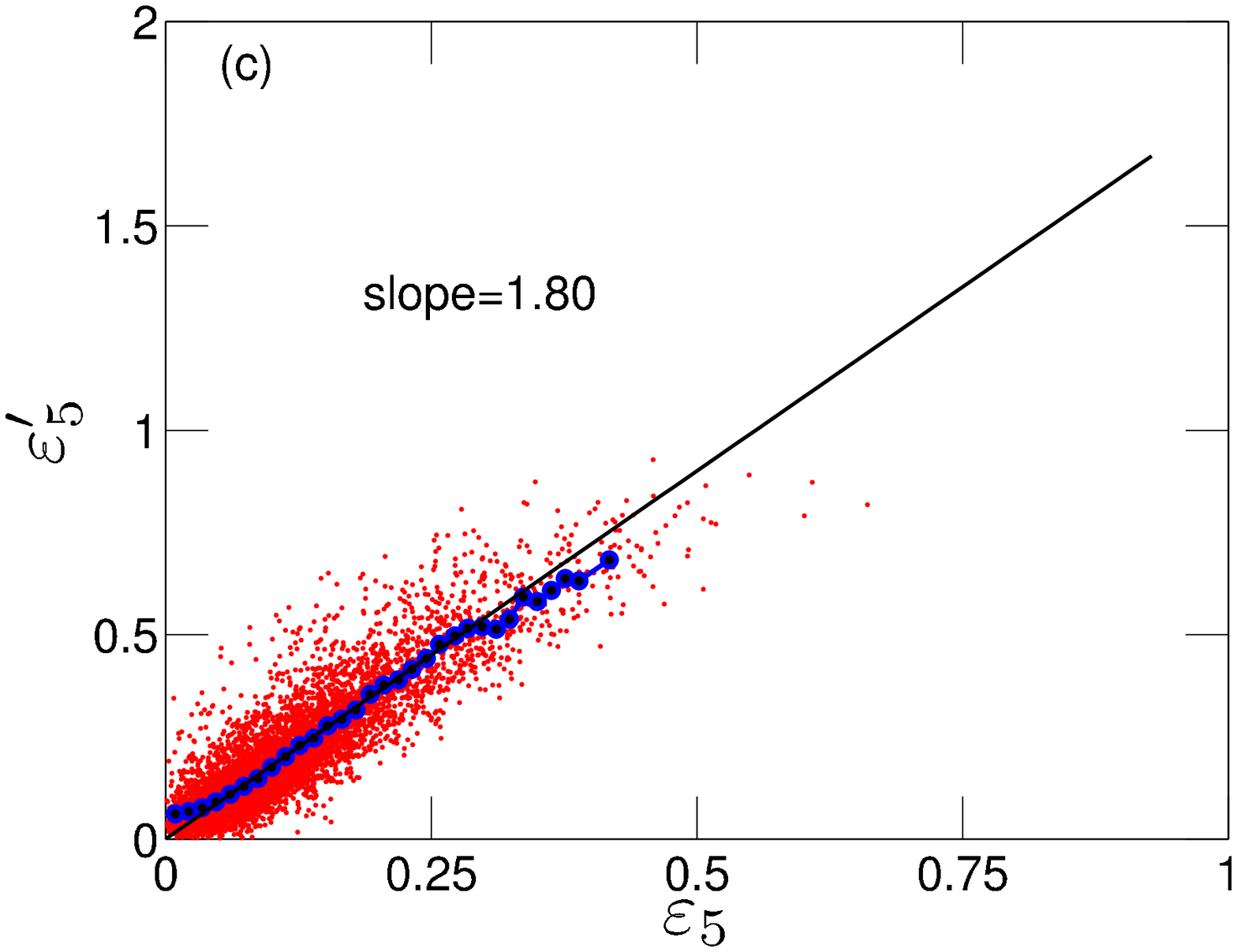}
 \caption{(Color online) Correlation between $\ecc_n$ and $\ecc'_n$,
   for $n\eq2,\,3,\,4$ (panels (a-c)). The blue dots are bin averages
   for bins that contain more than 10 events. The thick black lines are 
   linear fits.
 \label{F16}
 }
\end{figure*}
%
%
\begin{figure*}
 \includegraphics[width=0.32\linewidth]{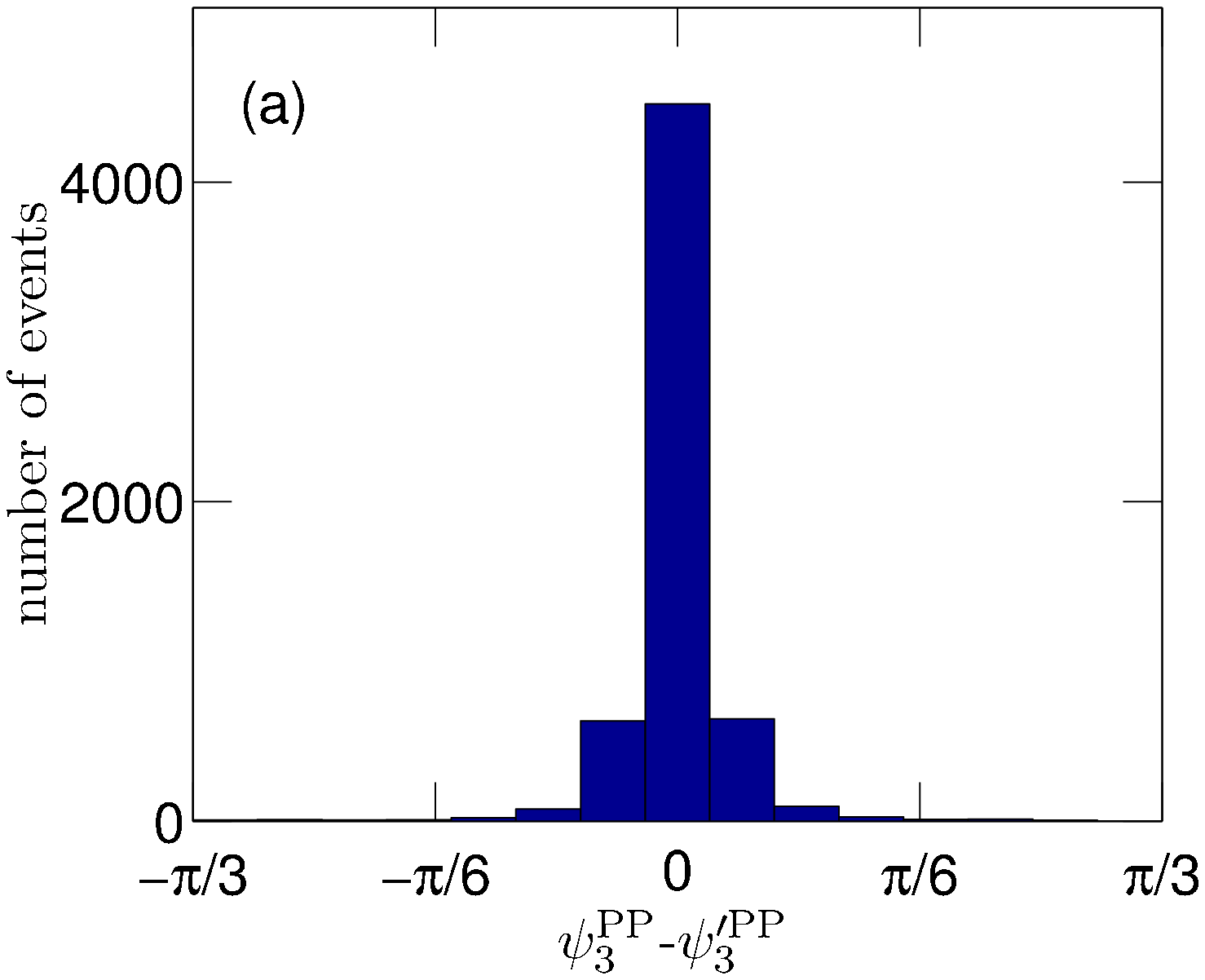}
 \includegraphics[width=0.32\linewidth]{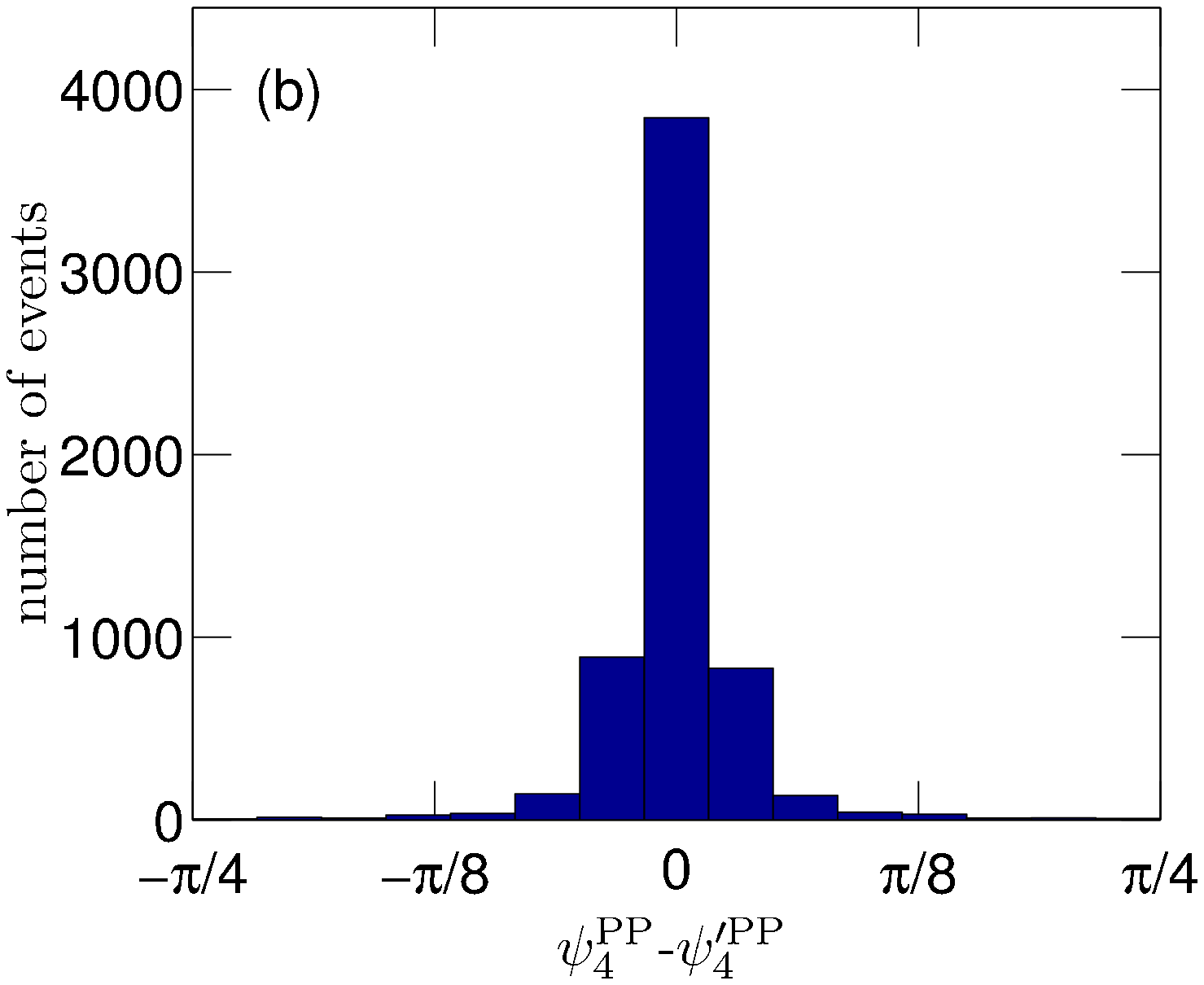}
 \includegraphics[width=0.32\linewidth]{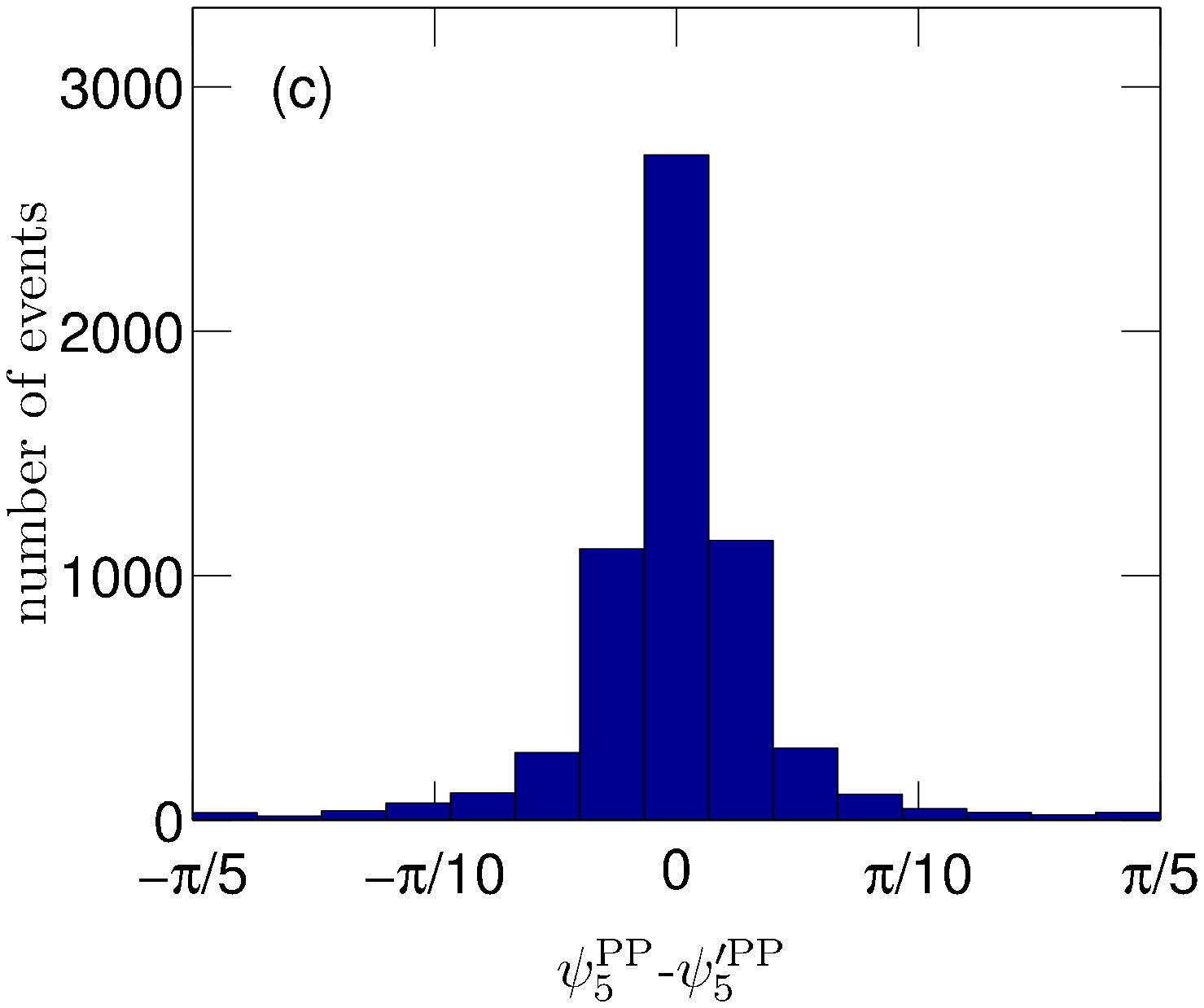}
 \caption{(Color online) Event-by-event correlation between the participant 
    plane angles associated with $r^2$- and $r^n$-weighted eccentricities
    for harmonic orders $n\eq3$, 4, and 5 (panels (a)-(c)).
 \label{F17}
 }
\end{figure*}
%

\acknowledgments{We gratefully acknowledge fruitful discussions with
and valuable comments from Mike Lisa, Matt Luzum, Jean-Yves Ollitrault, 
Art Poskanzer, Chun Shen, Raymond Snellings, Huichao Song, Sergei Voloshin, 
and especially Derek Teaney whose question at a recent workshop prompted 
the study of single-particle spectra presented in Sec.~\ref{sec5a}. 
Special thanks go to Matt Luzum and Art Poskanzer for pointing out
logical errors in our analysis as presented in the first version of this 
paper. This work was supported by the U.S.\ Department of Energy under 
Grants No.~\rm{DE-SC0004286} and (within the framework of the JET 
Collaboration) \rm{DE-SC0004104}.}

\appendix

\section{Comparison between eccentricities defined with 
         $\bm{r^2}$ and $\bm{r^n}$ weights}
\label{appa}

We here present a brief comparison between the $r^2$-weighted eccentricity 
coefficients $\ecc_n$ (Eq.~(\ref{eq15})) and the $r^n$-weighted $\ecc'_n$ 
(Eq.~(\ref{eq15})), as well as their associated angles $\psi_n^\mathrm{PP}$ 
and $\psi_n^\mathrm{'PP}$. Fig.~\ref{F16}
shows a scatter plot of $\ecc'_n$ vs. $\ecc_n$ for $n\eq3,4,5$. One observes
approximate proportionality ($\ecc'_3{\,\approx\,}1.22\,\ecc_3$, 
$\ecc'_4{\,\approx\,}1.48\,\ecc_4$, $\ecc'_5{\,\approx\,}1.80\,\ecc_5$)
over most of the eccentricity range, with slopes that increase with $n$.
So where Fig.~\ref{F5} shows a decrease of $\ecc_n$ with increasing $n$
at large impact parameters, the same is not true for the $\ecc'_n$
\cite{Qin:2010pf}. On the other hand, the linear relations between 
$\ecc'_n$ vs. $\ecc_n$ imply that the relations between $v_n$ and 
$\ecc'_n$ will look qualitatively the same as those between $v_n$ and
$\ecc_n$ in Fig.~\ref{F9}, with appropriately rescaled horizontal axes.

At the same time the participant plane angles associated with 
$r^2$-weighted and $r^n$-weighted eccentricities are tightly correlated, 
as shown in Fig.~\ref{F17}. 
For given $n$, the angles $\psi_n^\mathrm{PP}$ and $\psi_n^\mathrm{'PP}$
fluctuate around each other, with a variance that increases with 
$n$, on account of the decreasing values of $\ecc_n$. From a practical
point of view, we therefore consider both definitions as equivalent,
and choosing between them is a matter of personal preference.


\end{document}